\documentclass[preprint,12pt,3p]{elsarticle}

\usepackage{amssymb}
\usepackage{amsmath}
\usepackage{mathtools}
\usepackage{breqn}
\usepackage{multirow}
\usepackage{tablefootnote}
\usepackage{footnote}
\usepackage{caption}
\usepackage{subcaption}
\usepackage{array}
\usepackage{tabularx}
\usepackage{diagbox}
\usepackage{longtable}
\usepackage{float}
\usepackage[euler]{textgreek}
\usepackage{xcolor}
\usepackage{changepage}
\usepackage{tikz}
\usepackage{wasysym}
\usepackage{verbatim}

\journal{Journal of ...}
\usepackage{graphics}
\usepackage{booktabs}
\usepackage{siunitx}
\usepackage{tikz}
\usepackage[hyphens]{url}
\usepackage[hidelinks]{hyperref}
\hypersetup{breaklinks=true}

\usepackage{pifont}

\newcommand*{\priority}[1]{\begin{tikzpicture}[scale=0.15]%
    \draw (0,0) circle (1);
    \fill[fill opacity=0.5,fill=blue] (0,0) -- (90:1) arc (90:90-#1*3.6:1) -- cycle;
    \end{tikzpicture}}

\newcommand{\yes}{\checkmark}
\newcommand{\no}{\ding{55}}

\newcommand{\bigCI}{\mathrel{\text{\scalebox{1.07}{$\perp\mkern-10mu\perp$}}}}

\def\halfcheckmark{\tikz\draw[scale=0.4,fill=black](0,.35) -- (.25,0) -- (1,.7) -- (.25,.15) -- cycle (0.75,0.2) -- (0.77,0.2)  -- (0.6,0.7) -- cycle;}

\begin{document}

\begin{frontmatter}

  \title{Lightweight Blockchain Solutions: Taxonomy, Research Progress, and Comprehensive Review} 

\author[label1]{Khaleel Mershad}
\address[label1]{Department of Computer Science and Mathematics, School of Arts and Sciences, Lebanese American University (LAU), Beirut, Lebanon}
\ead{khaleel.mershad@lau.edu.lb}

\author[label2,labelMA]{Omar Cheikhrouhou\corref{cor1}}
\address[label2]{CES Lab, National School of Engineers of Sfax, University of Sfax, Sfax 3038, Tunisia.}
\address[labelMA]{Higher Institute of Computer Science of Mahdia, University of Monastir, Mahdia 5111,  Tunisia.}
\ead{omar.cheikhrouhou@isetsf.rnu.tn}

\cortext[cor1]{corresponding author}

\begin{abstract}
The proliferation of resource-constrained devices has become prevalent across various digital applications, including smart homes, smart healthcare, and smart transportation, among others. However, the integration of these devices brings many security issues. To address these concerns, Blockchain technology has been widely adopted due to its robust security characteristics, including immutability, cryptography, and distributed consensus. However, implementing blockchain within these networks is highly challenging due to the limited resources of the employed devices and the resource-intensive requirements of the blockchain. To overcome these challenges, a multitude of researchers have proposed lightweight blockchain solutions specifically designed for resource-constrained networks. 
In this paper, we present a taxonomy of lightweight blockchain solutions proposed in the literature. More precisely, we identify five areas within the “lightweight” concept, namely,  blockchain architecture, device authentication, cryptography model, consensus algorithm, and storage method.
We discuss the various methods employed in each “lightweight” category, highlighting existing gaps and identifying areas for improvement. Our review highlights the missing points in existing systems and paves the way to building a complete lightweight blockchain solution for networks of resource-constrained devices. 

\end{abstract}


\end{frontmatter}


\section{Introduction}

The number and types of networks that are created by connecting small and smart devices have increased significantly. Many of these networks fall under the Internet of Things (IoT) umbrella, such as those in home automation, healthcare, smart grid, etc. However, other types of networks also contain resource-constrained devices, such as fog/edge devices in cloud networks, onboard units in vehicular networks, etc. 
The proliferation of these networks was facilitated by the advancement of modern communications technologies such as 5G, computing paradigms such as cloud and edge computing, and evolutionary tools such as artificial intelligence \cite{zaabar2022intrusion}. These networks usually comprise small devices that are resource-constrained in terms of energy consumption, storage capacity, computational power, and battery lifetime \cite{cheikhrouhou2023lightweight, mershad2023blockchain}. 

Due to their small sizes, these devices can be deployed anywhere and managed remotely, or they can be mounted on machines such as wearables, vehicles, equipment, drones, etc. While this fact is considered one of the main advantages of these devices, it leaves them  susceptible to many physical and cyber threats. In addition, since these devices are mostly wireless, they are vulnerable to attackers who perform eavesdropping or 
man-in-the-middle attacks \cite{zaabar2022intrusion}. A taxonomy of the various attacks on such devices was presented in \cite{butun2019security}. 
In many applications, the resource-constrained network is part of a larger network such as a home, enterprise, or organization. This makes the security of the resource-constrained network essential to prevent attacks on the whole network. 
For these reasons, the security of resource-constrained devices and networks was studied from various perspectives, and a large number of security solutions were proposed with different objectives. 
A categorization of various security solutions for resource-constrained networks was presented in \cite{kouicem2018internet}.

The emergence of blockchain technology provides an interesting approach to securing many aspects of resource-constrained networks. The main reason to integrate blockchain into these networks is to remove centralization and automate the secure exchange of real-time data between these devices. Blockchain utilizes distributed algorithms to allow transactions to be verified and stored securely by all nodes in the network. In addition, the transactions saved in blockchain cannot be manipulated or tampered with without destroying the validity of the whole blockchain. When implemented within a resource-constrained network, the blockchain guarantees the integrity and immutability of the data produced by resource-constrained devices and provides an authentic history of this data. The authors of \cite{dai2019blockchain, abdelmaboud2022blockchain} summarize the main characteristics and features of blockchain when integrated into resource-constrained networks. 
 
Despite the benefits that blockchain brings to resource-constrained networks, there are still many obstacles that face its actual implementation at a full scale. The main hurdles are due to the limitations of resource-constrained devices. Issues such as storage, power consumption, and computational capability should be taken into consideration before attempting to implement a blockchain system in a resource-constrained network. To cope with these issues, many researchers developed various blockchain solutions that attempt to reduce the complexity and resource-demanding aspects of the blockchain. 
These solutions are usually presented as “lightweight” systems that allow for the integration of the blockchain into resource-constrained networks. Most of these solutions focus on a specific \textit{heavy} characteristic of the blockchain and provide a means to reduce its demands. For example, some of these systems focus on developing a low-processing, communications, and power-consuming consensus algorithm. Other systems focus on reducing the storage requirements of the blockchain, etc. 

At the time of writing this paper, there is no comprehensive classification of such solutions based on the “lightweight” features they offer. Hence, in this paper, we present a detailed taxonomy of lightweight blockchain solutions for resource-constrained networks. After studying the various solutions that were proposed in the literature, we divide them into five main categories: “lightweight Architecture”, 
“Lightweight Authentication”, 
“Lightweight Consensus”, 
“Lightweight Cryptography”, 
and “Lightweight Storage”. 
For each category, we analyze the various systems that have been proposed so far and categorize them into sub-categories based on their characteristics and main modes of operation. In addition, we study the objectives that were identified by the authors of the papers in each category and deduce from them the “lightweight” requirements of that category. Next, we scrutinize each system to identify which “lightweight” requirements it satisfies. Furthermore, we discuss the missing gaps and the future developments that are required in each category. Toward the end of the paper, we provide a roadmap of future directions and discuss the various aspects that future blockchain solutions need to take into consideration in order to create an efficient and comprehensive “lightweight blockchain” system for resource-constrained networks. 
To the best of our knowledge, our paper is the first that presents a full taxonomy of “lightweight blockchain” solutions and scrutinizes the state-of-the-art of each “lightweight” category. 

In summary, the main contributions of this survey are as follows:
\begin{itemize}
\item  We present a comprehensive taxonomy of lightweight blockchain solutions for resource-constrained networks.
\item We identify the requirements of resource-constrained networks related
to each “lightweight”  category.
\item We analyze the proposed lightweight solutions according to the identified requirements.
\item  Finally, we give future directions related to lightweight blockchain solutions.
\end{itemize}

\begin{figure*}[!t]
\centering
\includegraphics[width=6.7in]{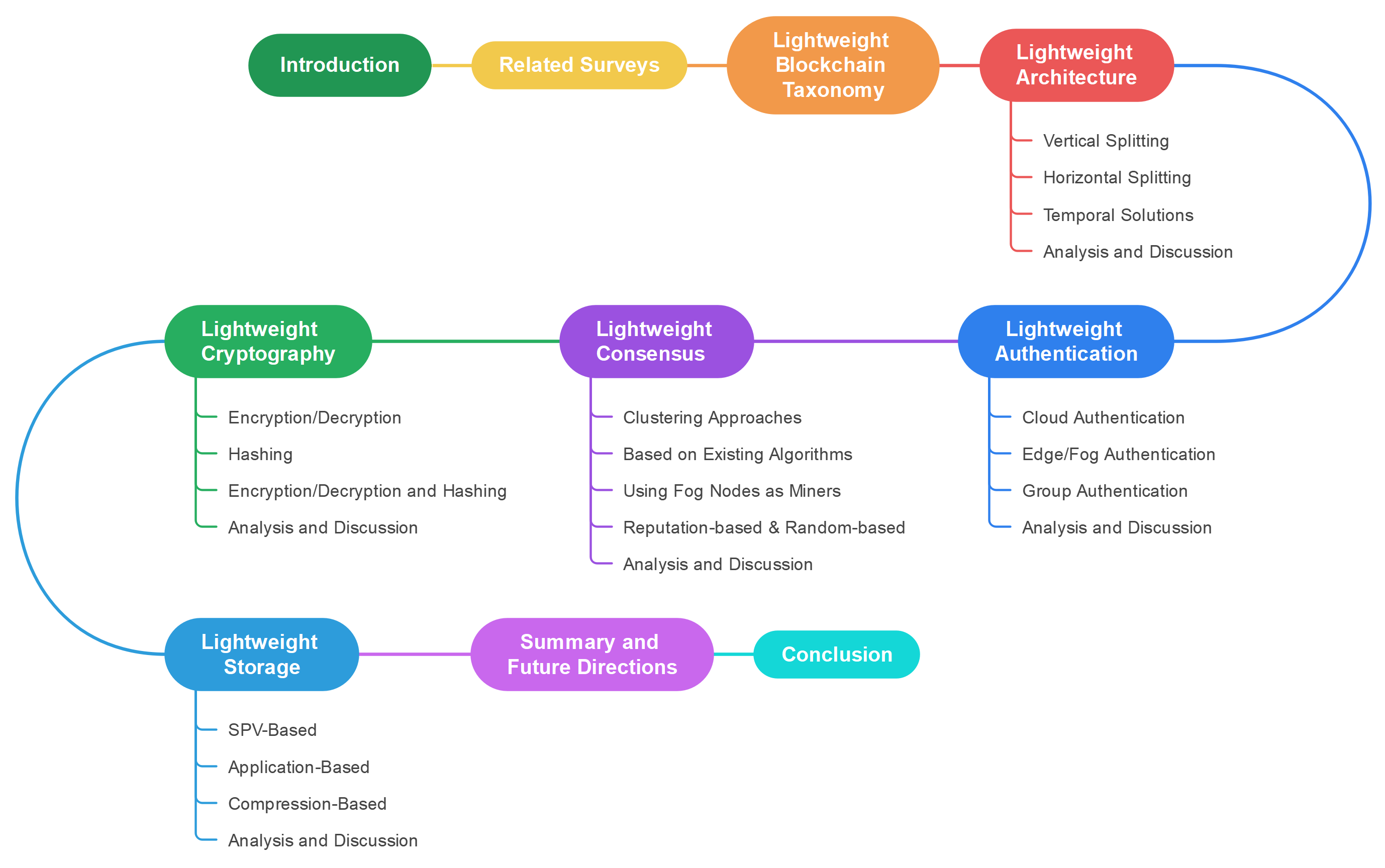}
\caption{Roadmap of the paper.}
\label{roadmap}
\end{figure*}

The remaining of this paper is organized as follows: in Section 2 we discuss briefly the differences between our paper and existing survey papers that studied topics related to blockchain in resource-constrained networks. Section 3 presents the classification of lightweight blockchain systems into five categories and the characteristics of the solutions in each category. Sections 4 to 8 analyze the previous works in each of the five lightweight categories. In Section 9, we discuss our findings and propose important guidelines and directions for future lightweight blockchain solutions. Finally, Section 10 concludes the paper and summarizes the findings. \autoref{roadmap} provides an overview of the paper roadmap. 

\section{Related Surveys and Motivation}

In this section, we discuss the previous surveys that share similar aims with this paper 
and illustrate the difference between each survey and our paper. 

A large number of research papers surveyed the various blockchain (BC) solutions for IoT networks and presented several taxonomies and summaries from different perspectives for these solutions \cite{dai2019blockchain, abdelmaboud2022blockchain, lao2020survey,  wang2019survey, pohrmen2019blockchain, ferrag2018blockchain, dwivedi2021blockchain, huo2022comprehensive, cheikhrouhou2022blockchain}. These papers reviewed the integration of BC into IoT in general and classified the solutions and systems in the literature using various classification models and methods. However, none of these papers focused on studying the lightweight aspects that must be considered when integrating the BC into IoT networks. This is a major difference between all these surveys and our paper, which is focused on categorizing BC solutions based on the lightweight category each solution belongs to. In addition, our paper reviews lightweight BC solutions in general, which could be implemented in any application (not necessarily IoT). IoT is the main field in which lightweight BC solutions have been implemented, but it is not the only field. Other fields include fog/edge networks, vehicular networks, unmanned aerial vehicles, vending machines, etc.  

Some surveys partially analyzed the lightweight BC solutions that were proposed for resource-constrained IoT networks. For example, the authors in \cite{darbandi2022blockchain} focused on analyzing the BC solutions for the Embedded Internet of Things. They classified these solutions based on their architecture into four types: double-blockchain architecture, SDN-based architecture, edge/fog-based architecture, and lightweight architecture. Hence, the authors considered only a limited aspect of lightweight BC which is the one related to architecture. On the other hand, the authors in \cite{salimitari2020survey} considered another lightweight aspect in their paper, which is consensus. They surveyed the various blockchain-based consensus methods that are applicable to resource-constrained IoT devices and networks and studied the BC type that is best suited to each IoT application.

Khor et al. \cite{khor2021public} focused on public blockchains for IoT applications and the prospects of using resource-constrained IoT devices with them. The authors identified the challenges that are faced by integrating the existing public blockchains with IoT devices and studied the effectiveness of the existing BC solutions in dealing with these challenges.  The authors identified four challenges, which are scalability, storage, computational resources, and energy consumption. The authors also discussed the characteristics a public BC should possess in order to be integrated with a resource-constrained IoT network. However, the paper did not present any clear taxonomy or categorization of the lightweight BC solutions that were analyzed by the authors. In addition, the authors limited their focus to only  IoT networks. In our paper, we provide a comprehensive review of lightweight BC solutions, encompassing various domains rather than focusing on a specific field.

A systematic literature review on the lightweight BC concept for IoT was presented in \cite{stefanescu2022systematic}. The authors answered four questions, namely: the definition of lightweight BC, the characteristics of the proposed lightweight solutions, the lightweight aspect of the proposed solutions, and how these solutions are evaluated. The authors analyzed the papers that proposed lightweight BC solutions for IoT and categorized them in terms of blockchain structure, consensus, storage, and cryptography.  The authors also suggested a lightweight BC definition that includes five characteristics: “low computational burden, low network overhead, low storage requirements, high throughput, high energy efficiency”. However, as the paper is a systematic review it just gives statistical information without giving descriptions of the proposed solutions. In addition, the authors did not consider the “lightweight authentication” aspect in their paper. Moreover, the authors defined the requirements of a “lightweight blockchain” in general. Unlike this survey, our work focuses on defining the lightweight BC requirements in each lightweight category and analyzing the existing papers in the category based on these requirements. Also, for each category, we pinpoint the missing aspects in the existing research and the improvements that are required in future research to fulfill the lightweight requirements of the category.

\begin{table}[!t]
\begin{adjustwidth}{-0.6cm}{}
\centering
\caption{Comparison between existing surveys that are related to lightweight blockchain and our paper.} 
\label{table_surveys} 
\begin{tabular}{|p {6.5 cm}|p {1.2 cm}|p {1.5 cm}|p {1.5 cm}|p {1.5 cm}|p {1.5 cm}|p {1.2 cm}|}
\hline
\diagbox{\textbf{Feature}}{\textbf{Reference}} & \cite{dai2019blockchain, abdelmaboud2022blockchain, lao2020survey, wang2019survey, pohrmen2019blockchain, ferrag2018blockchain,  dwivedi2021blockchain, huo2022comprehensive, cheikhrouhou2022blockchain} &  \hfil \cite{darbandi2022blockchain} & \hfil \cite{salimitari2020survey} & \hfil \cite{khor2021public} & \hfil \cite{stefanescu2022systematic} & This paper \\ \hline
Reviews blockchain solutions for resource-constrained networks & \hfil \priority{100} & \hfil \priority{100} & \hfil \priority{100} & \hfil \priority{100} & \hfil \priority{100} & \hfil \priority{100} \\
\hline
Focuses on lightweight-blockchain solutions	& \hfil \priority{0}	& \hfil \priority{50} & \hfil \priority{100} & \hfil \priority{100} & \hfil \priority{100} & \hfil \priority{100} \\ \hline 
Studies the lightweight-blockchain solutions that were published between 2021-2023 & \hfil \priority{0}	& \hfil \priority{0} & \hfil \priority{0} & \hfil \priority{0} & \hfil \priority{0} & \hfil \priority{100} \\ \hline 
Identifies categories of lightweight aspects in blockchain & \hfil \priority{0} & \hfil \priority{0} & \hfil \priority{0} & \hfil \priority{0} & \hfil \priority{50} & \hfil \priority{100} \\ \hline 
Identifies lightweight requirements of each lightweight category & \hfil \priority{0}	& \hfil \priority{0}	& \hfil \priority{0}	& \hfil \priority{0}	& \hfil \priority{0} & \hfil \priority{100} \\ \hline 
Maps existing solutions to lightweight requirements in each lightweight category & \hfil \priority{0}	& \hfil \priority{0}	& \hfil \priority{0}	& \hfil \priority{0}	& \hfil \priority{0}	& \hfil \priority{100} \\ \hline 
Identifies the overall requirements of lightweight blockchain & \hfil \priority{0}	& \hfil \priority{50} & \hfil \priority{50} & \hfil \priority{50} & \hfil \priority{50} & \hfil \priority{100} \\ \hline 
Suggests future research directions to enhance lightweight blockchain systems & \hfil \priority{0} & \hfil \priority{0} & \hfil \priority{50} & \hfil \priority{50} & \hfil \priority{50} & \hfil \priority{100} \\ \hline
\multicolumn{7}{l}{\priority{0} = No, \priority{50} = Partially, \priority{100} = Yes} \\
\end{tabular}
\end{adjustwidth}
\end{table}

Although the surveys mentioned in this section share commonalities with our paper, they lack a comprehensive categorization of lightweight BC solutions. In addition, none of these surveys presented a comprehensive taxonomy of the lightweight categories of the BC and defined the requirements of each category that should be satisfied in an efficient solution. Furthermore, as the “lightweight blockchain” is a very recent topic, a large number of related papers have been published in the last couple of years that have not been studied by any of the existing survey papers. In this paper, we analyze these solutions thoroughly, pinpoint the existing gaps, and propose future directions. More precisely, this paper has the following unique contributions: 

\begin{itemize}
\item 
We analyze all lightweight BC solutions that were proposed in the literature. Based on this analysis, we classified the solutions into five main categories based on the lightweight aspect addressed in the solution. These categories are: architecture, authentication, consensus, cryptography, and storage.
\item 
We review the objectives of the lightweight solutions in each category. Based on that, we define the lightweight requirements of each category, which are the conditions that should be met by a BC solution of this category.
\item 
We analyze the existing solutions in each category and identify the lightweight requirements that each solution satisfies. Based on the analysis of existing solutions in each category, we pinpoint the missing aspects that should be considered by future BC solutions.
\item 
We summarize and aggregate our findings by proposing a comprehensive lightweight BC system that satisfies the lightweight requirements of each category. Such a system would be the target of future research work in this area.
\end{itemize}

Table \ref{table_surveys} summarizes the differences between our paper and the existing surveys and illustrates the unique features that are present in our paper.

\section{Taxonomy of Lightweight Blockchain}
\label{Sec_BCcat}

Since the emergence of blockchain technology, a lot of efforts have been made to utilize it in various applications. The BC is very tempting to adopt as a solution that solves many security problems, due to its unique characteristics such as data immutability, transaction validation, and distributed consensus. However, researchers have realized that integrating the BC into resource-constrained devices is very challenging due to the high processing, communications, and power demands of traditional BC systems, and the limited resources of these devices. Hence, the research works that proposed BC-based solutions for resource-constrained networks have focused on modifying one or more of the conventional BC characteristics to make it suitable for such networks. 

In general, several aspects and operations of the BC are very heavy from the perspective of a light device. Mainly, the BC requires the node to securely maintain and update cryptographic keys and execute algorithms to digitally sign each transaction it creates, encrypt private transactions, and verify others' transactions. In addition, If the BC node is a miner, it should verify the signature of each transaction in the block, hash the block transactions to generate the block Merkle root, hash each block it creates, and participate in the BC consensus algorithm. The latter could require the node to spend a lot of resources. For example, Proof of Work (PoW) makes the miners engage in an extremely power-demanding competition to produce a block hash that has a leading number of zeros greater than or equal to a difficulty target. Practical Byzantine Fault Tolerance (PBFT) requires the node to send and receive a large number of messages. In addition, many traditional BC consensus models (such as Proof of Stake) require the existence of a cryptocurrency system, which hinders their applicability in resource-constrained networks. Finally, one of the main problems of BC is its high storage demand and everlasting growth. A BC node should have high and scalable storage capability to be able to store the BC locally. 

In this paper, we study the various “lightweight” aspects that the BC should incorporate in order to become suitable for resource-constrained environments. First, as mentioned before, resource-constrained devices are usually part of a larger network that comprises communication devices, edge nodes, and Internet and cloud servers. When implementing  BC in such wide network, the BC architecture should be modified in order to assign different roles in the BC based on the capabilities and needs of each node in the network. Several papers proposed lightweight BC architectures that aim to integrate the resource-constrained devices into the BC network without affecting their performance.

Furthermore, a lightweight node needs to authenticate itself once it joins the BC network. 
Subsequently, when a node requires BC data from another, the two nodes should mutually authenticate each other to start a session and exchange their data. The authentication process should be lightweight taking into consideration the nodes' limited resources. For this purpose, several researchers investigated different “lightweight authentication” mechanisms that allow the light node to authenticate itself within the BC network with little overhead. 

Moreover, while participating in the BC, the light node will perform several tasks that require cryptographic operations, as detailed before. Traditional cryptography systems, if implemented by a light node, will cause the node to deplete its resources quickly due to their high processing and energy requirements. Hence, “lightweight cryptography” becomes a necessity for the light node to be able to perform the cryptographic functions required by the BC without draining its energy quickly. 

Additionally, one of the main characteristics of a BC network is the distributed consensus process that allows all the BC nodes to agree on the legitimacy and correctness of a transaction before it can be added to the BC. A large number of consensus algorithms 
have been proposed. However, most of these algorithms are not suitable for resource-constrained nodes as described before. When the BC network contains light nodes that will participate in the consensus process, the latter should be designed such that it requires little communication and processing overhead. Hence, the term “lightweight consensus” has been used by many researchers to describe such consensus mechanisms.

Finally, a major demand of the BC is storage. Resource-constrained nodes are not able to store the whole BC. Hence, a solution is needed to allow these nodes to store part of the BC based on the demands of the applications they execute, while fetching the data of the other BC part from full BC nodes when needed. A “lightweight storage” mechanism should allow the light node to efficiently store the required BC part and verify the data that it obtains from the other nodes.

\begin{figure*}[!t]
\centering
\includegraphics[width=6.5in]{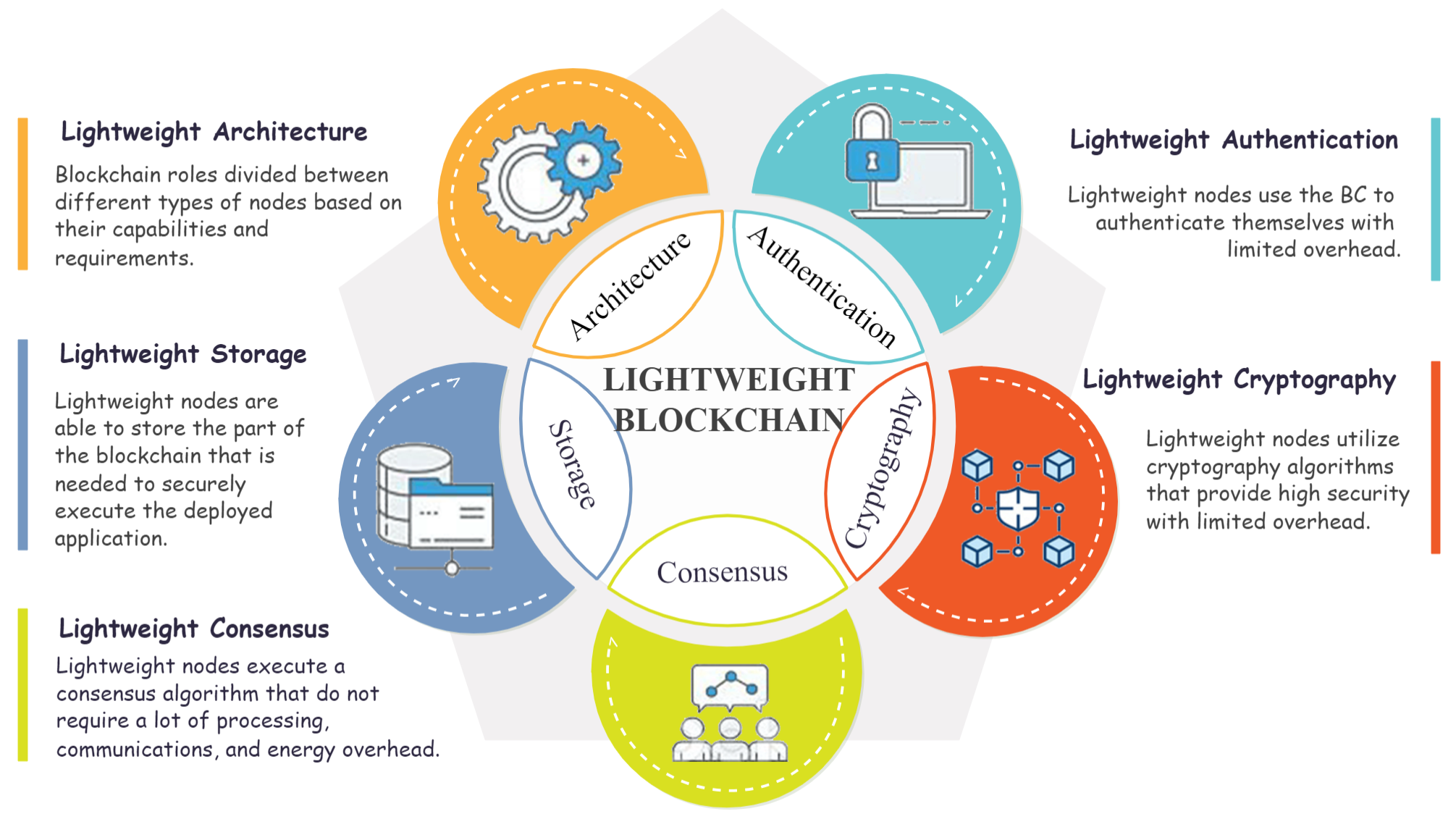}
\caption{Summary of the five lightweight blockchain categories.}
\label{System_Overview}
\end{figure*}

Based on the above description, this paper aims at reviewing the research works that have been proposed in each of the five lightweight categories. A summary of the five categories is illustrated in \autoref{System_Overview}. A large number of lightweight BC systems have been proposed so far. Most of these systems focus on one of the five categories discussed above. In order to deploy an effective “lightweight blockchain” system in a resource-constrained network, all five categories should be taken into consideration. Hence, our work will provide an overview of how a lightweight BC should be built by incorporating lightweight features within its architecture, authentication, cryptography, consensus, and storage subsystems and processes. In order to reach our objective, we study each of the five categories in depth: first, we outline the requirements that need to be satisfied by the BC to be “lightweight” in each category. Next, we summarize the research works that have been proposed so far in that category and identify the requirements each of them meets. Eventually, we present the research questions and challenges that should be taken into consideration by future research works in this area. 


Note that in order to make the study of “lightweight blockchain” effective and interesting, the research works that are summarized in each section have been selected based on the following criteria: 

\begin{itemize}
    \item \textit{Acceptance}: the paper should be published at a peer-reviewed, scientifically approved venue (journal/conference/book). 
    \item \textit{Importance}: the system should be relevant to the category with evidence in the form of implementation or experiments. 
    \item \textit{Freshness/recentness}: the paper should be published in the last five years. 
    \item \textit{Novelty}: the system should not have been replaced by a more recent similar system that provided much better performance. 
\end{itemize}

\section{Lightweight Architecture}
\label{Sec_LAr}

\begin{figure}[!t]
\centering
\includegraphics[width=6in]{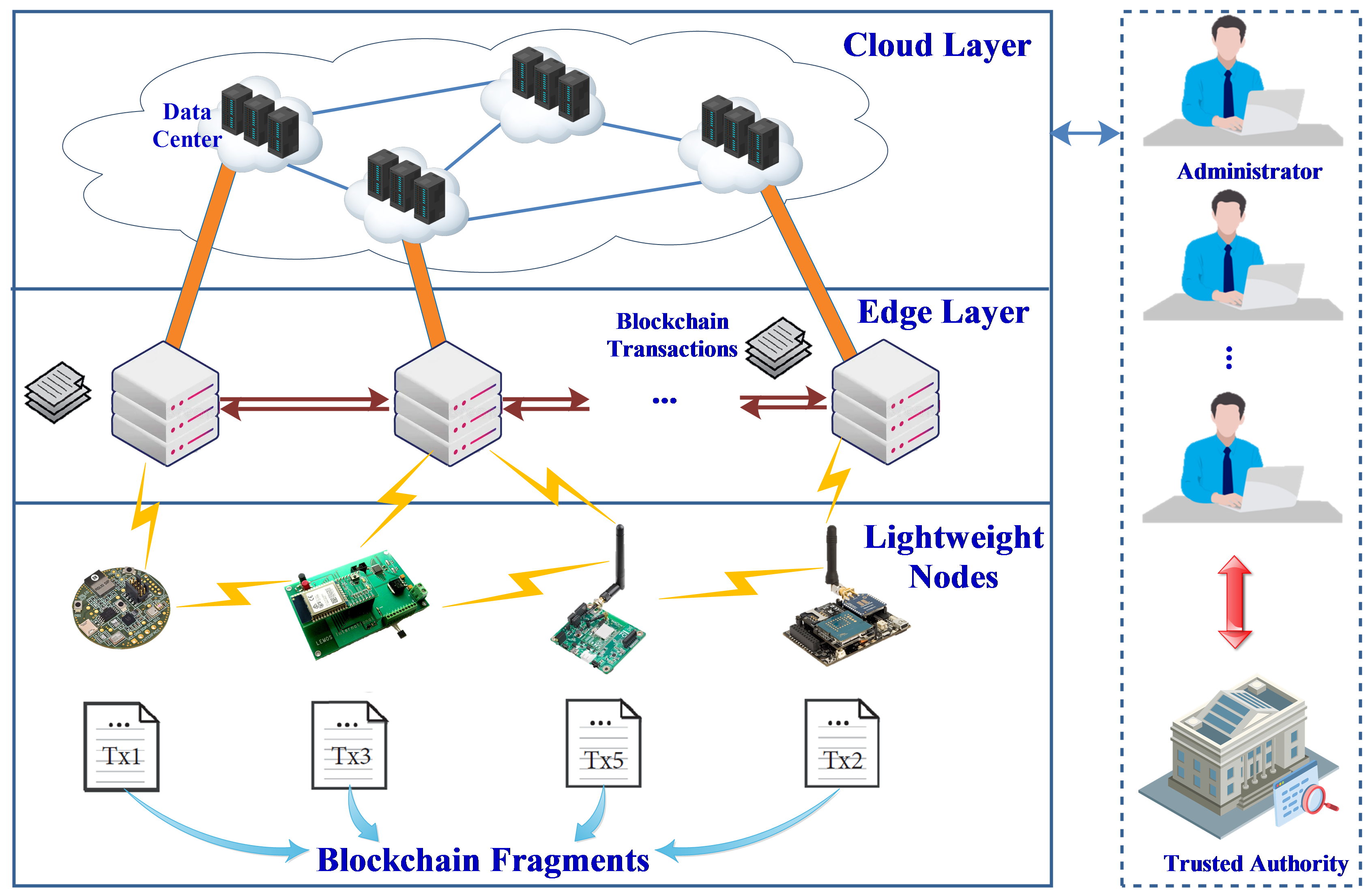}
\caption{A sample lightweight blockchain architecture that comprises three layers for lightweight nodes, edge servers, and cloud datacenters.}
\label{fig_SampLAr}
\end{figure}

The architecture of a traditional BC model is generally very heavy to be implemented by a resource-constrained node. Hence, several researchers studied the possibility of adjusting the BC architecture to accommodate the addition of resource-constrained nodes to the BC network. For example, a possible solution is to distribute the blockchain roles  
between the resource-constrained nodes. After studying the various solutions that proposed a lightweight BC architecture, we divided them into three subcategories, which can be described as follows:
\begin{itemize}
\item 
\textbf{Vertical splitting}: Using two or more blockchains where each BC stores part of the application data. Nodes store specific blockchains based on their roles and/or capabilities \cite{maftei2023massive, settipalli2023extended, hao2022stochastic, kang2022blockchain, na2022iot, yao2022accident, lee2021lightweight, xie2021eclb, le2019lightweight}. \autoref{fig_SampLAr} illustrates a sample lightweight BC architecture in which the BC is fragmented between the lightweight nodes.
\item 
\textbf{Horizontal splitting}: Dividing the resource-constrained network into sub-networks, where each sub-network stores part of the BC \cite{liu2023communitychain, gupta2022lightweight, mohapatra2022blockchain, yang2022lightweight, honar2021multi, kim2021autonomous, sunny2020towards, shahid2019sensor}. 
\item 
\textbf{Temporal solutions}: modifying the BC periodically to reduce its overhead 
\cite{corradini2022two, pyoung2019blockchain, shahid2019sensor}.
\end{itemize}

In this section, we discuss the different systems that were proposed in each of these three categories, focusing on analyzing the lightweight characteristics of each system and the lightweight requirements that it satisfies or lacks. At the end of the section, we highlight the missing aspects in the literature related to lightweight architecture, and provide insight into future directions.


\subsection{Vertical Splitting}

A lightweight blockchain architecture for Wireless Sensor Networks (WSNs) was adopted in \cite{maftei2023massive}. Here, blockchain-enabled gateways store the light BC that contains the WSN nodes’ identities and the hashes of the data records that are stored in the public BC. Gateways execute the smart contracts within the public BC to authenticate sensor nodes and verify their data. A similar BC architecture labeled Extended Lightweight Blockchain (ELB) healthcare system was proposed in \cite{settipalli2023extended}. The system utilizes several blockchains for the efficient processing of healthcare transactions. While the healthcare data is stored in the main BC, the claims of users are handled by BlockChain Managers (BCMs) and stored into one of two blockchains: valid claims BC and invalid claims BC. In addition, the system uses a Head BlockChain Manager which is accessed only by BCMs and contains the BC administrative data and events. 

One of the most important issues in resource-constrained networks is managing trust amongst the processing nodes. For this purpose, the authors of \cite{hao2022stochastic} propose a double blockchain (DBC) architecture that comprises an information blockchain (IBC) stored in the cloud layer and a reputation blockchain (RBC) stored in the fog layer. IoT data is processed at the fog nodes to generate new information blocks (IBs) and reputation blocks (RBs) that update the IBC and RBC. The RBC nodes are divided into full, light, and basic nodes based on their computation and storage capabilities. Nodes’ reputations were calculated based on past interactions with the miners and recommendation data from high-reputation nodes. 
Kang et al. \cite{kang2022blockchain} propose that lightweight nodes maintain a light chain. Here, lightweight nodes broadcast their transactions and create light blocks within the lightweight network. However, they offload the consensus process to the main network. Periodically, the lightweight nodes upload their light blocks to the main BC. When the blocks are successfully added to the main chain, the lightweight chain is reset. 

In order to enable lightweight nodes to participate in the BC, the authors in \cite{na2022iot} propose a multichain architecture in which resource-constrained nodes store a light chain and cloud nodes store a monitoring chain. Specific lightweight nodes, called Exporters are selected to participate in both chains and periodically export certain blocks from the light chain to the monitoring chain. When exporting a block, the lead Exporter collects the signatures of all nodes in the light chain on the exported block using the Schnorr signature method, which compresses the signature of all nodes and enables the verification of all signatures in a single step. The Exporter nodes validate the Schnorr signature before adding the block to the monitoring chain.

In \cite{yao2022accident}, a framework for determining the liability of accidents for autonomous vehicles was proposed. Here, the BC is divided into two chains: a “preservation chain” for storing the vehicles' data and an “accident identification chain” for storing the vehicles' secret information such as the real ID. The two chains interact using cross-chain technology. The second chain executes a smart contract when an accident occurs. While the “preservation chain” is public, the “accident identification chain” is private to security departments, courts, insurance companies, and car owners; who agree on the smart contracts that are executed to handle the accident. The preservation chain is stored by the roadside units (RSUs) and cloud servers, while the second chain is light and can be stored by all participants. 

The authors of \cite{lee2021lightweight} propose a permissioned blockchain for the network of smart meters within a smart city. The BC architecture comprises two blockchains: a data blockchain and a meta blockchain. The data blockchain, which is saved at fog and cloud nodes, contains all the smart meters' configurations and public data. The smart meters save only the meta blockchain which contains the smart meters' configuration and public data. On the other hand, the data blockchain, which is saved at fog and cloud nodes, contains the information stored in the meta blockchain in addition to data transactions.  

An Edge-Computing-Based Lightweight Blockchain (ECLB) protocol was proposed in \cite{xie2021eclb}. In ECLB, the full BC is saved on edge nodes, while the lightweight nodes store what the authors call the \textit{fragmented ledger structure}, which contains the block headers and some of the transactions in each block that are needed by the lightweight node. The main BC at the edge nodes is also divided into leader blocks and transactions blocks. The leader blocks contain the public keys of miners who generate blocks (i.e., leaders). Each resource-constrained node stores information about each block, such as the number of transactions it stores and the ID and location of each of these transactions. In \cite{le2019lightweight}, resource-constrained devices store only block headers and can verify BC transactions via the Simplified Payment Verification (SPV) method. In SPV, a lightweight device sends a query to a gateway to verify a transaction or the header of a new block. The authors propose to replace the gateway role with a group of BC nodes that are selected randomly by the lightweight node. 
The selection of witnesses, the confirmation, and the whitelist of witnesses are managed by Bloom filters. 
The lightweight node updates the trust weights of witnesses when it receives confirmations from them based on their correctness.

A summary of the solutions discussed in this section is shown in \autoref{table_ArchSum1}.

\begin{table}[!t]
\begin{adjustwidth}{-1.2cm}{}
\centering
\caption{Summary of lightweight architecture - vertical splitting solutions.} 
\label{table_ArchSum1} 
\centering
\begin{tabular}{|p {0.7 cm}|p {1.9 cm}|p {1.2 cm}|p {5.0 cm}|p {3.5 cm}|p {3.5 cm}|}
\hline
\textbf{Ref.} & \hfil \textbf{Application.} & \hfil \textbf{Year.} & \hfil{\textbf{Main idea}} & \hfil \textbf{Pros} & \hfil \textbf{Cons}  \\ 
\hline 
\hfil \cite{maftei2023massive} & \centering  WSN & \hfil 2023 & Light BC used by fog nodes to store IDs of light nodes and hashes of data records & - Scalable \newline- Secure identification  & - PoS Consensus for WSN \newline- High communication overhead \\
\hline
\hfil \cite{settipalli2023extended} & \centering  Healthcare & \hfil 2023 & BC Managers handle users' requests and save transactions into valid claims BC or invalid claims BC & - Local ledger maintenance \newline- Use of canals  & - Complex architecture  \newline- High latency \\
\hline
\hfil\cite{hao2022stochastic} & \centering  General & \hfil 2022 & Data is divided between information blocks and reputation blocks and saved on the information BC and reputation BC & - Node Classification \newline- High security & - Storage and processing overhead \\
\hline
\hfil\cite{kang2022blockchain} & \centering  Vending Machines & \hfil 2022 & Lightweight nodes upload the blocks they create and offload consensus to the main network & - Lightweight transaction model & - Several security vulnerabilities \newline- No performance testing \\
\hline
\hfil\cite{na2022iot} & \centering  General & \hfil 2022 & Exporters periodically export old blocks from the light chain to the monitoring chain & - Low latency \newline- High throughput   & - Limited testing \newline- Miss ratio not studied \\
\hline 
\hfil \cite{yao2022accident} & \centering  IoV & \hfil 2022 & Vehicles save data in “preservation” BC. Smart contracts in the “accident identification” BC determine accident liability & - Highly secure & - Limited lightweight \newline- High communication overhead \\
\hline
\hfil \cite{lee2021lightweight} & \centering  Smart grid & \hfil 2021 & Meta BC used by smart meters to store configurations and public data & - Lightweight and practical  & - Latency not studied \\
\hline
\hfil \cite{xie2021eclb} & \centering  General & \hfil 2021 & Light nodes store block headers and the transactions they need from each block only & - Edge nodes as miners\newline- CP-ABE for access control & - Latency not studied \newline- PoW for consensus \\
\hline
\hfil \cite{le2019lightweight} & \centering  General & \hfil 2019 & Group of BC nodes selected via Bloom filters to validate and confirm BC data & - Use of Bloom filters & - High latency \newline- High communication overhead \\
\hline
\end{tabular}
\end{adjustwidth}
\end{table}

\subsection{Horizontal Splitting}

Liu et al. \cite{liu2023communitychain} propose Communitychain, a customized BC architecture for smart homes based on sharding. The network is divided into subsets (shards) based on the types of devices. Each shard stores a BC that contains the transaction sets of the specific device type only. Devices in each shard are clustered and a single node is selected in each cluster to be a shard miner. Transactions can be local within the shard or involve nodes in multiple shards. In the latter case, an inter-shard routing protocol is applied by the miners of the corresponding shards to handle the transaction via cross-shard communication. A similar approach was proposed by Gupta et al. in \cite{gupta2022lightweight}. In this paper, the authors describe a branched blockchain for the IoV based on Chord protocol and distributed hash table technology. The vehicles are clustered based on geographic locations and the vehicles in each cluster store a blockchain branch. Each data slice is hashed and combined with the Chord data key to form a Chord data chunk. The Chord protocol organizes all devices as a ring network, and each data chunk is sent to the successor vehicles in the corresponding branches. Each vehicle augments the data chunks that it received to its branch blockchain.

In addition to dividing the BC into a group of local chains within the clusters, a software agent for blockchain formation and monitoring (SAB) was used by the authors in \cite{mohapatra2022blockchain} to govern the operations of the local BC. A SAB is installed at a fog node and communicates with the nearby lightweight devices. The SAB creates the genesis block of the local BC and determines the time at which each lightweight device can create a block. When its turn to create a new block arrives, the lightweight device aggregates its data, generates the block by solving a simplified PoW puzzle, and sends the block to a group of validators that are selected by the SAB for each new block. The authors in \cite{yang2022lightweight} propose clustering the resource-constrained nodes based on how often they interact with each other. Hence, nodes are divided into high-frequency and low-frequency groups. This permits scaling the network such that each group contains a number of nodes that is proportional to the frequency of communications between the group members. 

A private multi-layer BC model in which self-clustering is applied based on the Genetic Algorithm and Simulated Annealing (SA) is proposed in \cite{honar2021multi}. The BC network is divided into three layers. At the first layer, resource-constrained nodes are divided into clusters. At the second layer, cluster heads (CHs) store the local copy of the BC. At the third layer, the cellular base stations (BSs) store the full global BC. Nodes in Layers 2 and 3 collaborate to create new blocks and execute the consensus algorithm. 
Some resource-constrained nodes at layer 1 can be peers that maintain a copy of the local ledger and act as Hyperledger endorsers or committers, while CHs and BSs act as Hyperledger orderers. 

In \cite{kim2021autonomous}, the authors propose a lightweight architecture in which deep clustering is applied to group IoT nodes into clusters that dynamically change based on the changes in the network. The characteristics of IoT nodes are saved in 2D tensors and inputted to a convolutional neural network (CNN) that extracts the features' weights of each node. The results are inputted into a k-means clustering algorithm in order to group the IoT nodes into clusters that will minimize the overall load on the network. The nodes in each cluster are divided into clients and validators using graph neural network (GNN) node classification. The GNN classifies nodes into classes based on their loads, and only the nodes in the highest class are assigned the task of dApp spreading.

A lightweight BC for critical infrastructure protection was proposed by Sunny \textit{et al.} in \cite{sunny2020towards}. Here, the authors utilize a consortium BC for the infrastructure monitoring application. A network of \textit{Supervisory Computers} (SCs) is formed between all organizations that participate in the application: each organization creates its own IoT network, divides it into clusters, and connects the CHs to its SCs. The CHs gather the application data from the IoT nodes, create new blocks, and send them to the SCs. The latter execute the consensus protocol to add the block to the BC. On the other hand, the authors in \cite{shahid2019sensor} utilized the \textit{Voronoi} model to divide the IoT network area into geographic regions, where the nodes in each region store a local BC that contains only the transactions needed in that region. The overall BC is formed by combining the local blockchains of all regions. All IoT nodes in a region act as equal peers (i.e., no CHs). In addition, each IoT node replaces its copy of the BC as it moves from one region to another.

A summary of the solutions discussed in this section is shown in \autoref{table_ArchSum2}.

\begin{table}[!t]
\begin{adjustwidth}{-1.2cm}{}
\centering
\caption{Summary of lightweight architecture - horizontal splitting solutions.} 
\label{table_ArchSum2} 
\centering
\begin{tabular}{|p {0.7 cm}|p {1.9 cm}|p {1.2 cm}|p {5.0 cm}|p {3.5 cm}|p {3.5 cm}|}
\hline
\textbf{Ref.} & \hfil \textbf{Application.} & \hfil \textbf{Year.} & \hfil{\textbf{Main idea}} & \hfil \textbf{Pros} & \hfil \textbf{Cons}  \\ 
\hline 
\hfil \cite{liu2023communitychain} & \centering Smart Home & \hfil 2023 &  Clustering is based on Types of IoT devices. A single miner is selected per cluster  & - Low latency \newline- Efficient consensus & - Single point of failure \newline- High communication overhead \\
\hline 
\hfil \cite{gupta2022lightweight} & \centering IoV & \hfil 2022 &  Geographic Clustering in which the Chord protocol is used to distribute data between the branch BCs  & - High security \newline- Branched blockchain & - Very high latency \newline- Considerable communication overhead \\
\hline 
\hfil \cite{mohapatra2022blockchain} & \centering General & \hfil 2022 & Software agents are used for block creation and BC monitoring & - Efficient clustering \newline- High security & - IoT nodes use PoW \newline- High latency \\
\hline 
\hfil \cite{yang2022lightweight} & \centering Payment Systems & \hfil 2022 &  IoT nodes are clustered based on their interaction frequency & - Scalability \newline- High security & - No testing \newline- High communication overhead \\
\hline 
\hfil \cite{honar2021multi} & \centering General & \hfil 2021 & CHs store local BC that contains the cluster data and create local blocks & - Efficient clustering method \newline- Scalability   & - High load on CHs \newline - Limited testing \\
\hline
\hfil \cite{kim2021autonomous} & \centering General & \hfil 2021 & CNN is used to cluster IoT nodes dynamically based on their loads to maximize dApp spreading & - High dApp deployment and communication efficiency & - High latency \newline - Limited scalability \\
\hline 
\hfil \cite{sunny2020towards} & \centering Critical infrastructure & \hfil 2020 & Supervisory Computers monitor and administer the CHs. The latter create blocks and them to SCs who execute consensus & - Highly lightweight \newline- Simple architecture & - Several attacks not considered \newline - High control latency \\
\hline 
\hfil \cite{shahid2019sensor} & \centering IoT of mobile devices & \hfil 2019 & IoT network is divided into clusters via the \textit{Voronoi} model & - Lightweight and scalable & - High load on IoT nodes \newline- Weak security \\
\hline 
\end{tabular}
\end{adjustwidth}
\end{table}

\subsection{Temporal Solutions}

The authors in \cite{corradini2022two} propose a “Trust Blockchain” that smart IoT objects use to check the trust of each other. Smart objects are grouped into communities, where each community stores a local trust chain. In addition, all the main devices store a global trust chain. During each period, each smart device randomly selects a group of devices in its community and tests the services they offer via probing. Based on the results, the smart device assigns a reputation score to each device it tested and saves it in the local chain. At the end of the period, the trust values of each device are aggregated and saved in the global chain, while the local chain is reset. Smart objects that do not meet the minimum reputation level are removed from the community.

In \cite{pyoung2019blockchain}, the authors propose that each transaction in a block should have a lifetime. 
The block expires when all the transactions within it expire. The authors assume a permissioned BC in which only edge servers with permissions can participate in the BC. In order to solve cases where the BC becomes disconnected because a block in the middle has been deleted, the authors propose two data structures: the \textit{Endtime Ordering Graph} (EOG), which is a tree based on the order of expiration of lifetimes, and \textit{Arrival Ordering Graph} (AOG), which is a list based on the order of creation of transactions. The proposed system, LiTiChain, combines the two data structures and adds to each block one or two links based on its relationships in the two structures.

In Sensor-chain \cite{shahid2019sensor}, IoT nodes are grouped into clusters based on their geographic locations. Periodically, a certain IoT node in each region is assigned the role of the aggregator. This node takes the current copy of the local chain and aggregates all data in the blocks using aggregation functions (such as mean, average, etc.). The aggregation results are inserted into a single block which becomes the genesis block of the new copy of the local BC. This operation is repeated periodically to keep the local BC within a small-size limit and allow IoT nodes to store it.

A summary of the solutions discussed in this section is shown in \autoref{table_ArchSum3}.

\begin{table}[!t]
\begin{adjustwidth}{-1.2cm}{}
\centering
\caption{Summary of lightweight architecture - temporal solutions.} 
\label{table_ArchSum3} 
\centering
\begin{tabular}{|p {0.7 cm}|p {1.9 cm}|p {1.2 cm}|p {5.0 cm}|p {3.5 cm}|p {3.5 cm}|}
\hline
\textbf{Ref.} & \hfil \textbf{Application.} & \hfil \textbf{Year.} & \hfil{\textbf{Main idea}} & \hfil \textbf{Pros} & \hfil \textbf{Cons}  \\ 
\hline 
\hfil \cite{corradini2022two} & \centering General & \hfil 2022 & Trust scores in the local chain are aggregated and moved to the global chain & - High security \newline- Acceptable latency & - Limited scope \newline- Limited experiments \\
\hline
\hfil \cite{pyoung2019blockchain} & \centering General & \hfil 2020 & When all transactions in a block expire, the block body is deleted from the BC & - High scalability \newline- Reduced storage & - High latency \newline - No security considerations \\
\hline
\hfil \cite{shahid2019sensor} & \centering IoT of mobile devices & \hfil 2019 & BC data is periodically aggregated into a block that becomes the Genesis of the next blockchain & - Scalable \newline- Reduced storage & - Single point of failure (aggregator) \newline - High latency \\
\hline 
\end{tabular}
\end{adjustwidth}
\end{table}

\begin{figure}[!t]
\begin{adjustwidth}{-0.6cm}{}
\centering
\includegraphics[width=7.0in]{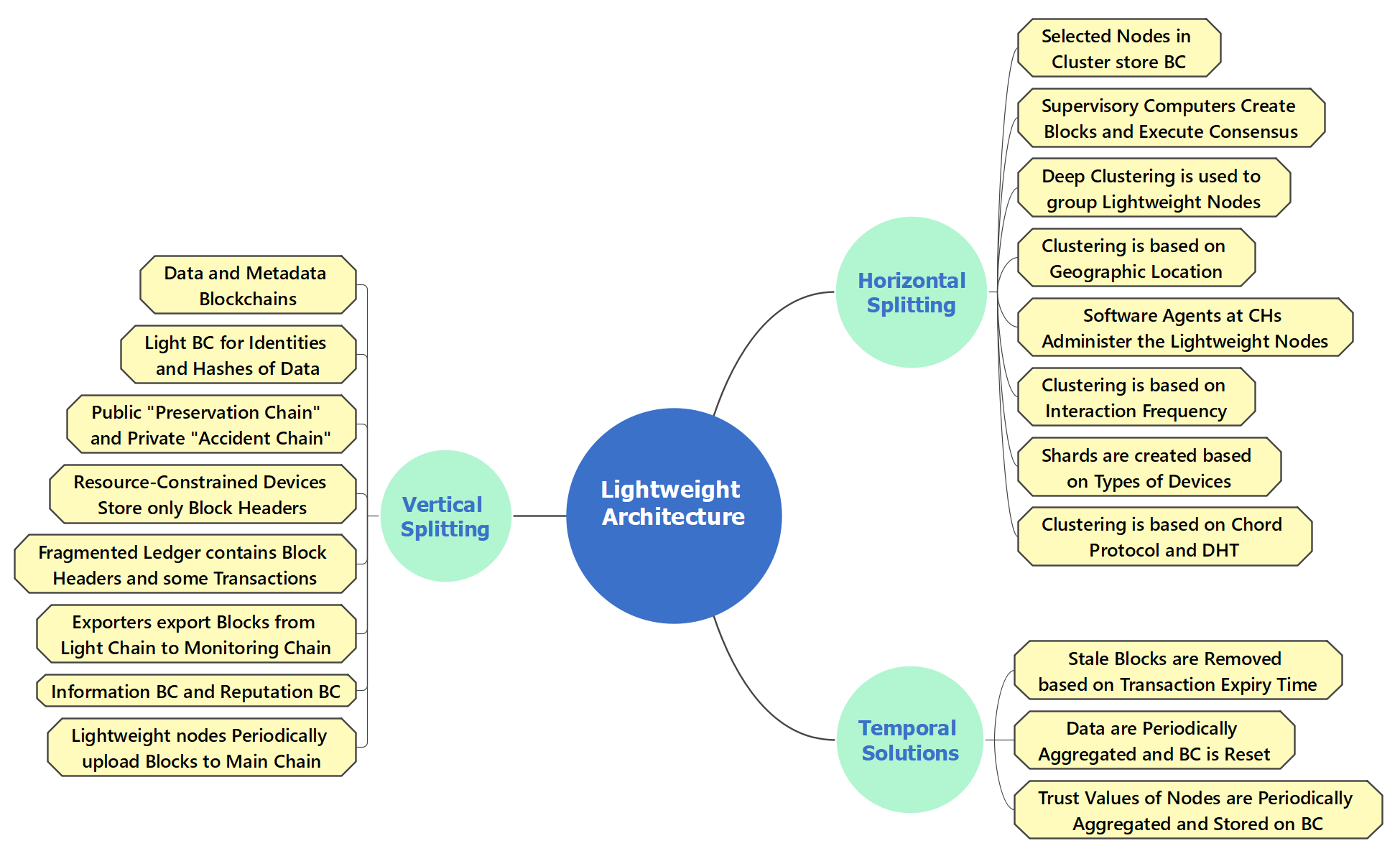}
\caption{Methods for lightweight blockchain architecture.}
\label{fig_LA}
\end{adjustwidth}
\end{figure}

\autoref{fig_LA} illustrates the basic idea of each solution in each of the three Lightweight Architecture categories.

\subsection{Analysis and Discussion}

Various approaches for creating a lightweight BC architecture for resource-constrained networks were proposed in the literature and described in this section. 
In order to evaluate the effectiveness of each approach, we studied the characteristics of each system intensively. 
From the papers that we studied in this category, we identify the requirements of a lightweight BC architecture as follows:

\begin{enumerate}
\item 
\textbf{Applicability:} A resource-constrained node should be able to implement the lightweight BC architecture to provide the required security features to the deployed application. 
\item 
\textbf{Latency:} The lightweight architecture must be able to support strict delay requirements of the deployed application and result in fast integration of data into the BC.
\item 
\textbf{Scalability:} The lightweight architecture must support scalability, which means that lightweight nodes should be able to keep implementing the architecture as the BC grows.
\item 
\textbf{Lightweight:} The interactions between the BC components within the lightweight architecture must not add significant overhead (processing, communication, and power consumption) to the deployed application.
\item 
\textbf{Security:} The interactions between the BC components within the lightweight architecture must be secure. Attackers should not be able to compromise the BC by exploiting an architecture vulnerability.
\item 
\textbf{Usability:} Lightweight architecture must be user-friendly and support efficient usability.
\end{enumerate}

Each of the studied systems was analyzed to discover which lightweight architecture requirements it satisfies. \autoref{table_LA} presents the results of the study. For each requirement that the system satisfies, a “\yes” symbol is inserted in the corresponding column. If the system does not satisfy that requirement, or if the authors did not consider that requirement in their paper, a “\no” symbol was added. In case a system satisfies a requirement partially, the symbol “\halfcheckmark” was used (for example, for the \textit{Applicability} requirement, a system is partially applicable when it requires specific conditions to be implemented in a resource-constrained network). For certain requirements, the symbol “\CheckedBox” indicates that the corresponding paper studied that requirement for a specific application only. 
Note that a similar approach will be used in the next sections for other lightweight properties. 
Furthermore, in each section, we will focus on discussing the BC aspects that are related to the lightweight feature of that section only. Toward the end of the paper, we combine the findings from all sections to discuss the lightweight BC challenges and future directions as a whole.

\begin{table}[!t]
\begin{adjustwidth}{-0.6cm}{}
\centering
\caption{Mapping of lightweight architecture solutions to the lightweight architecture requirements.} 
\label{table_LA} 
\begin{tabular}{|p {0.8 cm}|p {2.7 cm}|p {1.8 cm}|p {2.2 cm}|p {2.7 cm}|p {2.0 cm}|p {2.4 cm}|}
\hline
\textbf{Ref.} & \hfil \textbf{Applicability} & \hfil \textbf{Latency} & \hfil \textbf{Scalability} & \hfil \textbf{Lightweight} & \hfil \textbf{Security} & \hfil \textbf{Usability} \\ \hline 
\hfil \cite{maftei2023massive} & \hfil \halfcheckmark & \hfil \yes & \hfil \yes & \hfil \no & \centering \CheckedBox & \hfil \yes \\
\hline 
\hfil \cite{settipalli2023extended} & \hfil \CheckedBox & \hfil \no & \hfil \no & \hfil \halfcheckmark & \hfil \yes & \hfil \CheckedBox \\
\hline 
\hfil \cite{hao2022stochastic} & \hfil \yes & \hfil \no & \hfil \yes & \hfil \no & \hfil \yes & \hfil \CheckedBox \\
\hline 
\hfil \cite{kang2022blockchain} & \hfil \CheckedBox & \hfil \no & \hfil \no & \hfil \yes & \hfil \no & \hfil \yes \\
\hline 
\hfil \cite{na2022iot} & \hfil \yes & \hfil \yes & \hfil \no & \hfil \halfcheckmark & \hfil \no & \hfil \CheckedBox \\
\hline 
\hfil \cite{yao2022accident} & \centering \CheckedBox & \hfil \no & \hfil \no & \centering \CheckedBox & \hfil \yes & \centering\arraybackslash \CheckedBox \\
\hline 
\hfil \cite{lee2021lightweight} & \hfil \yes & \hfil \no & \hfil \yes & \hfil \yes & \centering \no & \hfil \yes \\
\hline 
\hfil \cite{xie2021eclb} & \hfil \yes & \hfil \no & \hfil \halfcheckmark & \hfil \yes & \hfil \yes & \hfil \yes \\ 
\hline
\hfil \cite{le2019lightweight} & \hfil \yes & \hfil \no & \hfil \no & \hfil \yes & \hfil \no & \hfil \yes \\
\hline 
\hfil \cite{liu2023communitychain} & \hfil \CheckedBox & \centering \yes & \hfil \yes & \hfil \no & \centering \no & \hfil \halfcheckmark \\ 
\hline 
\hfil \cite{gupta2022lightweight} & \hfil \CheckedBox & \centering \no & \hfil \no & \hfil \no & \centering \yes & \hfil \CheckedBox \\ 
\hline 
\hfil \cite{mohapatra2022blockchain} & \hfil \yes & \centering \no & \hfil \yes & \hfil \no & \centering \yes & \hfil \halfcheckmark \\ 
\hline 
\hfil \cite{yang2022lightweight} & \hfil \yes & \centering \no & \hfil \yes & \hfil \no & \centering \yes & \hfil \no \\ 
\hline 
\hfil \cite{honar2021multi} & \hfil \yes & \hfil \halfcheckmark & \hfil \yes & \hfil \yes & \hfil \halfcheckmark & \hfil \yes \\
\hline 
\hfil \cite{kim2021autonomous} & \hfil \CheckedBox & \hfil \no & \hfil \no & \hfil \halfcheckmark & \centering \no & \hfil \CheckedBox \\
\hline
\hfil \cite{sunny2020towards} & \centering \CheckedBox & \hfil \halfcheckmark & \centering \no & \hfil \yes & \hfil \no & \hfil \yes \\
\hline 
\hfil \cite{shahid2019sensor} & \hfil \halfcheckmark & \centering \no & \hfil \yes & \hfil \halfcheckmark & \centering \no & \hfil \yes \\ 
\hline 
\hfil \cite{corradini2022two} & \hfil \halfcheckmark & \hfil \yes & \hfil \no & \hfil \yes & \centering \yes & \hfil \CheckedBox \\
\hline 
\hfil \cite{pyoung2019blockchain} & \hfil \yes & \hfil \no & \hfil \yes & \hfil \halfcheckmark & \centering \no & \hfil \yes \\
\hline 
\multicolumn{7}{l}{\yes = Satisfied, \no = Not satisfied, \halfcheckmark = Partially satisfied, \CheckedBox= Satisfied only for certain application} \\
\end{tabular}
\end{adjustwidth}
\end{table}

After examining the various lightweight BC architectures, analyzing their characteristics, and identifying the lightweight architecture requirements that each of them satisfies, we summarize our findings as follows. 
\begin{itemize}
    \item 
\textbf{Latency Requirements:} From \autoref{table_LA}, we notice that many researchers who proposed lightweight BC architectures have missed some crucial requirements that are essential in resource-constrained environments. For example, many applications, such as healthcare; real-time streaming; emergency response; etc. have stringent requirements in terms of latency. A blockchain-based security solution must keep the average end-to-end latency less than a certain threshold (depending on the application) in order to be applicable. Most of the lightweight BC architectures that have been proposed so far did not consider this issue or considered it but failed to prove that the architecture satisfies the latency requirements of the application. Very few solutions proved that they produce low latencies that are acceptable in resource-constrained networks.
    \item 
\textbf{Blockchain Security:} Another issue is related to the architecture security. While many solutions focused on analyzing the security from the general BC perspective, they overlook the fact that modifying the BC architecture imposes new security requirements related to that modified architecture. For example, several solutions proposed dividing the BC into two parts: a local BC within the resource-constrained network and a global BC between the edge/cloud nodes or between the edge/cloud nodes and CHs. In such cases, the two chains communicate via cross-chain bridges. A lightweight BC architecture should provide a cross-chain bridging method that is both efficient and secure. If this security aspect is not guaranteed, attackers could exploit vulnerabilities in the communication between the two chains to compromise the resource-constrained network. 
Another security concern arising from a modified BC architecture is related to the BC 51\% attack (also called the majority attack). A local chain within a cluster will contain a reduced number of verifiers who validate a local block. One of the strengths of BC technology is high decentralization, which means that a very large number of verifiers should exist such that an attacker cannot compromise more than 50\% of them. However, in a local chain, this property becomes questionable. None of the lightweight BC architectures that have been proposed studies this issue sufficiently and provides constraints on the number of verifiers in a local chain to guarantee its security. 
    \item 
\textbf{Blockchain Scalability:} 
Finally, an important characteristic of an efficient BC architecture is its scalability. In this regard, resource-constrained nodes should be able to continue implementing the BC as its size increases. Very few architectures, such as that in \cite{shahid2019sensor}, provided an efficient scalability solution. However, none of the proposed systems were able to prove the scalability of their approach in a real network setup and with a real-life BC size. For example, even when applying the aggregation method proposed in \cite{shahid2019sensor}, the size of the aggregated data may become large as the BC grows, which makes the resource-constrained node unable to store it anymore (more about this will be discussed in \autoref{Sec_LSt}). Another example is related to the cluster-based lightweight approaches: the authors of these systems assumed that the resource-constrained node should be able to support the local chain with no problems. However, the size of the local chain could become very large with time such that the resource-constrained node is no longer able to store it.  
\end{itemize}

\section{Lightweight Authentication}

In order to securely participate in a BC network, a resource-constrained node should authenticate itself to the network participants and mutually authenticate with each other node that it needs to exchange data with. In some applications, a group of resource-constrained nodes collaborate to do a certain task. Hence, a group authentication process is executed so that each node confirms the identity of each other node in the group. The authentication process must be executed before any data communication takes place in order to ensure that the communicating parties trust each other and also to exchange cryptographic keys that will be used to secure the exchanged data. 
Authentication is an essential process in any network application, which is usually followed by an access control check to determine the operations that the node can perform in the network. 

When integrating the BC into a resource-constrained network, each lightweight node needs to authenticate itself to the BC network so that it will be allowed to create BC transactions, query the BC data on full nodes, and participate in other BC operations if needed (such as verifying transactions, executing the consensus algorithm, etc.). Several researchers proposed systems for lightweight authentication via the BC. Most of these systems consider the resource-constrained node as a light BC node, while other nodes (usually cloud and/or edge/fog nodes) will be full BC nodes. Some of these systems utilize a private or consortium BC framework and add the authentication data and results as BC transactions. Other systems rely on smart contracts that execute the authentication functions. 
In addition to single-node authentication, many systems considered the group authentication process and proposed methods for lightweight group authentication using the BC. After studying the various systems that proposed BC-based lightweight authentication mechanisms for resource-constrained nodes, we divide these systems into three main categories: 

\begin{itemize}
\item 
\textbf{Cloud Authentication:} A cloud-based BC network is used to authenticate resource-constrained devices \cite{wu2023blockchain, badshah2022lake, hao2022blockchain, tan2022blockchain, tao2022b, qin2021lbac, wang2020blockslap, danish2019lightweight, yao2019bla}.
\item 
\textbf{Edge/Fog Authentication:} A network of edge or fog nodes deploy the BC and use it to authenticate resource-constrained devices. In some systems, the BC is deployed by edge/fog nodes only. In other systems, cloud servers also participate in the authentication BC \cite{gowda2023bskm, harbi4149706lightweight, islam2022fbi, wang2022dag, alkhazaali2020lightweight, cui2020hybrid, khalid2020decentralized}. 
\item 
\textbf{Group Authentication:} A group of resource-constrained nodes apply a common process to authenticate each other \cite{naresh2022provably, tan2022blockchain, wang2022lightweight, lin2019homechain}. 
\end{itemize}

In what follows, we review and analyze the lightweight authentication systems that have been proposed in each category.

\subsection{Cloud Authentication}

An authentication scheme between patients and medical servers using a consortium BC was proposed in \cite{wu2023blockchain}. It merges BC with biometric fuzzy extraction to create a secret session key. Members of the consortium BC comprise medical servers and the registration center (RC). 
When the patient sends an authentication request to the medical server, the latter calls the smart contract to verify the patient’s information. If the patient is authentic, the server and the patient agree on a shared session key. The server then sends the authentication result to the accounting node and saves it to the BC. 

A lightweight authentication and key exchange model for a blockchain-enabled Smart Grid environment was proposed in \cite{badshah2022lake}. In this system, service providers (SPs) operate a private BC to authenticate smart meters (SMs). The system also contains a trusted registration authority that registers the SMs and SPs in the BC before the system starts. An SM establishes mutual authentication with the SP via a key reproduction algorithm. After that, both parties generate a shared secret that will be used in future communications and a session key. When an SP receives a request from an SM, it creates a BC transaction and adds it to a global transaction pool. When the transaction threshold is reached in the pool, a leader SP is elected via round-robin to create the next block. 

Another authentication system that uses consortium BC was proposed in \cite{hao2022blockchain}. The BC stores the access policies, provisions of authentication services, and trust scores of access request nodes. The system facilitates access control within and between different domains. The authors add new opcodes for authentication purposes to the locking and unlocking scripts of BC transactions. Full BC nodes process the opcodes and execute the corresponding authentication actions on the BC. The authors tested the proposed system using four cryptographic algorithms, with the Chacha20 cipher yielding the best performance.

The authors in \cite{tan2022blockchain} use a permissioned BC to save authentication data of UAVs, while smart contracts are used to perform authentication-related operations. The BC is saved by peer nodes in the cloud. The nodes are divided into endorsers; who endorse UAVs authentication requests, orderers, who create new blocks by combining authentication transactions, and validators, who validate new blocks and execute the consensus protocol to add them to the BC. 
UAVs register with the ground control station to get a pseudonym and a public/private key pair, and then publish their public key (PK) and Dynamic Parameter (DP) in the BC. 
When UAVs want to communicate with each other, they use the key agreement process to generate a session key. 

The authors in \cite{tao2022b} propose a method for obfuscating the access control policy with fuzzy attribute set to enhance the authentication security. In addition, the proposed system outsources complex operations to proxy servers to reduce the computing complexity of system users. The framework contains five algorithms: \textit{Setup}, \textit{KeyGen}, \textit{Enc}, \textit{ProxyDec}, and \textit{Dec}. During Setup, the public key is generated such that it contains the attributes set that is selected from generic attributes. \textit{KeyGen} is used to output the attribute private key of each user. \textit{Enc} takes an access control predicate, the shared messages, and the public key to generate the ciphertext. \textit{ProxyDec} is executed by the proxy server to preliminary decrypt the ciphertext and outputs a proxy ciphertext only if the user attribute set satisfies the access control policy. Finally, \textit{Dec} is executed by the data users to get the shared message. 

A consortium BC is proposed to authenticate resource-constrained nodes via an attribute-based encryption (ABE) scheme in \cite{qin2021lbac}. The system contains a certificate authority (CA) that manages the attribute sets that are assigned by attribute authorities to users, and generates the attribute tokens that are used by users to access the BC. 
The resource-constrained nodes that generate data are called Data Owners (DO). DOs encrypt their data using ABE, save the ciphertext at cloud service providers (CSPs), and store the encrypted symmetric keys and data hash in the BC. In order to request a data transaction, a resource-constrained node submits an access request to the BC smart contract, obtains the encrypted ABE key and hash, downloads the ciphertext from the CSP, and decrypts the ciphertext with the ABE key to obtain the data. Although the approach is interesting, it puts significant overhead on the resource-constrained nodes in terms of latency. 
In addition, the degree of decentralization is reduced. 

The authors of \cite{wang2020blockslap} propose a lightweight authentication model for smart grid systems. The network contains a Smart Meter (SM), a utility center (UC) that manages groups of SMs, in addition to a set of registration authorities (RA). 
The permissioned BC within the RA network includes a smart contract for registering the SMs and UCs and authenticating them. 
When the SM and UC want to communicate with each other, the SM sends an authentication request to the UC that contains the SM’s ID and signature. The UC authenticates the SM via the RA. If the SM is authentic, the UC generates the session key and sends it to the SM along with the UC’s signature. The SM verifies the UC’s signature via the smart contract before it starts using the session key to exchange data with the UC. The authors do not consider the case when the RA is compromised. In addition, the proposed system depends on the existence of a secure channel between the SM/UC and the RA.

Danish \textit{et al.} \cite{danish2019lightweight} propose a permissioned BC system for the authentication of resource-constrained devices when joining a LoRaWAN. The authors propose a separate BC network, such as a cloud-based network formed from the vendors’ servers, that is used to store the authentication and private data of resource-constrained devices. The BC is used to securely provide the device authentication data to the network server when the device needs to authenticate before starting a new session. 
When the device connects to the network server, it sends an authentication request that contains the certificate. The network server authenticates the device by sending the certificate to the BC network. The problem with this approach is that whenever the device wants to start a new session it needs to authenticate, which generates a lot of delay at the start of the session. In addition, the system considers only device-to-network server communication but not device-to-device communication. 

Vehicular fog services (VFS) are systems that provide cloud services to vehicles by exploiting fog nodes such as RSUs. The authors of \cite{yao2019bla} propose a system in which a vehicle registers with the BC network, authenticates itself, and sends an authentication token to the RSU each time it needs to consume a new VFS. The system consists of the Audit Department (AD), the Service Managers (SM), which are central servers that manage a group of data centers, the Witness Peers (WP, one per cluster) who are responsible for the BC consensus, and the RSUs. The SMs and WPs create the consortium BC that is used to store the authentication data of vehicles. Each vehicle authenticates itself by sending the authentication data to the SM via the RSU. The SM validates the authentication data and adds the result as a new transaction in the BC. 

A summary of the solutions discussed in this section is shown in \autoref{table_AuthSum1}.

\begin{table}[!t]
\begin{adjustwidth}{-1.2cm}{}
\centering
\caption{Summary of lightweight cloud authentication solutions.} 
\label{table_AuthSum1} 
\centering
\begin{tabular}{|p {0.7 cm}|p {1.9 cm}|p {1.2 cm}|p {5.0 cm}|p {3.5 cm}|p {3.5 cm}|}
\hline
\textbf{Ref.} & \hfil \textbf{Application.} & \hfil \textbf{Year.} & \hfil{\textbf{Main idea}} & \hfil \textbf{Pros} & \hfil \textbf{Cons}  \\ 
\hline 
\hfil \cite{wu2023blockchain} & \centering Healthcare & \hfil 2023 & Merging BC with biometric fuzzy extraction to create the session key & - Use of XOR and hash algorithms for authentication & - Limited testing \newline- Poor scalability \\
\hline 
\hfil \cite{badshah2022lake} & \centering Smart grid & \hfil 2022 & Service providers operate a private blockchain to authenticate smart meters & - Key reproduction algorithm \newline- Dynamic addition of SMs and SPs & - Very high latency \newline- Poor scalability \\
\hline 
\hfil \cite{hao2022blockchain} & \centering General & \hfil 2022 & Consortium BC for cross-domain authentication; new opcodes for authentication in transaction script & - Scalability \newline- Token accumulation mechanism & - High latency \newline- BC network is isolated \\
\hline 
\hfil \cite{tan2022blockchain} & \centering UAV networks & \hfil 2022 & BC stores the authentication data of UAVs; UAVs and GCSs use APIs to call smart contract & - Use of Dynamic Parameter (DP) \newline- Automatic session key generation & - Requires secure channel \newline- Scalability not tested \\
\hline
\hfil \cite{tao2022b} & \centering General & \hfil 2022 & Obfuscating the access control policy with fuzzy attribute set & - Outsourcing complex operations to proxy server & - High storage overhead \newline- Data owner could be compromised \\
\hline 
\hfil \cite{qin2021lbac} & \centering General & \hfil 2021 & DOs encrypt their data using ABE; resource-constrained nodes accesses data using attribute tokens & - Use of ABE \newline- DOs control data access & - High latency \newline- Single point of failure \\
\hline 
\hfil \cite{wang2020blockslap} & \centering Smart grid & \hfil 2020 & Smart Meters and Utility Centers mutually authenticate each other via BC smart contract & - Use of smart contract \newline- Authentication via RA & - Requires secure channel \newline- RA could be compromised \\
\hline
\hfil \cite{danish2019lightweight} & \centering LoRaWAN & \hfil 2019 & Authentication of resource-constrained devices with network server when joining a LoRaWAN & - Security \newline- Scalability   & - High delay \newline- No device-to-device authentication \\
\hline
\hfil \cite{yao2019bla} & \centering Vehicular fog services & \hfil 2019 & SMs authenticate vehicles and send authentication tokens to enable them to consume fog services & - Witness Peers \newline- Consuming multiple services via single token & - High authentication delay \newline- RSUs are trusted \\
\hline
\end{tabular}
\end{adjustwidth}
\end{table}

\subsection{Edge/Fog Authentication}

A decentralized system for a fog computing environment was proposed by Gowda \textit{et al.} in \cite{gowda2023bskm}. Here, the authors utilize a one-way hash chain for the generation of private and public key pairs of the resource-constrained edge devices (ED). Upon successful authentication of the two EDs that want to communicate, the session key generation at both EDs is based on the key pair provided by the fog server and stored in the private BC. The fog server issues the key pairs to respective EDs in an encrypted format based on their public key. When the EDs receive the shared key pair and other credentials, they will validate them based on the time of arrival and then decrypt them. Finally, each of the two EDs calculates the session key based on the key pairs and random numbers of both EDs.

Harbi \textit{et al.} \cite{harbi4149706lightweight} discuss a mechanism for maintaining the privacy of IoT data in the cloud. 
The private authentication BC is maintained by fog nodes. 
Each user registers with a single fog node by sending to it the hash of its ID and password. The fog node authenticates the user via the smart contract and then sends to it a user trust token. The user inputs the received token and its biometric information to a fuzzy extractor function to generate its certificate. When the user connects to the cloud, it signs the request using its certificate and the current timestamp. The fog node authenticates the user by inputting the signature to the BC smart contract. 

A blockchain-based Federated Learning-assisted data collection scheme, in which drones are deployed to aid resource-constrained devices, was proposed in \cite{islam2022fbi}. The proposed system uses differential privacy (DP) to ensure privacy during data gathering. In addition, a two-phase authentication mechanism including a cuckoo filter (CF) and timestamp-based nonce verification is used. The verification method involves a Hampel filter (HL) and loss checks. Resource-constrained nodes store their data in the private BC and share a trained model with the drones. 
Each drone stores a light copy of the BC that includes the participants’ information and collected data. Drones are used to deploy training models at resource-constrained devices and collect the results. A drone will authenticate each device before it is allowed to contribute to the training process. After two invalid authentication attempts, the drone adds the resource-constrained node to the Vulnerable List, and after a certain number of invalid attempts, the drone adds the node to the Malicious List.

Fog nodes implement the distributed ledger technology IOTA to authenticate resource-constrained nodes in \cite{wang2022dag}. Each fog node supervises a collection of resource-constrained devices adjacent to it. In order to assure the legitimacy, non-repudiation, and non-forgery of the identity information of resource-constrained devices, IOTA is utilized as an identity management technique to reach a consensus on the identity of devices that want to participate in network activities. Each device registers with the fog node. The latter validates the device identity on the directed acyclic graph (DAG) and issues a public certificate to the device, after obtaining the consensus of the IOTA members. After that, the device access control attributes are generated by the fog node and added to the ledger. Two resource-constrained devices can mutually authenticate each other via their supervisor fog nodes.

In \cite{alkhazaali2020lightweight}, the authors describe a BC-based authentication system that 
contains four components: a trusted Registration Node (RN) that registers the other nodes in the BC and tracks illegal activities, Fog Computing Nodes (FCN) that act as full BC nodes; each FCN manages the resource-constrained nodes in its section, Organizer Peer (OP) that administers the BC consensus algorithm, and resource-constrained devices. 
Each entity exchanges messages with the RA to authenticate itself using its secret identity and obtain its credentials and certificate. 
A resource-constrained node signs a request using its certificate and sends it to the FCN. The latter validates the signature and timestamp before executing the request on the smart contract. When the endorsement policy is fulfilled, the resource-constrained node collects the result from the FCN and validates it before processing the data. 

The authors in \cite{cui2020hybrid} propose a blockchain-based authentication mechanism that depends on clustering. The resource-constrained nodes are divided into clusters with CHs. Each CH should have a direct link to at least one base station. The system comprises a hybrid BC approach: the base stations and cloud users engage in a public BC, while the CHs within each subnet (WSN) create a private local BC. The public BC stores the identities of all WSN nodes and the smart contracts that are used to authenticate CHs and authorize communications between nodes in different subnets. The local BC is used to authenticate the local nodes and execute the smart contracts that authorize the communication between two WSN nodes in the local subnet. 
One of the disadvantages of this system is that the real IDs of nodes are stored in the public BC. Another problem is that in a small WSN, the number of clusters is not large, which makes the local BC less secure. 

In \cite{khalid2020decentralized}, the authors propose a BC network composed of fog nodes that store a private authentication BC. Each group of resource-constrained devices within a certain system (home, hospital, company, etc.) is considered a single group and associated with a single fog node. Each group is registered with the BC network by utilizing a secure channel with the fog node. The latter binds each device ID with a blockchain ID and saves them in the BC. Also, the fog node sends an authentication certificate to the admin who shares it with all devices in the group. Next, each device authenticates itself with the fog node and receives a device-dedicated certificate. When a device needs to communicate with another, the device sends a communication request that contains the IDs of the two devices and the device certificate to the fog node. 

A summary of the solutions discussed in this section is shown in \autoref{table_AuthSum2}.

\begin{table}[!t]
\begin{adjustwidth}{-1.2cm}{}
\centering
\caption{Summary of lightweight edge/fog authentication solutions.} 
\label{table_AuthSum2} 
\centering
\begin{tabular}{|p {0.7 cm}|p {1.9 cm}|p {1.2 cm}|p {5.0 cm}|p {3.5 cm}|p {3.5 cm}|}
\hline
\textbf{Ref.} & \hfil \textbf{Application.} & \hfil \textbf{Year.} & \hfil{\textbf{Main idea}} & \hfil \textbf{Pros} & \hfil \textbf{Cons}  \\ 
\hline 
\hfil \cite{gowda2023bskm} & \centering General & \hfil 2023 & Session key generation at edge devices is based on the key pair provided by fog server and stored in the BC & - Automatic calculation of session key \newline-  Low computation and storage cost & - High latency \newline- Fog node is fully trusted \\
\hline
\hfil \cite{harbi4149706lightweight} & \centering General & \hfil 2022 & Trust token is sent by fog node to user who uses it with biometric information to generate certificate & - Use of  fuzzy extractor function \newline-  Resistance against several attacks & - Limited to user authentication \newline- Limited testing \\
\hline
\hfil \cite{islam2022fbi} & \centering UAV networks & \hfil 2022 & Two-phase authentication scheme via cuckoo filter (CF) and timestamp-based nonce verification & - Use of differential privacy \newline-  FL-based data accumulation & - High computational overhead \newline- Poor scalability \\
\hline
\hfil \cite{wang2022dag} & \centering IOTA DAG only & \hfil 2022 & Fog nodes implement IOTA DAG to reach consensus on the identity of lightweight devices & - Use of IPFS \newline-  Non-repudiation & - No smart contracts \newline- Limited testing \\
\hline 
\hfil \cite{alkhazaali2020lightweight} & \centering General & \hfil 2020 & Lightweight node signs request using certificate and sends it to FCN who executes it on smart contract & - Simple and lightweight \newline-  Low latency & - Single point of failure (OP) \newline- Weak security \\
\hline
\hfil \cite{cui2020hybrid} & \centering General & \hfil 2020 & Each lightweight node authenticates with its CH via smart contract on the local or global BC & - Clustering \newline- High security & - High communication overhead \newline- Poor decentralization \\
\hline 
\hfil \cite{khalid2020decentralized} & \centering General & \hfil 2020 & Lightweight nodes register with a fog node and obtain a dedicated certificate that is used for authentication & - Creation of communication transaction \newline- Lightweight & - Fog nodes are fully trusted \newline- Requires secure channel \\
\hline 
\end{tabular}
\end{adjustwidth}
\end{table}

\subsection{Group Authentication}

Naresh \textit{et al.} \cite{naresh2022provably} propose a group key agreement mechanism that utilizes the BC smart contract. The latter plays the role of the Group Controller (GC) that is responsible for generating the shared key and sharing it with the group members. The proposed system utilizes the \textit{ECDSA} algorithm to generate the required keys, which are the public/private keys for each node and the GC, the two-party common key between the GC and each node, and the shared key between the GC and the group members. Each node generates its two-party common key with the GC by using ECC multiplication of its private key and the GC’s public key. Similarly, the GC generates the two-party common key with each node via the ECC multiplication of its private key with the node’s public key. Next, the GC sends to each node \textit{N} the first part of the shared key (encrypted with the two-party key), which is calculated by multiplying the two-party keys of all nodes in the group except the two-party key of \textit{N}. When nodes join or leave the group, the GC recalculates and sends the first part of the group shared key to each node in the group. 

Tan \textit{et al.} \cite{tan2022blockchain} propose a method for group authentication between groups of UAVs via BC smart contracts. Here, a set of session keys are generated, one for each pair of nodes in the group; 
The session key is generated based on the two members’ public keys and Dynamic Parameters (DPs). The authors show that the proposed group authentication mechanism produces a much smaller authentication delay as compared to other approaches. However, the communication overhead of the proposed system is high. 

A five-tier blockchain-based architecture for authentication in the smart grid was proposed in \cite{wang2022lightweight}. The BC saves the key distribution process and system parameters that are used by the dispatching center 
to secure the smart grid and avoid tampering with the system. The proposed system uses a certificateless group key agreement technique that enables the key distribution process to be executed without bilinear pairing. In addition, the system includes a key reconstruction mechanism based on the joining and leaving of gateway nodes. The dispatching center acts as a super node in the proposed consortium BC, while the servers in the smart grid centers act as full BC nodes. 
During each session, the dispatching center generates a shared secret key between each group of communicating gateways. 

Lin \textit{et al.} \cite{lin2019homechain} propose a BC-based system for remote mutual authentication between users and smart home gateways. 
The system uses a key pair and a message authentication code (MAC) to ensure that only the requester is able to receive the response from the legitimate home gateway. All devices in a home, including smart devices and user equipment, constitute a group, from which a group manager is selected. The group manager registers the group devices in the BC. After registration, each device in the group obtains its own group private key, while the whole group receives a single group public key. When a group member wants to publish an access or control request with the home gateway, a new public/private key pair is generated to avoid replay attacks and profiling. Once the request is executed, the home gateway encrypts the result using the group public key and computes the corresponding MAC using its private key. The result is uploaded into the BC smart contract.

A summary of the solutions discussed in this section is shown in \autoref{table_AuthSum3}.

\begin{table}[!t]
\begin{adjustwidth}{-1.2cm}{}
\centering
\caption{Summary of lightweight group authentication solutions.} 
\label{table_AuthSum3} 
\centering
\begin{tabular}{|p {0.7 cm}|p {1.9 cm}|p {1.2 cm}|p {5.0 cm}|p {3.5 cm}|p {3.5 cm}|}
\hline
\textbf{Ref.} & \hfil \textbf{Application.} & \hfil \textbf{Year.} & \hfil{\textbf{Main idea}} & \hfil \textbf{Pros} & \hfil \textbf{Cons}  \\ 
\hline 
\hfil \cite{naresh2022provably} & \centering General & \hfil 2022 & Smart contract is the Group Controller that generates the shared key and shares it with the group members & - High security \newline- Dynamic update of group key & - High storage and computation requirements \\
\hline
\hfil \cite{tan2022blockchain} & \centering UAV networks & \hfil 2022 & Dedicated session key is generated for each group member when communicating with another member  & - Use of Dynamic Parameter (DP) \newline- High security & - High computation and communication overhead  \\
\hline
\hfil \cite{wang2022lightweight} & \centering Smart grid & \hfil 2022 & Dispatching center generates shared secret key between each group of communicating gateways  & - Certificateless agreement \newline- Dynamic update of shared key & - No testing \newline- Single point of failure (DC)  \\
\hline
\hfil \cite{lin2019homechain} & \centering Smart home & \hfil 2019 & Device uses group private key to authenticate with home gateway and obtain session key pair  & - Privacy protection of access policy \newline- Behavior auditing via BC & - Very high latency \newline- High overhead at system setup  \\
\hline
\end{tabular}
\end{adjustwidth}
\end{table}

\begin{figure*}[!t]
\noindent

\makebox[\textwidth]{\includegraphics[width=7.2in]{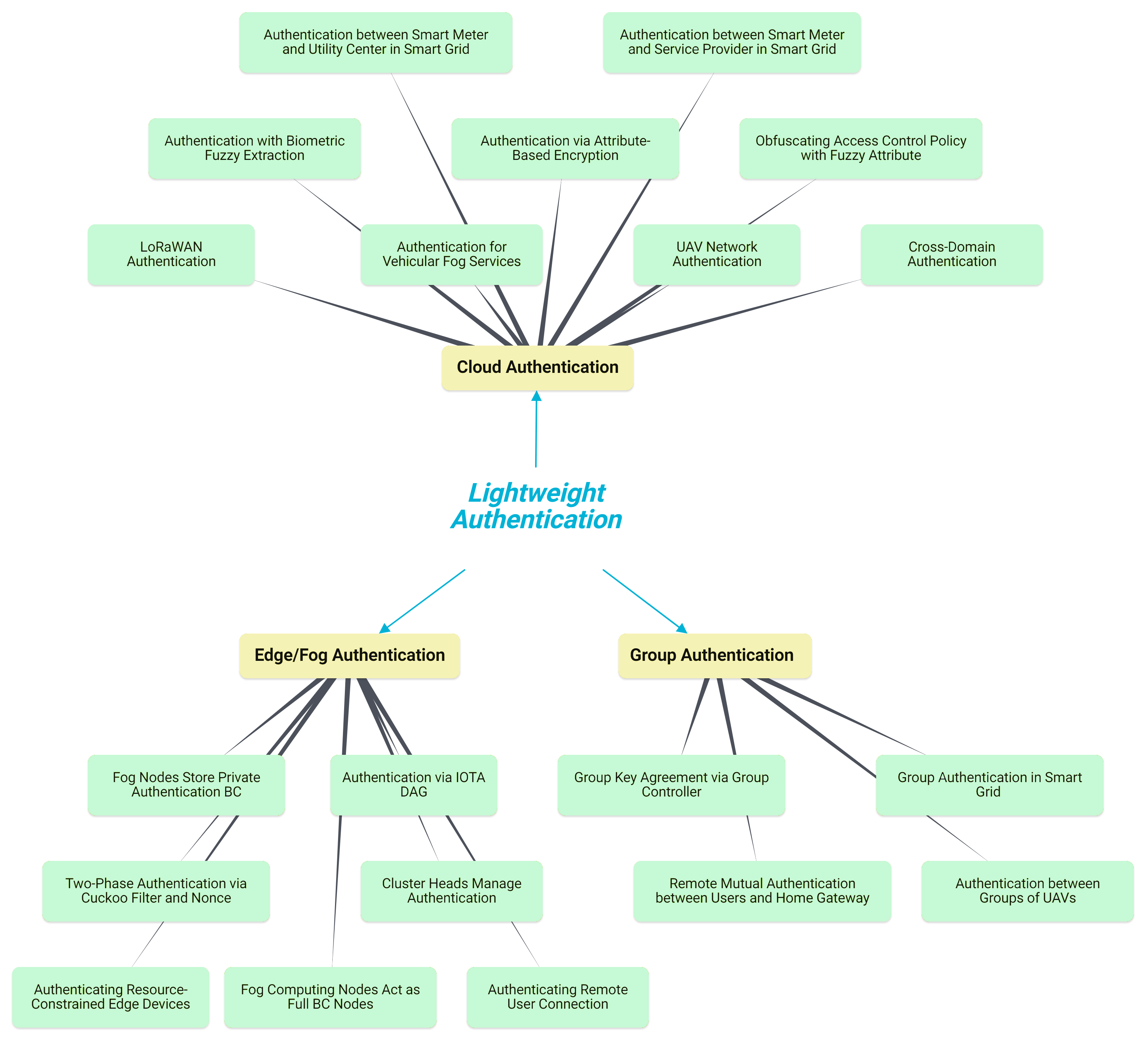}}
\caption{Methods for lightweight authentication via the blockchain.}
\label{fig_LAu}
\end{figure*}

The main ideas of the lightweight authentication systems discussed in this section are summarized in \autoref{fig_LAu}.

\subsection{Analysis and Discussion}

After studying the different methods that have been proposed for lightweight BC authentication in resource-constrained networks, 
we outline the requirements for lightweight authentication. These requirements were deduced by examining the objectives that were stated by the authors of the studied papers. 
\begin{enumerate}
\item 
\textbf{Confidentiality:} guaranteeing that the authentication data of the resource-constrained node will not be stolen by attackers.
\item 
\textbf{Integrity:} guaranteeing that the authentication data of the resource-constrained node will not be modified by attackers.
\item 
\textbf{Anonymity:} enabling each resource-constrained device to use a pseudonym or virtual identification instead of its real ID and change it when needed.
\item 
\textbf{Traceability:} guaranteeing that authorities can trace a malicious resource-constrained node, revoke its authentication/authorization credentials, and stop it from accessing the network. 
\item 
\textbf{Non-repudiation:} guaranteeing that when a malicious resource-constrained node is detected, its pernicious activities can be strongly linked to its real ID. 
\item 
\textbf{Lightweight:} guaranteeing that the authentication/authorization process produces low 
overhead on the resource-constrained node.
\end{enumerate}

Next, we evaluate the reviewed solutions by checking which lightweight authentication requirements they satisfy. 
The results are shown in \autoref{table_LAu}. Note that all these approaches provide confidentiality and integrity for the authentication data of resource-constrained nodes via the BC. However, some of these systems contain security issues as previously discussed. Because of these issues, the confidentiality and integrity provided by these systems are questionable. Hence, the symbol “$\mathsf{?}$” was used in \autoref{table_LAu} to represent uncertainty. On the other hand, the symbol “$\otimes$” was used when the analyzed system did not study the corresponding requirement, which makes it difficult to determine whether the system satisfies or not the requirement. With respect to the “Lightweight” requirement, we used the “$\uparrow$”, “$\leftrightarrow$”, and “$\downarrow$” symbols to depict that the system satisfies this requirement to a “high”, “medium”, or “low” degree, respectively. 

\begin{table}[!t]
\begin{adjustwidth}{-1.3cm}{}
\centering 
\caption{Mapping of lightweight authentication solutions to the lightweight authentication requirements.} 
\label{table_LAu} 
\begin{tabular}{|p {0.8 cm}|p {3.0 cm}|p {1.8 cm}|p {2.2 cm}|p {2.4 cm}|p {3.5 cm}|p {2.4 cm}|} 
\hline 
\textbf{Ref.} & \textbf{Confidentiality} & \textbf{Integrity} &  \textbf{Anonymity} &  \textbf{Traceability} &  \textbf{Non-repudiation} &  \textbf{Lightweight} \\ \hline 
\hfil \cite{wu2023blockchain} & \hfil \yes & \hfil \yes & \hfil \yes & \hfil \no & \hfil \no & \hfil $\leftrightarrow$ \\
\hline 
\hfil \cite{badshah2022lake} & \hfil \yes  & \hfil \yes & \hfil \yes & \hfil \no & \hfil \no & \hfil $\downarrow$ \\
\hline 
\hfil \cite{hao2022blockchain} & \hfil \yes  & \hfil \yes & \hfil \no & \hfil \yes & \hfil \no & \hfil $\leftrightarrow$  \\
\hline 
\hfil \cite{tan2022blockchain} & \hfil $\mathsf{?}$ & \hfil  $\mathsf{?}$ & \hfil \yes & \hfil \yes & \hfil $\otimes$ & \centering\arraybackslash $\downarrow$ \\ 
\hline
\hfil \cite{tao2022b} & \hfil $\mathsf{?}$ & \hfil $\mathsf{?}$ & \hfil \no & \hfil \no  & \hfil \no  & \hfil $\downarrow$ \\
\hline 
\hfil \cite{qin2021lbac} & \hfil \yes & \hfil \yes & \hfil \no & \hfil \no & \hfil $\otimes$ & \centering\arraybackslash $\downarrow$ \\
\hline
\hfil \cite{wang2020blockslap} & \hfil $\mathsf{?}$ & \hfil $\mathsf{?}$ & \hfil \yes & \hfil \no & \hfil $\otimes$ & \hfil $\leftrightarrow$ \\
\hline 
\hfil \cite{danish2019lightweight} & \hfil \yes & \hfil \yes & \hfil \no & \hfil \no & \hfil $\otimes$ & \hfil $\leftrightarrow$ \\ 
\hline 
\hfil \cite{yao2019bla} & \hfil \yes & \hfil \yes & \hfil \yes & \hfil \yes & \hfil \yes & \centering\arraybackslash $\leftrightarrow$ \\ 
\hline 
\hfil \cite{gowda2023bskm} & \hfil $\mathsf{?}$  & \hfil $\mathsf{?}$  & \hfil \no & \hfil \no & \hfil \no & \hfil $\leftrightarrow$ \\
\hline 
\hfil \cite{harbi4149706lightweight} & \hfil $\mathsf{?}$ & \hfil $\mathsf{?}$ & \hfil \yes & \hfil \yes  & \hfil \no & \hfil $\downarrow$ \\
\hline 
\hfil \cite{islam2022fbi} & \hfil \yes  & \hfil $\otimes$ & \hfil \no & \hfil \no  & \hfil \no  & \hfil $\downarrow$ \\
\hline 
\hfil \cite{wang2022dag} & \hfil \yes & \hfil \yes & \hfil \yes & \hfil \no & \hfil \yes & \hfil $\mathsf{?}$ \\
\hline 
\hfil \cite{alkhazaali2020lightweight} & \hfil \yes & \hfil \yes & \hfil \no & \hfil \yes  & \hfil \no & \hfil $\uparrow$ \\
\hline 
\hfil \cite{cui2020hybrid} & \hfil \yes & \hfil \yes & \hfil \no & \hfil $\otimes$ & \hfil \yes & \centering\arraybackslash $\downarrow$ \\ 
\hline
\hfil \cite{khalid2020decentralized} & \hfil \yes & \hfil \yes & \hfil \no & \hfil \no & \hfil \yes & \centering\arraybackslash $\leftrightarrow$ \\
\hline
\hfil \cite{naresh2022provably} & \hfil \yes & \hfil \yes & \hfil \yes & \hfil $\otimes$ & \hfil $\otimes$ & \centering\arraybackslash $\leftrightarrow$ \\
\hline
\hfil \cite{wang2022lightweight} & \hfil \yes  & \hfil \yes & \hfil \no & \hfil \no  & \hfil \no & \hfil $\mathsf{?}$ \\
\hline 
\hfil \cite{lin2019homechain} & \hfil \yes & \hfil \yes & \hfil \yes & \hfil \yes  & \hfil \no & \hfil $\downarrow$ \\
\hline 
\multicolumn{7}{l}{\yes = Satisfied, \no = Not satisfied, $\otimes$=Not studied,  $\uparrow$=High, $\leftrightarrow$=Medium, $\downarrow$=Low, $\mathsf{?}$= Uncertainty } \\
\end{tabular}
\end{adjustwidth}
\end{table}

After examining the lightweight authentication systems, we deduce the following insights:
\begin{itemize}
    \item 
\textbf{Authentication Overhead:} We notice that most of the studied systems 
do not add significant overhead in normal conditions. However, some of them could lead to increased overhead that may disrupt the job of the resource-constrained node in certain situations. For example, some systems require a resource-constrained node to send a registration request to the BC. The fog/cloud node that receives the request creates a transaction that contains the request and adds it to the pool of pending transactions. When the consensus algorithm is executed and the next block is added to the BC, the node is authenticated. This process results in a considerable authentication delay. Other approaches require that all the authentication-related updates to the BC are broadcast to all resource-constrained nodes, which could result in high communication overhead in large networks. A similar problem appears in some group authentication systems, with nodes joining or leaving the group resulting in recalculating the group shared key and broadcasting it to all nodes in the group. On the other hand, some systems 
produce an average transaction latency greater than 2 seconds, 
which is not acceptable in many delay-sensitive applications. 
\item 
\textbf{Node Anonymity:} Another important characteristic that should exist in the lightweight BC authentication system is related to conditional anonymity. Here, the resource-constrained node should be able to use alternative identification, such as pseudonyms, instead of its real ID. At the same time, the real IDs should be stored securely and associated with the alternative IDs such that authorities are able to trace the real ID of any node when it behaves maliciously and remove it from the network. Several studied systems did not consider these two important properties (i.e., anonymity and traceability) in the authentication mechanism. Finally, another important property is non-repudiation, which states that all actions that are done by the node are stored in the BC and associated with the node ID, which removes the possibility that the node can deny any operation that it did. The authentication system should include a method for ensuring non-repudiation. However, we notice from \autoref{table_LAu} that this property was considered by only four of the studied systems. 
\item 
\textbf{Cryptography System:} Another issue is related to using lightweight cryptography in the authentication of resource-constrained nodes. Note that a detailed discussion on lightweight cryptography for the BC will be presented in a later section. With respect to authentication, we notice that all authentication systems that were discussed in this section use traditional cryptography algorithms (such as RSA, Diffie–Hellman, ECC, Keccak, etc.) during the authentication process. A lightweight authentication system for the BC should exploit the capabilities of a highly lightweight cryptography mechanism to provide high security and less overhead during the authentication and key exchange processes. More details about the lightweight cryptography systems will be presented in \autoref{Sec_LCr}. 
\end{itemize}

To conclude, although the systems that were examined provide interesting approaches for lightweight BC authentication, 
more effort should be made to create a system that is highly secure and satisfies the \textbf{anonymity}, \textbf{traceability}, and \textbf{non-repudiation} properties, while producing acceptable processing, communication, and delay \textbf{overhead}.

\section{Lightweight Consensus}

The consensus algorithm forms the backbone of a BC system, ensuring agreement among participants on the validity and ordering of transactions. Consensus ensures also that all participants reach an agreement on the current state of the BC. This guarantees that every participant has an identical copy of the ledger, reflecting the same set of blocks and transactions. 
However, the resource-intensive nature of traditional consensus mechanisms, such as PoW and PBFT, poses challenges to scalability and efficiency, particularly in resource-constrained environments. To address these limitations, researchers have explored lightweight consensus solutions that aim to achieve consensus with reduced computational overhead and energy consumption. In this section, we review and analyze the mechanisms proposed in the literature and examine their fundamental principles and performance characteristics. After studying the various existing lightweight consensus solutions, we classified them into the following four categories:


\begin{itemize}

\item \textbf{Clustering approaches:} These systems apply clustering to the resource-constrained network and implement the consensus algorithm within each cluster to reduce its overhead \cite{ekanayake2021lightweight, mohanty2020efficient, uddin2019lightweight}. 

\item  \textbf{Based on existing algorithms:} These systems attempt to modify existing consensus algorithms such as PoW, PoS, PBFT, etc. to make them suitable for resource-constrained networks  \cite{Qi2023LightPoW, kara2022proof, kong2022lap, wang2022airbc, chai2021cyberchain, li2021lightweight, andola2020poewal, su2020lvbs}. 

\item \textbf{Using fog nodes as miners:} Several systems make use of fog/edge nodes that connect the resource-constrained network to the main network and make fog/edge nodes execute the consensus algorithm \cite{arifeen2022autoencoder, xu2022microchain, mershad2021proof, zhang2020ldc, zheng2020lightweight}.

\item \textbf{Reputation-based and Random-based consensus:} These systems apply a method to select the BC miners either based on reputation or randomly. In reputation-based consensus, resource-constrained nodes are ranked based on their reputations, and the block creation task is assigned to the node(s) with the highest reputation. In random-based consensus, the miner is selected based on a randomization approach \cite{chen2022lblco, islam2022integrating, xi2022crowdlbm, dorri2020TreeChain, worley2020scrybe}.
\end{itemize}

In the following sections, we discuss the systems that were proposed in each category.

\subsection{Clustering Approaches} 

A new consensus protocol, Proof of equivalent Work (PeW), is proposed in \cite{ekanayake2021lightweight}. 
The system works by dividing the BC into separate shards where each shard contains a master node and a group of slave nodes, hence creating a master-slave blockchain (MSB). The transactions to be included in the next block are distributed between master nodes. Each master node assigns the task of finding the nonce to its slave nodes. Each slave node calculates the nonce and sends the results to the master. The latter collects the individual works of its slaves and creates the block. 

The authors of \cite{mohanty2020efficient} propose a lightweight consensus model for smart-home networks. The IoT nodes in each home form a cluster in which a CH is selected to play the role of the Overlay Block Manager (OBM). The OBMs create and mine the BC blocks. Each OBM saves the local blocks in the local storage space while the complete BC is stored at the cloud service providers. An OBM receives the data transactions from the IoT nodes in its cluster and other OBMs, saves them to create its block, and broadcasts the block to other OBMs. The consensus operation is conducted as follows: Each OBM generates an arbitrary time value \textit{T} after which the OBM creates its block. 
If the OBM receives a new block from another OBM while it is waiting, it resets the process by generating a new random \textit{T}. If the OBM waits for \textit{T} without receiving a block from another OBM, it generates its block and broadcasts it to the other OBMs. 

In \cite{uddin2019lightweight}, the authors propose a BC system for underwater IoT networks (IoUT). The IoUT nodes are clustered geographically and the CH is selected as the most capable node in the cluster. Gateways are placed at the edge of the network and play the role of fog nodes that connect to the cloud service providers and the near IoUT nodes. Nodes in a cluster forward their data to the CH. The latter creates transactions and forwards them to the higher-level CH, and so on until the transactions reach one of the gateways. The latter orders the transactions it receives, creates the block, and selects a group of BC miners to execute a modified PoS algorithm 
that is based on the Technique for Order of Preference by Similarity to Ideal Solution (TOPSIS) mechanism. The latter considers several criteria such as geometric distance, reputation, amount of stake, attack vulnerability, and throughput. 

A summary of lightweight consensus solutions based on clustering approaches is given in \autoref{table_ConSClust}.

\begin{table}[!t]
\begin{adjustwidth}{-1.2cm}{}
\centering
\caption{Summary of lightweight consensus solutions - clustering approaches.} 
\label{table_ConSClust} 
\centering
\begin{tabular}{|p {0.7 cm}|p {1.9 cm}|p {1.2 cm}|p {5.0 cm}|p {3.5 cm}|p {3.5 cm}|}
\hline
\textbf{Ref.} & \hfil \textbf{Application.} & \hfil \textbf{Year.} & \hfil{\textbf{Main idea}} & \hfil \textbf{Pros} & \hfil \textbf{Cons}  \\ 
\hline 
\hfil \cite{ekanayake2021lightweight} & \centering  General & \hfil 2021 & Dividing the blockchain into two layers: master node(s) and slave agents (SAs). & - Distribution of consensus tasks \newline- Partial rewards mechanism  & - Very high delay in transactions addition \newline- Weak security \\
\hline 

\hfil \cite{mohanty2020efficient} &  \centering  Smart home &  \hfil 2020 &  CH plays the role of BC manager and is responsible for collecting transactions and mining blocks &  - Good scalability \newline- Fair block generation distribution &  - Forking problem \newline- Possibility of discarding blocks \\
\hline

\hfil \cite{uddin2019lightweight} &  \centering  Underwater IoT &  \hfil 2019 &  IoUT nodes are clustered geographically and the CH is selected as the most capable node in the cluster &  - High throughput \newline- Several criteria in TOPSIS &  Single point of failure (gateway) \newline- Unrealistic simulation environment \\
\hline 

\end{tabular}
\end{adjustwidth}
\end{table}

\subsection{Based on Existing Algorithms}

The authors in \cite{Qi2023LightPoW} propose a lightweight Proof-of-Work (PoW) consensus mechanism for IoT that aims to improve security and reduce energy consumption. The proposed consensus algorithm called  Trust-based Time-Constrained PoW (TBTC-PoW)  allocates reasonable time windows to nodes based on computing resources, trust values, and mining difficulty. For trust calculation, the scheme includes a Heterogeneity Considered Trust Mechanism (HCTM)  that evaluates the trust value according to the capacity and the quality of the resource-constrained node. The capacity of a node is evaluated according to the available computing, storage, and communication resources, and quality is evaluated according to its security and reliability. The mining time of the nodes is dynamically limited according to the trust value based on the basic time window. However, the proposed scheme requires significant storage overhead which creates a problem that was not addressed by the authors.

A novel consensus protocol called Proof of Chance (PoCh) was proposed in \cite{kara2022proof}, which is designed for the industrial Internet of Things (IIoT). PoCh uses chance rather than computing power to reach a consensus. First, a set of candidates (to be elected as miners) are chosen and they broadcast their identity to other nodes. Candidate nodes must have a chance value ($Cval$) that satisfies the following equation $hash(Cval+Id) < Target$. Next, the miner identity is computed as the sum of the block number and candidate identities (modulo the number of candidates). PoCh requires that the number of candidates must be between $6$ and $15$ otherwise the consensus sends a fail message. PoCh also defines a set of timers that must be respected by nodes at each step of the algorithm. Moreover, the value of \textit{chance} and \textit{target} are updated at each iteration of the algorithm. 
The protocol is partially synchronous, meaning that nodes can only move to the next consensus round after a new block is broadcast or a consensus failure message is broadcast. 
However, PoCh presents several limits. First, as it is based on timers and requires that nodes must be synchronized. Moreover, the algorithm contains several conditions that must be satisfied otherwise a fail message will be generated. 

The authors in \cite{kong2022lap} proposed LAP-BFT, a Lightweight Asynchronous Provable Byzantine Fault-Tolerant Consensus Mechanism for achieving atomic broadcast of the latest state assessment of UAV nodes in an asynchronous UAV network. 
The consensus method aims to minimize the communication overhead by dividing the set of local trusted transactions and distributing them through reliable broadcast control transmission. It incorporates vector commitments to achieve multivalue Byzantine consensus (PMVBA) for identity and data, ensuring provability while reducing computational complexity.  
Although the proposed protocol reduces the consensus latency of the original PBFT protocol, the obtained latency is still very high for resource-constrained applications (average of 10s). 

Another improvement of the PBFT algorithm was proposed by \cite{wang2022airbc} in the context of UAVs network. The authors propose to reduce the communication overhead of the original PBFT by constructing a small-scale committee of miners to perform the consensus. Therefore, the consensus mechanism is performed by miners rather than the entire network. The miners are selected based on their trustworthiness and therefore, a reputation mechanism was introduced based on the recording of UAVs behavior. Simulation results show that the proposed scheme can reduce 69\% of the consensus latency. However, the security issues that arise from limiting the number of consensus nodes were not studied by the authors. 

A Diffused Practical Byzantine Fault Tolerance (DPBFT) algorithm was proposed for the IoV in \cite{chai2021cyberchain}. The proposed system decouples the consensus process in the cyberspace from vehicles’ actions in the physical world. In order to achieve that, a digital replica is created for each vehicle in the cyberspace. The corresponding BC, labeled CyberChain (CC), is constructed in the virtual world between the cybertwins (CTs) of vehicles and edge servers. The authors state that many tasks in the IoV, such as a vehicle handover, require the consensus process to be conducted in a fast manner. Hence, the proposed DPBFT algorithm executes in a small range around each cross-region of vehicles (CRV). 
The authors show that the proposed system reduces the consensus latency by an average of 47\%. However, the major disadvantage of DPBFT is its susceptibility to malicious attacks. The authors illustrate that DPBFT is 40\% more vulnerable to security attacks and consensus failure. 

\begin{figure}[!t]
\centering
\includegraphics[width=4.5in]{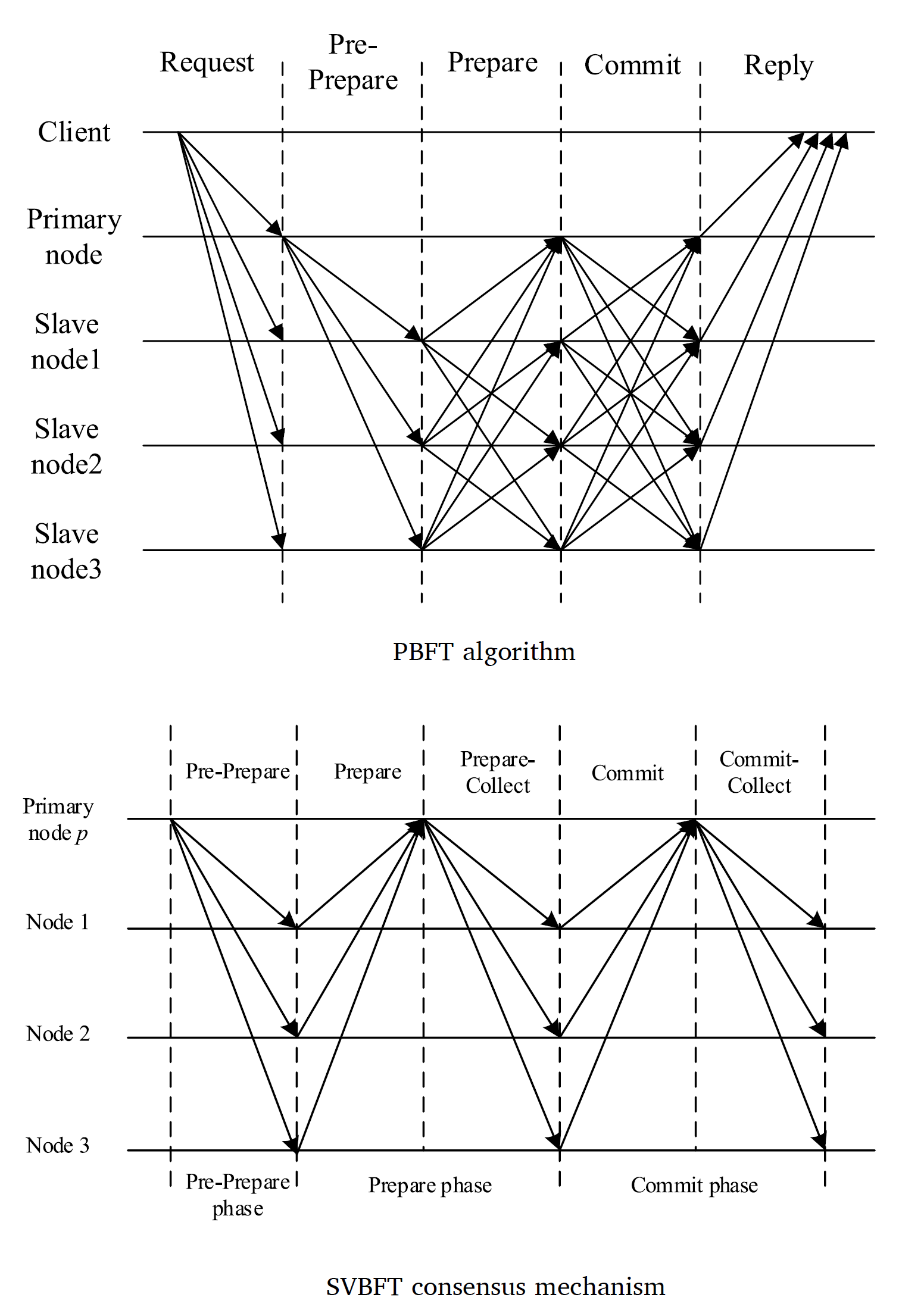}
\caption{Comparison between the original PBFT algorithm and the lightweight version labeled SVBFT. (Adapted from \cite{li2021lightweight})}
\label{fig_SLCs}
\end{figure}

Li \textit{et al.} \cite{li2021lightweight} propose another modified PBFT consensus algorithm. 
In the original PBFT, the consensus operation is conducted in five phases: request, pre-prepare, prepare, commit, and reply. The third and fourth phases require each of the miners to broadcast a message to all other miners, which is not suitable for resource-constrained nodes that 
cannot handle the large number of resulting communications. In order to resolve this issue, the authors propose to transform the consensus into five stages: pre-prepare, prepare, prepare-collect, commit, and commit-collect that replace broadcast with unicast between the master to slaves and vice versa, as shown in \autoref{fig_SLCs}. In order to achieve this, the master must be a trusted node. Hence, the authors propose a reward/punishment addition to the PBFT in which each node maintains the status scores of all other nodes and updates them after each consensus operation and after detecting/receiving malicious behavior reports. The status score of each node changes based on the node's legitimate or malicious behavior. 

In order to avoid the huge resource consumption required by PoW, Andola \textit{et al.} \cite{andola2020poewal} propose the Proof of Elapsed Work and Luck (PoEWAL) protocol for IoT networks. In the proposed approach, the IoT nodes generate data transactions, broadcast them to other nodes, and group the transactions to create new blocks. Each node is required to solve a hashing puzzle, similar to PoW, to mine the new block. However, the hashing puzzle does not demand that the node must find a hash value with a specified starting number of zeros. Rather, at the start of each mining round, the nodes are given a specified period during which each node attempts to find the best hash result that starts with the maximum possible number of zeros. At the end of the specified period, each node selects its best hash result and broadcasts it with the corresponding block to the whole network. Each node waits for a small time 
to receive the results of other nodes, then it selects the result with the best hash 
and adds it to the BC. 
The major disadvantage of PoEWAL is that it poses a huge communication overhead on the nodes, especially in very large networks. 

A lightweight version of Delegated Proof of Stake labeled LVBS is proposed in \cite{su2020lvbs} to allow UAVs to participate in the BC consensus. The authors assume that ground vehicles assist the UAVs in the disaster/relief scenario and that mobile emergency communication vehicles (ECVs) are deployed to send commands and receive updates from UAVs. 
The BC network is divided into full nodes and light nodes where the latter store only the block headers. ECVs are full nodes while ordinary ground vehicles and UAVs can select to be full or light nodes based on their resources. In LVBS, 
each voter has a certain stake that increases/decreases based on the voter's behavior and voting results. 
After all voters broadcast their votes, the score of each node is calculated. 
A set of nodes that achieve the highest scores are selected as witnesses who execute the consensus algorithm. 

Table \ref{table_ConSExist} provides an overview of the lightweight consensus solutions that are based on traditional consensus protocols.

\begin{table}[!t]
\begin{adjustwidth}{-1.2cm}{}
\centering
\caption{Summary of lightweight consensus solutions - based on existing algorithms.} 
\label{table_ConSExist} 
\centering
\begin{tabular}{|p {0.7 cm}|p {1.9 cm}|p {1.2 cm}|p {5.0 cm}|p {3.5 cm}|p {3.5 cm}|}
\hline
\textbf{Ref.} & \hfil \textbf{Application.} & \hfil \textbf{Year.} & \hfil{\textbf{Main idea}} & \hfil \textbf{Pros} & \hfil \textbf{Cons}  \\ 
\hline 

\hfil \cite{Qi2023LightPoW} &  \centering  IoT &  \hfil 2023 &  Trust based Time-Constrained PoW adapts the workload according to the capacity of the node & - Low computation overhead \newline- Good security & - High storage overhead \newline- Scalability not tested \\ 
\hline 

\hfil \cite{kara2022proof} &  \centering  IIoT &  \hfil 2022 &  Miners are selected based on their chance scores that are calculated based on several hash values & - Low computation overhead \newline- Enforcing time limits & Possibility of consensus failure \newline- Unbalanced miner selection \\ 
\hline 

\hfil \cite{kong2022lap} &  \centering  UAVs &  \hfil 2022 &  Consensus method that achieves atomic broadcast of the latest state assessment of UAV nodes & - Less communication overhead than PBFT \newline- Simple algorithm & - High consensus latency \newline- Not tested in real scenarios \\ 
\hline 

\hfil \cite{wang2022airbc} &  \centering  UAVs &  \hfil 2022 &  An improved version of PBFT where a small number of nodes are selected to participate in the consensus process as miners & Reducing the communication overhead and latency of original PBFT &  - Miners are vulnerable to resource depletion \newline- Weak security \\ 
\hline 

\hfil \cite{chai2021cyberchain} &  \centering IoV &  \hfil 2021 & Diffused PBFT executes consensus at the Sub-CyberSpace level and then at the global network level & - Low consensus latency \newline- Low signaling overhead & - Weak security \newline- Possibility of forks \\ 
\hline 

\hfil \cite{li2021lightweight} &  \centering  IoT &  \hfil 2021 & A modified version of the PBFT that replaces broadcasting with unicasting to primary node & - Reduced communication overhead \newline- High throughput &  DoS attack and single point of failure (primary node) \\
\hline

\hfil \cite{andola2020poewal} &  \centering  IoT &  \hfil 2020 &  Each node attempts to produce the best hash result during the mining period & Much less processing overhead than PoW &  - Huge communication overhead \newline- Limited testing \\
\hline 

\hfil \cite{su2020lvbs} &  \centering  UAVs &  \hfil 2020 &  A lightweight version of DPoS in which a UAV can be full or light node based on its resources & - Less overhead than original DPoS \newline- Including positive and negative votes &  - Vulnerable to attacks \newline- Scalability not tested \\
\hline 

\end{tabular}
\end{adjustwidth}
\end{table}

\subsection{Using Fog Nodes as Miners}

The authors in \cite{arifeen2022autoencoder} address the problem of verifying the originality of data generated by IoT devices. For this purpose, they propose an autoencoder (AE)-integrated Chaincode (CC)-based consensus mechanism in which the autoencoder differentiates normal data from anomalous data. The solution employs the concept of Hyperledger Fabric and uses edge servers to host the BC. 
The autoencoder is integrated into the chaincode (the smart contract) which is invoked by each endorsing peer. It uses an unsupervised machine learning algorithm that takes the data sent by IoT devices as input and attempts to detect any malicious information based on the training data and the previous nodes' behaviors. During the consensus process, each endorsing node invokes the CC to validate the transactions that it received from IoT nodes. 

Xu \textit{et al.} \cite{xu2022microchain} proposed a hierarchical 
microchained fabric for cross-devices decentralized federated learning (DFL) at the edge network. The proposed architecture allows for auditable and privacy-preserving data sharing in local model training time. The proposed consensus protocol is called microchain consensus. It is a partially decentralized BC that uses a Proof-of-Credit (PoC) consensus mechanism. Each node in the network has an initial credit stake, which is periodically updated according to its contribution to the training model and consensus protocol. The consensus protocol is executed by a small subset of nodes, called the validator committee, which is elected randomly using a Verifiable Random Function (VRF) based cryptographic sortition scheme. 

The authors of \cite{mershad2021proof} propose a lightweight consensus protocol for the IoV based on accumulated trust. The BC is maintained and updated by the RSUs. Certain RSUs are selected to become BC miners based on their trust ratings. Each RSU validates the transactions that it receives from vehicles before forwarding them to the set of RSUs that act as miners. A set of RSUs, called Trusted Nodes (TNs) assume the responsibility of mining new blocks, while all RSUs participate in the process of block validation. Each RSU accumulates trust points in order to become a TN. Each time an RSU validates a transaction correctly, it receives positive trust points, and vice versa. When the RSU accumulates enough trust points over a sufficiently long period, it is selected by the TNs to become a BC miner. Each time a new block should be mined, one of the TNs is selected to act as the Current Miner (CM) that is responsible for 
mining the new block. 

A lightweight consensus algorithm for the industrial Internet of Things was proposed in \cite{zhang2020ldc}. The consensus protocol is executed by multiple edge gateways that are deployed based on a square virtual grid. Each IIoT node sends its data to the nearest edge gateway and also to a Verification Edge Gateway that validates the accuracy of the data. The destination edge gateway creates the new block and shares it with the verification edge gateway. The latter validates the data in the block by checking it with the ledger and sends the verification result to the destination edge gateway. The data is considered to be accurate if the number of verification edge gateways that send confirmed validation is greater than 50\% of the total number of verified edge gateways. However, the proposed consensus algorithm is vulnerable to 51\% attacks. 

Zheng \textit{et al.} \cite{zheng2020lightweight} presented a lightweight consensus protocol for vehicular social networks. The main idea is to divide the BC into a private local chain that is maintained and administered by vehicles within a certain group, and a public global chain that is maintained and administered by RSUs and cloud servers. Vehicles form groups within the social network such that each vehicle is able to communicate directly with the other vehicles in the group. 
The vehicles within each group execute a Byzantine-based algorithm to add their transactions to the local chain. 
The consensus decisions that are created in the local chain are uploaded to the global chain by making each vehicle in the group randomly upload the decisions to the nearest RSU when it passes near it. 
However, the approach imposes a huge communication overhead on the vehicles due to the large number of rounds that are needed to reach a consensus for each event. 

The lightweight consensus solutions, employing fog nodes as miners, are summarized in Table \ref{table_ConSFog}.

\begin{table}[!t]
\begin{adjustwidth}{-1.2cm}{}
\centering
\caption{Summary of lightweight consensus solutions - Using fog nodes as miners.} 
\label{table_ConSFog} 
\centering
\begin{tabular}{|p {0.7 cm}|p {1.9 cm}|p {1.2 cm}|p {5.0 cm}|p {3.5 cm}|p {3.5 cm}|}
\hline
\textbf{Ref.} & \hfil \textbf{Application.} & \hfil \textbf{Year.} & \hfil{\textbf{Main idea}} & \hfil \textbf{Pros} & \hfil \textbf{Cons}  \\ 
\hline 

\hfil \cite{arifeen2022autoencoder} &  \centering Industrial IoT &  \hfil 2022 &  An autoencoder is used with the Hyperledger chaincode (CC) in the consensus process for detecting anomalous data & - Verifying the originality of data \newline- Strong transaction validation & - High latency \newline- Very low throughput \\ 
\hline 

\hfil \cite{xu2022microchain} &  \centering  IoT &  \hfil 2022 &  A Proof-of-Credit (PoC) consensus mechanism in which each node has an initial credit stake that is periodically updated  & -Efficient update of stake \newline- Good scalability \newline-Fork handling & - High latency \newline- Weak security \\ 
\hline 

\hfil \cite{mershad2021proof} & \centering IoV &  \hfil 2021 & RSUs in the IoV accumulate trust by correctly validating transactions in order to become miners & - Highly scalable \newline- Eliminates the forking problem & - Communication overhead \newline- High redundancy \\
\hline 

\hfil \cite{zhang2020ldc} &  \centering  IIoT &  \hfil 2020 &  Edge gateways are divided into destination edge gateways that receive data and create blocks and verification edge gateways that validate blocks & - Simple consensus algorithm \newline- Acceptable delay & - Vulnerable to attacks \newline- Scalability not tested \\ 
\hline 

\hfil \cite{zheng2020lightweight} &  \centering  Vehicular social networks &  \hfil 2020 &  Dividing the BC into a private local chain maintained by vehicles and public global chain maintained by RSUs and cloud servers & - Dividing consensus into local then global \newline- Good scalability & - Huge communication overhead \newline- High delay \newline- Security not tested \\ 
\hline 

\end{tabular}
\end{adjustwidth}
\end{table}

\subsection{Reputation-Based and Random-Based Consensus}

The authors in \cite{chen2022lblco} proposed a reputation-based consensus called proof of repute score based on hidden block (HBPORS). The algorithm selects a miner based on its reputation score, and the latter is calculated based on the node's contribution to the BC. At the start of each consensus round, each node 
calculates the reputation scores of the other nodes. Based on the reputation scores, nodes are classified into honest nodes and ordinary nodes. Next, a lottery pool is created to select the miner. Each honest node that has a reputation score higher than a certain threshold is added to the lottery pool, and a random seed is applied to select the next miner. 
HBPORS also includes a hidden block mechanism in which the miner broadcasts a hidden block to the resource-constrained nodes instead of the full block. The hidden block contains only data with low communication overhead. The miner selects to broadcast a hidden block with a certain probability that depends on several factors such as the network performance and the block details. 

Motivated by the supply chain context, the authors in \cite{islam2022integrating} proposed a reputation-based BC consensus algorithm. From the total nodes, (10\% + 1) nodes are selected as representative nodes based on their reputation. One of these nodes plays the role of the leader, while others are backup nodes. The leader is selected by rotation. All transactions are sent to the representative nodes. The leader verifies transactions and constructs the new block. Simultaneously, the backup nodes verify the block and vote on its correctness. Any node in the network will add the block to its own copy of the BC if the proposed block has 100\% votes from the backup nodes. 
The results show that the proposed system performs well when the percentage of malicious nodes is low. However, as this percentage increases above 10\%, the consensus time highly increases to become in the range of several seconds.

BCGR (Blockchain Consensus based on Global Reputation)\cite{xi2022crowdlbm} is a lightweight consensus mechanism based on Byzantine reputation to improve the consensus algorithm’s scalability. The BCGR consensus contains the following phases; \textit{Initialize} (in which the Master node of the
current view is selected), \textit{Request} (in which the transaction is signed and broadcast to the entire network), \textit{Verify} (in which the legitimacy of the transaction is verified), \textit{Prepare} (in which the master node signs the transaction and broadcasts it to the slave nodes), \textit{Commit} (in which slave nodes broadcast a confirmation message), and \textit{Reply} (in which the new block is written, or the view switching protocol is executed to select a new Master node). The algorithm is based on a local reputation and a global reputation.  The local reputation between two nodes $i$ and $j$ represents the degree of satisfactory transactions $i$ had with $j$. The global reputation is calculated as an aggregation of local reputation with neighbors. 

The authors in \cite{dorri2020TreeChain} proposed Tree-chain that relies on hash function outputs for randomization among validators. Tree-Chain introduces two levels of randomization: transaction level and BC level. At the transaction level, the validator for each transaction is defined randomly based on the most significant bits of the hash of the transaction, known as the consensus code. At the BC level, each validator is dedicated to storing transactions with a particular consensus code. 
Each validator only commits blocks in a particular ledger for a particular time frame known as the consensus period. 
At the beginning of each consensus period, validators are allocated to a particular ledger based on their public key. 
The consensus algorithm does not demand the validators to solve any puzzle or provide proof of work before storing a new block, which reduces delay and computational overhead. 

In \cite{worley2020scrybe}, a random miner is selected from the group of miners each time a new block is added. The authors develop a method that guarantees that all miners select the same random miner to create the block. This is done by making each miner generate a random number and broadcast its hash and signature to all other miners. The hash values are used to prevent each miner from knowing the random numbers that were selected by other miners before it shares its random number. After each miner receives the hash values from all other miners, it broadcasts the random number that it selected at the beginning. Next, after each miner receives the random numbers of all other miners, it adds them and calculates the \textit{mod} base \textit{N} (where \textit{N} is the number of miners) to obtain the ID of the next block miner. 

An overview of reputation-based and random-based lightweight consensus solutions is presented in Table \ref{table_ConSRep}.

\begin{table}[!t]
\begin{adjustwidth}{-1.2cm}{}
\centering
\caption{Summary of lightweight consensus solutions - Reputation-based and random-based consensus.} 
\label{table_ConSRep} 
\centering
\begin{tabular}{|p {0.7 cm}|p {1.9 cm}|p {1.2 cm}|p {5.0 cm}|p {3.5 cm}|p {3.5 cm}|}
\hline
\textbf{Ref.} & \hfil \textbf{Application.} & \hfil \textbf{Year.} & \hfil{\textbf{Main idea}} & \hfil \textbf{Pros} & \hfil \textbf{Cons}  \\ 

\hline 
\hfil \cite{chen2022lblco} &  \centering  IoT &  \hfil 2022 &  Miner is selected based on its reputation score that is calculated based on the node's contribution to the BC  & - Hidden block mechanism \newline- Lower communication overhead than PoW & - Very high consensus delay \newline- Security not tested \\ 
\hline 

\hfil \cite{islam2022integrating} &  \centering  supply chain &  \hfil 2022 & Reputation based consensus; consensus decision is taken by (10\%+1) representative nodes & - Selection of representative nodes \newline- Low consensus latency for small network & - Weak security \newline- High consensus delay when \% of malicious nodes is high\\ 
\hline 

\hfil \cite{xi2022crowdlbm} &  \centering  IoT &  \hfil 2022 &  Modified version of PBFT in which the Master node is selected based on its reputation  & - Low latency \newline- Good throughput & - Security not tested \newline - Limited simulations (maximum of 100 nodes) \\ 
\hline 

\hfil \cite{dorri2020TreeChain} &  \centering  IoT &  \hfil 2020 &  Tree-chain relies on hash function outputs for randomization among validators & - Randomization based on Key Weight Metric \newline- Low computation overhead & - Does not guarantee load balancing between validators \newline- Consensus delay and scalability not tested\\ 
\hline 

\hfil \cite{worley2020scrybe} &  \centering  General &  \hfil 2020 & Nodes exchange secret information to select a random miner & - Forking handling \newline- Less computational overhead than PoW  & - High communication overhead \newline- System not tested \\ 
\hline 

\end{tabular}
\end{adjustwidth}
\end{table}

\begin{figure}[!t]
\centering
\includegraphics[width=6.7in]{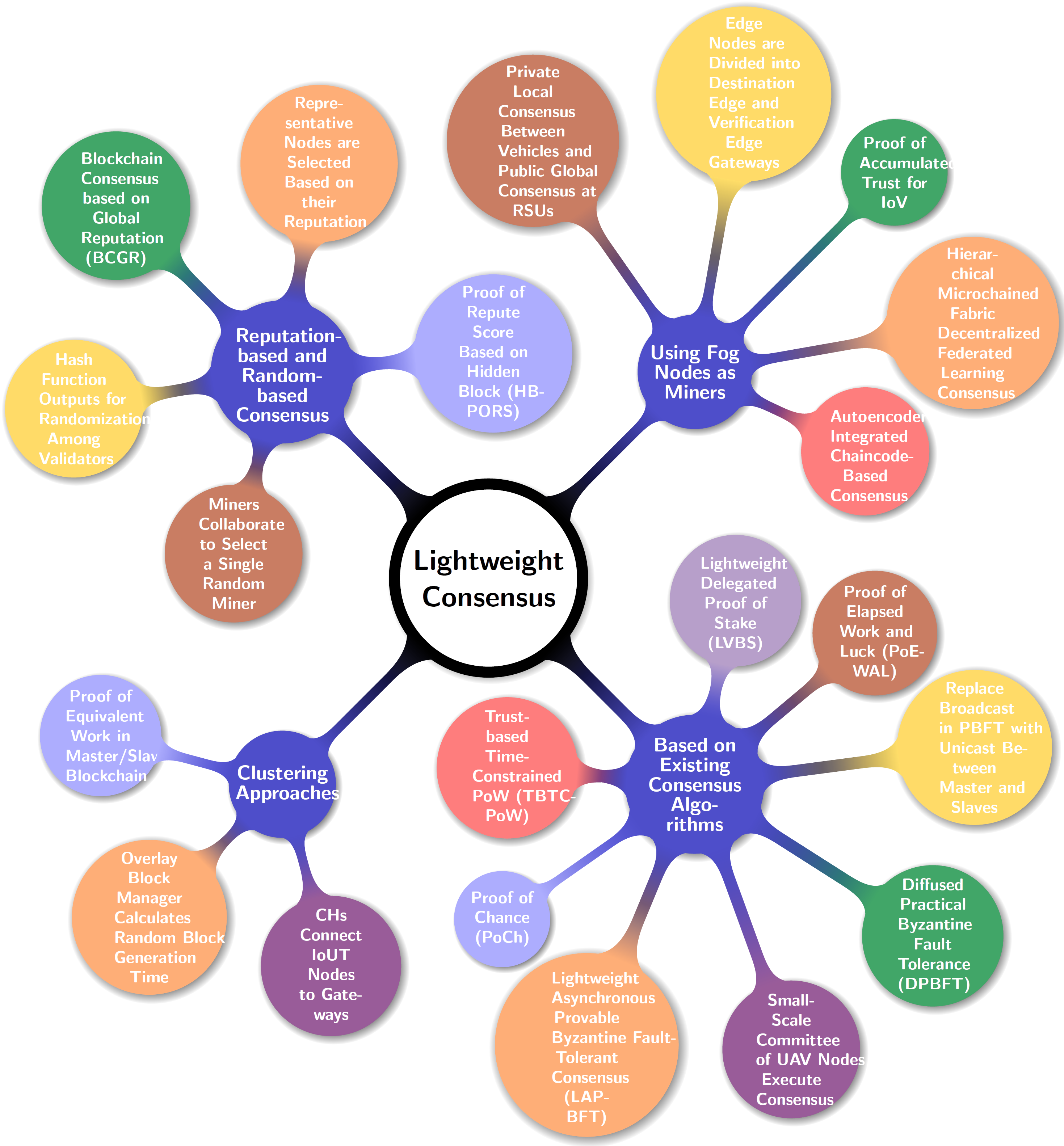}
\caption{Methods for lightweight consensus in the blockchain.}
\label{fig_Con}
\end{figure}

A general overview of the various consensus protocols in the four categories is presented in \autoref{fig_Con}.

\subsection{Analysis and Discussion}

A wide variety of lightweight consensus mechanisms with different flavors were discussed in the previous subsections. 
We notice from the description of the studied consensus protocols that most of them assign light roles during the consensus process to resource-constrained nodes, taking into consideration their limited resources. Some of these systems assign the block mining task to edge nodes, 
other systems select powerful cluster heads to perform or assist in mining, 
while other systems attempt ot reduce the mining load on the resource-constrained nodes. 
In certain consensus mechanisms, 
the selection of miners is based on the reputation of nodes. In other systems, a miner is selected randomly after checking its reputation. The reputation of nodes can be determined through various criteria, such as their past performance, consistency, reliability, and behavior. Nodes that have consistently demonstrated their trustworthiness are ranked higher and given priority in the mining process. However, reputation-based protocols are prone to scalability challenges. As the network grows, evaluating and maintaining reputation information for a large number of nodes becomes more computationally intensive and resource-demanding. Several schemes opted for the modification of existing consensus algorithms 
to reduce their overhead.

First, we outline the requirements of lightweight consensus, which we collected from the objectives that were stated in the studied papers, as follows:

\begin{enumerate}
\item 
\textbf{Correctness:} The BC consensus algorithm should produce correct results that reflect the decision that is agreed on by the majority of the BC nodes that participate in the algorithm. This property can be measured by the percentage of wrong transactions among all transactions inserted into the BC. The lower this percentage is, the higher the consensus correctness (and vice versa). 
\item 
\textbf{Lightweight:} In order to be lightweight, the consensus algorithm should not pose significant 
processing or communication overhead on the resource-constrained node.
\item 
\textbf{Energy consumption:} Many traditional consensus algorithms are known to be energy intensive. A lightweight algorithm should not cause the resource-constrained node to consume a lot of its energy.
\item 
\textbf{Consensus delay:} Latency is a delicate factor in resource-constrained networks, especially in delay-sensitive applications. Hence, the consensus algorithm should be executed quickly in order to add the data of the resource-constrained nodes into the BC and enable users to make use of them within the time threshold of the application. 
\item 
\textbf{Throughput:} A BC consensus algorithm should have high transaction throughput, which is measured by the number of transactions that are added to the BC per time unit.
\item 
\textbf{Resilience to cyberattacks:} Several attacks have been discovered that threaten BC consensus algorithms, such as the 51\%, Sybil, and replay attacks. A secure consensus algorithm should resist various attacks that target the consensus process. 
\item 
\textbf{Scalability:} The consensus algorithm should maintain high performance when the number of nodes that participate in the consensus increases. A consensus algorithm that suffers from any type of performance degradation when new nodes are added to the consensus network is said to be less scalable.
\item 
\textbf{Fork handling:} During the consensus process, the BC could fork into multiple branches. 
An effective consensus algorithm should handle forking correctly and provide a solution that selects the correct branch based on valid reasons.  
\end{enumerate}

In order to evaluate the reviewed solutions from the perspective of the lightweight consensus requirements, we analyze the operations and results of each of the protocols thoroughly. 

The results of our analysis are shown in \autoref{table_Consen}. In the table, we use symbols similar to those in Tables \ref{table_LA} and \ref{table_LAu}. With respect to the \textit{Resilience to Cyberattacks} requirement, we will use the symbol $\ominus$ to denote that the system studied a limited set of consensus-related attacks and did not consider many other attacks. The symbol $\oslash$ is used to illustrate that the system focused on a specific consensus attack only. Finally, the symbol $\odot$ is used to represent that our analysis showed that the system is vulnerable to a specific attack.

\begin{table}[!t]
\begin{adjustwidth}{-1.2cm}{}
\centering
\caption{Mapping of lightweight consensus protocols to the lightweight consensus requirements.} 
\label{table_Consen} 
\centering
\begin{tabular}{|p {0.7 cm}|p {1.8 cm}|p {1.6 cm}|p {2.0 cm}|p {1.9 cm}|p {1.9 cm}|p {2.0 cm}|p {1.4 cm}|p {1.7 cm}|}
\hline
\textbf{Ref.} & \hfil \textbf{Correct.} & \hfil \textbf{Light.} & \centering  \textbf{En. Cp.} & \hfil \textbf{Delay} & \hfil \textbf{Thr.} & \hfil \textbf{Att. Rs.} & \hfil \textbf{Scal.} & \textbf{Fk. Hd.} \\ 
\hline 
\hfil \cite{ekanayake2021lightweight} & \centering  $\otimes$ & \hfil $\leftrightarrow$ & \hfil $\otimes$ & \centering  $\uparrow$ & \centering $\otimes$ & \hfil $\oslash$ & \hfil $\uparrow$ & \centering\arraybackslash $\otimes$ \\
\hline
\hfil \cite{mohanty2020efficient} &  \centering  $\otimes$ &  \hfil $\leftrightarrow$ &  \hfil $\leftrightarrow$ &  \centering  $\leftrightarrow$ &  \hfil $\uparrow$ & \centering $\ominus$ &  \hfil $\uparrow$ &  \hfil \no \\
\hline 
\hfil \cite{uddin2019lightweight} &  \centering  \yes &  \hfil$\uparrow$ &  \hfil $\leftrightarrow$ &  \hfil $\downarrow$ &  \hfil $\uparrow$ &  \hfil $\oslash$ &  \hfil $\uparrow$ &  \hfil \no \\
\hline 
\hfil \cite{Qi2023LightPoW} &  \centering  $\otimes$ &  \hfil $\uparrow$ &  \hfil $\leftrightarrow$ &  \centering  $\leftrightarrow$ & \centering  $\leftrightarrow$ & \centering  $\oslash$ &  \hfil$\otimes$  &  \hfil \no \\
\hline 
\hfil \cite{kara2022proof} &  \centering  \no &  \hfil $\downarrow$ &  \hfil $\otimes$ &  \centering  $\leftrightarrow$ & \centering  $\leftrightarrow$ & \centering  $\odot$ &  \hfil $\otimes$  &  \hfil \yes \\
\hline 
\hfil \cite{kong2022lap} &  \centering  \yes &  \hfil $\uparrow$ &  \hfil $\uparrow$ &  \centering  $\otimes$ & \centering  $\leftrightarrow$ & \centering $\oslash$ &  \hfil \yes  &  \hfil$\otimes$ \\
\hline 
\hfil \cite{wang2022airbc} &  \centering  $\otimes$ &  \hfil $\uparrow$ &  \hfil $\otimes$ &  \centering  $\downarrow$ & \centering  $\uparrow$ & \centering $\oslash$ &  \hfil $\otimes$ &  \hfil $\otimes$ \\
\hline 
\hfil \cite{chai2021cyberchain} &  \centering $\downarrow$  &  \hfil $\leftrightarrow$ &  \hfil $\otimes$ &  \centering $\uparrow$  &  \hfil $\otimes$ &  \centering  $\oslash$ &  \centering $\otimes$ &  \centering\arraybackslash $\otimes$ \\
\hline 
\hfil \cite{li2021lightweight} &  \centering  $\otimes$ &  \hfil $\leftrightarrow$ &  \hfil $\otimes$ &  \centering  $\leftrightarrow$ &  \hfil $\uparrow$ &  \centering  $\ominus$ &  \centering $\leftrightarrow$ &  \centering\arraybackslash $\otimes$ \\
\hline 
\hfil \cite{andola2020poewal} &  \centering  $\otimes$ &  \hfil $\downarrow$ &  \hfil $\leftrightarrow$ &  \hfil $\uparrow$ &  \hfil $\leftrightarrow$ &  \centering  $\otimes$ &  \hfil $\otimes$ &  \hfil \yes \\
\hline 
\hfil \cite{su2020lvbs} &  \centering  $\otimes$ &  \hfil $\uparrow$  &  \hfil $\otimes$ &  \centering $\otimes$ &  \centering $\otimes$ & \centering $\ominus$ &  \centering $\otimes$ &  \hfil \yes \\
\hline 
\hfil \cite{arifeen2022autoencoder} &  \centering  \yes &  \hfil $\leftrightarrow$ &  \hfil $\otimes$ &  \centering  $\leftrightarrow$ & \centering  $\leftrightarrow$ & \centering  $\oslash$ &  \hfil$\otimes$  &  \hfil$\otimes$ \\
\hline 
\hfil \cite{xu2022microchain} &  \centering  \no &  \hfil $\uparrow$ &  \hfil $\otimes$ &  \centering  $\downarrow$ & \centering  $\uparrow$ & \centering $\odot$ &  \hfil $\uparrow$ &  \hfil \yes \\
\hline 
\hfil \cite{mershad2021proof} & \centering \yes &  \hfil $\leftrightarrow$ &  \hfil $\otimes$ &  \centering  $\leftrightarrow$ &   \centering $\otimes$ & \centering $\ominus$ &  \centering $\otimes$ &  \centering\arraybackslash $\otimes$ \\
\hline 
\hfil \cite{zhang2020ldc} &  \centering  \no &  \hfil $\uparrow$ &  \hfil $\otimes$ &  \centering  $\leftrightarrow$ & \centering  $\otimes$ & \centering  \no &  \hfil$\otimes$  &  \hfil$\otimes$ \\
\hline 
\hfil \cite{zheng2020lightweight} &  \centering  $\otimes$ &  \hfil $\downarrow$ &  \hfil $\otimes$ &  \centering  $\uparrow$ & \centering  $\otimes$ & \centering $\ominus$ &  \hfil$\uparrow$ &  \centering\arraybackslash $\otimes$ \\
\hline 
\hfil \cite{chen2022lblco} &  \centering  $\otimes$ &  \hfil $\uparrow$ &  \hfil $\otimes$ &  \centering  $\otimes$ &  \centering $\uparrow$ & \centering $\otimes$  &  \hfil $\downarrow$ &  \hfil $\otimes$ \\
\hline 
\hfil \cite{islam2022integrating} &  \centering  \yes &  \hfil $\uparrow$ &  \hfil $\otimes$ &  \centering  $\otimes$ & \centering  $\uparrow$ & \centering 
$\oslash$ &  \hfil $\otimes$ &  \hfil $\otimes$ \\
\hline 
\hfil \cite{xi2022crowdlbm} &  \centering  \yes &  \hfil $\uparrow$ &  \hfil $\otimes$ &  \centering  $\downarrow$ & \centering  $\uparrow$ & \centering $\oslash$ &  \hfil $\uparrow$ &  \hfil $\otimes$ \\
\hline 
\hfil \cite{dorri2020TreeChain} &  \centering  $\otimes$ &  \hfil $\uparrow$ &  \hfil $\otimes$ &  \centering  $\otimes$ & \centering  $\otimes$ & \centering $\oslash$ &  \hfil$\otimes$  &  \hfil$\otimes$ \\
\hline 
\hfil \cite{worley2020scrybe} &  \centering  $\otimes$ &  \hfil $\downarrow$ &  \hfil $\otimes$ &  \centering  $\otimes$ & \centering  $\otimes$ & \centering $\ominus$ &  \hfil $\otimes$  &  \hfil \no \\
\hline 
\multicolumn{9}{l}{\small Correct. = Correctness, Light. = Lightweight, En. Cp.= Energy Consumption, Thr. = Throughput,} \\
\multicolumn{9}{l}{\small Att. Rs. = Attack Resilience, Scal. = Scalability, Fk. Hd. = Fork Handling} \\
\multicolumn{9}{l}{\yes = Satisfied, \no = Not satisfied, $\otimes$=Not studied,  $\uparrow$=High, $\leftrightarrow$=Medium, $\downarrow$=Low} \\
\multicolumn{9}{l}{$\ominus$ = Did not consider many attacks, $\oslash$= Consider specific attacks, $\odot$ = Vulnerable to a specific attack } \\
\end{tabular}
\end{adjustwidth}
\end{table}

Based on our analysis of the lightweight consensus algorithms that have been proposed so far, we notice several aspects that need to be improved in this area to reach an efficient and secure consensus protocol that allows resource-constrained nodes to participate in the BC without significant overhead: 
\begin{itemize}
    \item 
\textbf{Correctness:} The first important characteristic of a consensus protocol, in general, is its Correctness. In the BC, the main aim of consensus is to verify the correctness and legitimacy of new transactions before adding them to the BC. 
From \autoref{table_Consen}, we notice that very few authors considered testing the correctness of their proposed consensus algorithms. The same note can be said about \textit{Energy Consumption}: although this factor is one of the most important to test in the consensus protocol, only very few authors 
tested it in their experiments or simulations. 
\item 
\textbf{Resource Consumption:} 
We notice that several authors compared their consensus protocols with traditional ones, such as PoW and PBFT, and concluded that their proposed systems are suitable for resource-constrained environments based on the fact that their protocols produce better performance than traditional ones. However, resource-constrained nodes cannot afford to perform continuous processing and engage in frequent communications, even if these operations are done at a smaller scale as compared to a normal BC node. For example, the PoEWAL protocol \cite{andola2020poewal} requires the resource-constrained node to continuously hash the block during a specified period and select the best hash result at the end of the period as its mining result. In many resource-constrained networks, the resources of a node are very precious to be spent in traditional mining, even at a smaller scale. Only very few protocols 
showed \textit{Lightweight} aspects in their consensus results. For example, in \cite{uddin2019lightweight}, the resource-constrained nodes verify transactions only without participating in the mining process. 
As for the consensus delay, several authors 
considered a consensus delay of a few seconds acceptable, which is not true in many resource-constrained networks. As mentioned before, many applications are delay-sensitive and require the transaction to be added to the BC in a very fast manner so that users can make use of it. 
\item 
\textbf{Throughput:} Another important factor of the consensus process is the throughput, which is measured by the number of transactions that are added to the BC per second. An efficient consensus protocol should achieve high throughput. Several proposed lightweight consensus mechanisms achieved good throughput. However, many papers did not test this requirement. 
\item 
\textbf{Resilience to Attacks:} With respect to attack resilience, we notice that most authors focused on specific attacks that are related to the application they study, while neglecting important attacks that could occur in BC networks, such as Sybil, replay, and 51\% attacks. In addition, several authors 
proposed a consensus protocol without studying the attacks that could occur due to the specific characteristics of their system, as we discussed in the previous subsections. 
\item 
\textbf{Scalability:} One of the important requirements of a consensus algorithm is scalability, which ensures that as the number of participants in the consensus increases, the performance of the consensus algorithm is minimally affected. Consensus scalability should be tested by significantly increasing the number of consensus nodes in the experiments/simulations and observing the variation in the consensus results (in terms of accuracy, latency, processing, communication, and energy consumption). 
\item 
\textbf{Fork Handling:} Finally, fork handling is a major aspect that should be considered by a BC consensus protocol, since the BC decentralization property makes it susceptible to forking. 
A consensus protocol must specify the cases in which the BC may branch, and propose an approach to resolve the forking by selecting the correct branch based on justified reasons. From \autoref{table_Consen}, we notice that few papers studied this requirement.
\end{itemize}

In order to reach a general lightweight consensus protocol that can be implemented by resource-constrained nodes as part of their BC software, we need to take the following into consideration: 
 
\begin{itemize}
\item 
The consensus protocol should guarantee that a minimum percentage of wrong and malicious transactions should be added to the BC. 
The more effective the consensus protocol is, the smaller the number of invalid transactions that can sneak their way into the BC.
\item 
The consensus algorithm should be lightweight in terms of processing and communication overhead, end-to-end latency, and energy consumption. 
All these resources are considered vital to the continuity of a resource-constrained node. Hence, a protocol should be light in all these aspects for the node to be able to implement it successfully. 
\item 
A consensus protocol should achieve an acceptable transaction throughput. The “acceptable” threshold depends on the size of the data that is produced by the application.
\item 
A consensus protocol should be scalable. In many protocols, \textit{Scalability} and \textit{Lightweight} create a trade-off such that when the number of participants in the protocol increases, it becomes less \textit{Lightweight}. 
An efficient consensus protocol should be minimally affected in terms of performance as the number of participants increases (note that the number of participants can mean the number of miners, validators, or both). 
\item 
Finally, a major concern of a consensus protocol is its security. Malicious participants could perform a variety of attacks during the consensus process. 
A secure consensus protocol should be able to resist well-known consensus attacks, and be designed such that it does not include vulnerabilities that could be exploited to launch new attacks. 
In addition, the consensus protocol should comprise a solution to forking, in case one of the operations in the consensus algorithm could result in a BC fork. 
\end{itemize}

\section{Lightweight Cryptography}
\label{Sec_LCr}

The blockchain relies on strong cryptography algorithms to safeguard the stored data and secure the underlying communications. Cryptography algorithms that are implemented in traditional BC applications require the node to have powerful computational capabilities in order to perform the required cryptographic operations (such as encryption, decryption, and hashing). In addition, these algorithms usually consume a lot of power as compared to other software programs. Resource-constrained nodes, such as IoT sensors, are characterized by limited processing capability and utilize small-size batteries that provide a limited energy supply. For these reasons, “lightweight cryptography” is concerned with providing cryptographic solutions that utilize less memory, less computing resources, and less power supply to perform the required cryptographic operations, while maintaining strong security measures. A large number of lightweight cryptography solutions for BC have been proposed in the literature. These solutions can be classified into the following categories:
\begin{itemize}
\item 
\textbf{Lightweight encryption/decryption:} These systems propose models that use \linebreak lightweight encryption/decryption in BC operations such as transaction data encryption and digital signature \cite{avula2023efficient, jnss2023novel, khashan2023efficient, li2023privacy, hameedi2022improving, xu2022vmt, zhang2022fruit, singh2020internet, dwivedi2019decentralized}. 
\item 
\textbf{Lightweight hashing:} These systems propose lightweight hashing functions for BC operations such as generating the block hash and Merkle root \cite{ahamed2022distributed, alshehri2022aac, abed2021analysis, apriani2021performance, nabeel2021security, fu2020study}.
\item 
\textbf{Lightweight encryption/decryption and hashing:} These systems propose both lightweight encryption/decryption and hashing schemes, which could be via the same cryptographic algorithm or via different algorithms \cite{mershad2022proact, revanesh2022dag, khan2021aechain, guruprakash2020ec, yan2020pcbchain}. 
\end{itemize}

Next, we discuss the systems that were proposed in each category.

\subsection{Lightweight Encryption/Decryption}

A system that uses the BC for securing electronic health records (EHR) in the Internet of Medical Things (IoMT) was proposed in \cite{avula2023efficient}. The authors utilize lattice-based cryptography with the homomorphic proxy re-encryption scheme (LBC-HPRS) to encrypt BC data. The three main participants in proxy re-encryption (PRE) are the proxy, delegator, and delegate. The proxy converts a delegator’s ciphertext into a delegatee’s ciphertext and calculates the plain and re-encrypted ciphertext of any user. Firstly, the HPRE produces the public parameters. Next, keys are produced to support the security of sensitive data. For the encryption operation, public keys are utilized, and the secret key is used during the decryption operation by the user. 

The authors in \cite{jnss2023novel} propose integrating the BC with the Tiny Lightweight Symmetric Encryption (TLSE) scheme which selects the best encryption key by applying the Aquila Optimization Algorithm (AOA). The system utilizes the fogbus to link several hardware devices such as resource-constrained, fog, broker, and repository nodes. In the utilized BC, the TLSE model is used to dynamically increase key confusion in each encryption round. 
In order to optimize the number of rounds and the selected keys, the AOA algorithm is applied in four phases: expanded exploration of potential solutions, narrowed exploration of selected solutions, expanded exploitation of iteration parameters, and narrowed exploitation of iteration parameters. 

A hybrid centralized and decentralized architecture for heterogeneous resource-constrained networks based on BC and edge computing is proposed in \cite{khashan2023efficient}. The system utilizes both lightweight symmetric and asymmetric encryption. Symmetric encryption is implemented using the SPECK algorithm, while ECC160 is used for asymmetric encryption. A symmetric key is generated every time a resource-constrained node sends data on the network. This key is used to encrypt the data via SPECK. When two resource-constrained nodes want to communicate with each other, a secure communication channel is created between them via asymmetric encryption. 

Lightweight multi-receiver encryption (MRE) was used in \cite{li2023privacy} to protect the privacy of smart grid users when they send their energy data to energy companies. 
In the proposed system, the key generation center (KGC) 
generates public and private keys to other entities. During setup, the KGC sends a public-private key pair and a signing key pair to the energy company (EC) and smart meter (SM). After that, the SM stores its energy data off-chain after encrypting it with a symmetric key \textit{k}. Next, the EC and SM execute a transaction application procedure that takes the EC’s identity and purchase contract as input and outputs transaction data. The SM calls the BC smart contract by inputting to it the transaction data, and the smart contract generates a transaction that contains the agreement between the SM and the EC. Next, the MRE procedure is performed by the SM, which takes as input the symmetric key \textit{k}, the identities of all participant ECs, and their public keys and produces the ciphertext \textit{CT}. Finally, the EC decrypts \textit{CT} via MRE to obtain \textit{k}, and then decrypts the off-chain data to obtain the user energy data. 

Multipart AES encryption was used in \cite{hameedi2022improving} to preserve the security of the BC. Multiple data encryption is a technique that is used to decrease the total computations required for encryption while preserving a high security level. Using this technique, the original data block is divided into multiple parts and each part is encrypted using a distinct small-size key. The authors use four AES keys to encrypt the data of the resource-constrained node. 
The authors also use the TinyECC algorithm to make each resource-constrained node generate its ECC key pair and periodically use it to obtain the four AES keys from the server. 

In \cite{xu2022vmt}, a lightweight encryption technique, named VMT, is proposed for securing the transmission of VANET messages. The system utilizes Encryption-Decryption Servers (EDS) that provide encryption/decryption services for RSUs and vehicles. In addition, a Cryptographic Auxiliary Server (CAS) assists the EDS during the encryption/decryption process. A higher security level is achieved by encrypting messages via the mutual cooperation of the EDS and CAS. A vehicle sends a message \textit{M} to the EDS authenticating itself via its encrypted pseudo-ID and password. The EDS and CAS encrypt \textit{M} by generating and exchanging random numbers and hashing functions. The EDS sends the encrypted message to the vehicle which transmits it to the nearest RSU. The latter fetches the corresponding password from the BC and sends it along with the encrypted message to the EDS. The latter collaborates with the CAS to decrypt the message and send it to the RSU. 

The authors in \cite{zhang2022fruit} propose a lightweight encryption scheme that combines matrix decomposition with proxy re-encryption. The proposed system comprises a blockchain-based privacy-preserving incentive scheme that uses smart contracts to implement payment management and reputation prediction. Lightweight encryption is used with task allocation to calculate the data quality and truthful knowledge in a privacy-preserving manner. The key generation center (KGC) selects a lower triangular random matrix for each node to generate its encryption keys. Each data node selects its interested tasks, extends each task to a lower triangular matrix, encrypts it, and sends the results to the decentralized database. A buyer identifies the tasks he/she requires, generates a polynomial function from them, extracts its coefficients to build a lower triangular random matrix, and encrypts it. 

In \cite{singh2020internet}, the authors implemented the SPECK algorithm for the encryption/decryption of blocks in the BC of the pharmaceutical supply chain. The proposed system utilizes the \textit{bloXroute} server concept, which is a Blockchain Distribution Network (BDN) that enhances the BC scalability at the network layer via a large number of \textit{bloXroute} servers that propagate the BC transactions among BC nodes on behalf of users. In order to preserve the confidentiality of the transactions and prevent the \textit{bloXroute} servers from tampering with the block, the block creator encrypts it before sending it to the \textit{bloXroute} servers. 
The tests performed on a network of 10,000 nodes show that SPECK-128—256 is 55\% faster than AES-128 and 20\% faster than LEA-256. In addition, the energy consumption of SPECK is 39\% less than AES and 7\% less than LEA. 

In \cite{dwivedi2019decentralized}, a double encryption technique is applied for the BC network of healthcare wearable devices. The authors use a distinct lightweight symmetric key to encrypt the BC transactions data, and encrypt this lightweight key with the receiver’s public key. Hence, each BC transaction will contain three parts: encryption of data, encryption of symmetric key, and signature. The authors utilize a permissioned BC in which patients’ data are shared only with health providers who are authorized to access them. The BC administration is maintained by an overlay network that is divided into clusters, where each cluster contains a CH who is a member of the overlay network and oversees the BC operations of its cluster members via smart contracts. 

A summary of the solutions discussed in this section is shown in \autoref{table_Crypto1}.

\begin{table}[!t]
\begin{adjustwidth}{-1.2cm}{}
\centering
\caption{Summary of lightweight cryptography - encryption/decryption solutions.} 
\label{table_Crypto1} 
\centering
\begin{tabular}{|p {0.7 cm}|p {1.9 cm}|p {1.2 cm}|p {5.0 cm}|p {3.5 cm}|p {3.5 cm}|}
\hline
\textbf{Ref.} & \hfil \textbf{Application.} & \hfil \textbf{Year.} & \hfil{\textbf{Main idea}} & \hfil \textbf{Pros} & \hfil \textbf{Cons}  \\ 
\hline 
\hfil \cite{avula2023efficient} & \centering Healthcare & \hfil 2023 & Lattice-based cryptography with homomorphic proxy re-encryption is used to encrypt BC data & - Low encryption time \newline- Use of Enhanced Merkle tree & - High latency \newline- Use of PoW \\
\hline
\hfil \cite{jnss2023novel} & \centering General & \hfil 2023 & Utilizing Tiny Lightweight Symmetric Encryption and selecting the best encryption key via the Aquila Optimization Algorithm & - Low encryption/decryption, processing, and waiting delays & - Susceptible to advanced security attacks \newline- High computation overhead \\
\hline
\hfil \cite{khashan2023efficient} & \centering General & \hfil 2023 & SPECK algorithm to encrypt BC data and ECC160 to create secure channel & - Low delay and power consumption \newline- High security & - High computation overhead and memory storage \\
\hline
\hfil \cite{li2023privacy} & \centering Smart grid & \hfil 2023 & Smart meter executes multi-receiver encryption using the public keys of energy companies & - Low latency \newline- Good security & - Limited testing \newline- Single point of failure (KGC) \\
\hline
\hfil \cite{hameedi2022improving} & \centering General & \hfil 2022 & Data block is divided into parts and each part is encrypted using a distinct small-size key & - Strong against brute-force attacks \newline- Use of TinyECC & - Susceptible to advanced security attacks  \newline- Low Scalability \\
\hline
\hfil \cite{xu2022vmt} & \centering VANET & \hfil 2022 & Data is encrypted via the mutual cooperation of the EDS and CAS & - High security \newline- Anonymity & - High communication overhead   \newline- High latency \\
\hline
\hfil \cite{zhang2022fruit} & \centering General & \hfil 2022 & Lightweight encryption with task allocation to calculate the data quality and truthful knowledge in a privacy-preserving manner & - Use of Dirichlet distribution \newline- Good incentive scheme & - Limited testing \newline- Single point of failure (KGC) \\
\hline 
\hfil \cite{singh2020internet} & \centering Supply chain & \hfil 2020 & SPECK algorithm for the encryption of blocks in the BC of the pharmaceutical supply chain & - High encryption speed \newline- Good security & - Unsuitable for real-time data \newline- High energy consumption \\
\hline
\hfil \cite{dwivedi2019decentralized} & \centering Healthcare & \hfil 2019 & BC data is encrypted with lightweight symmetric key which is encrypted with the receiver’s public key & - Scalability \newline- High confidentiality & - Distinct key for each transaction \newline- High computational overhead \\
\hline
\end{tabular}
\end{adjustwidth}
\end{table}

\subsection{Lightweight Hashing}

The KECCAK256 hash function was used in \cite{ahamed2022distributed} to hash the transactions in the lightweight BC stored by UAV nodes to secure their communications during data collection and transmission. In the proposed system, UAVs are managed by a GCS that collects and processes all the UAVs’ data. UAVs store Reformatory BC blocks, which are created by replacing each transaction in the block body with its KECCAK256 hash. In addition, the block header contains the Merkle root of a reputation tree, where each node in the tree is also hashed using KECCAK256. Moreover, the authors propose a mechanism to compress the 160-bit final hash output into 80 bits to conserve the UAV memory. 

In \cite{alshehri2022aac}, an attribute-based access control (AAC) model for resource-constrained networks using the Hyperledger Fabric BC was proposed. In this system, the number of access attributes is chosen based on the data sensitivity. The attribute count is calculated via a fuzzy logic scheme using data type and preference. The BC is utilized to store meta-data and security credentials of data owners and data users using a lightweight hashing algorithm. In addition, the authentication of several users is handled in parallel by a neural network that can take multiple requests together. A data user is authorized using their identity, certificate, signature, and Physical Unclonable Function (PUF); then a credibility score is calculated for the user. The system utilizes the QUARK hash functions in all operations that require hashing. 

The authors of \cite{abed2021analysis} test three lightweight hash functions (SPONGENT, PHOTON, and QUARK) on FPGA nodes that execute the BC operations. 
The authors compare the three functions from the perspectives of throughput, resource utilization, and energy consumption. The results show that SPONGENT is by far better than the other two in terms of throughput. However, SPONGENT has the highest hardware resource requirements and energy consumption. PHOTON utilizes less area than the other two by approximately 50\%, while QUARK has lower power and energy consumption than PHOTON. 
The authors recommend SPONGENT for the BC of applications that require high security and throughput, and PHOTON for the BC of applications that are very resource-limited. 
Similarly, The authors of \cite{apriani2021performance} compared SPONGENT, PHOTON, and KECCAK in terms of processing time and memory requirements. The results in \cite{apriani2021performance} show that the delay of SPONGENT is less by 30\% than that of KECCAK and by 77\% than that of PHOTON. On the other hand, SPONGENT requires 11\% more memory than PHOTON and 23\% more than KECCAK. 

A lightweight hash function named Lightweight New Mersenne Number Transform (LNMNT) for resource-constrained devices was proposed in \cite{nabeel2021security}. The proposed function builds on the sponge technique to decrease the internal memory size. It uses a Mersenne number, which is a number that is less than a power of two by one, as the building block of a Number Theoretic Transform (NTT) that improves the diffusion property of the proposed function. In addition, the variable transform length of NMNT enables the hash function to have a variable block size based on the security requirements. 
When a suitable prime Mersenne Number is used, the function can be calculated with variable-length hash digests. 

A technique for optimizing BC hashing based on the Proactive Reconfigurable Computing Architecture (PRCA) was proposed in \cite{fu2020study}. The authors utilize the pipeline hashing algorithm to enhance the calculation performance of the BC hashing function. In the proposed system, the proactive reconfigurable computer, which consists of an ATOM microprocessor; four reconfigurable FPGAs; and DDR3 memory, plays the role of a full BC node. It uses a dynamic hash algorithm that selects the most suitable hash function and calculates the hash by means of pipelines. In order to increase the security of the transactions, a proactive reconfigurable hash is used that consists of several structures of hash algorithms that can be used separately or in series. While hashing, the optimal number of pipeline series is calculated using the Carry-Save Adders (CSA) strategy which helps in reducing the delay of critical paths. 

A summary of the solutions discussed in this section is shown in \autoref{table_Crypto2}.

\begin{table}[!t]
\begin{adjustwidth}{-1.2cm}{}
\centering
\caption{Summary of lightweight cryptography - hashing solutions.} 
\label{table_Crypto2} 
\centering
\begin{tabular}{|p {0.7 cm}|p {1.9 cm}|p {1.2 cm}|p {5.0 cm}|p {3.5 cm}|p {3.5 cm}|}
\hline
\textbf{Ref.} & \hfil \textbf{Application.} & \hfil \textbf{Year.} & \hfil{\textbf{Main idea}} & \hfil \textbf{Pros} & \hfil \textbf{Cons}  \\ 
\hline 
\hfil \cite{ahamed2022distributed} & \centering UAV networks & \hfil 2022 & UAVs store Reformatory BC blocks, in which each transaction is replaced with its KECCAK256 hash & - Use of reputation tree \newline- Compression of hashes & - High delay overhead \newline- Low throughput \\
\hline 
\hfil \cite{alshehri2022aac} & \centering General & \hfil 2022 & Using the QUARK hash functions in all BC operations that require hashing & - Acceptable latency \newline- Good throughput & - Limited testing \newline- Low scalability \\
\hline 
\hfil \cite{abed2021analysis} & \centering General & \hfil 2021 & Testing three lightweight hash functions (SPONGENT, PHOTON, and QUARK) on FPGA nodes  & - Finding the best function in terms of throughput, energy consumption, and space & - No delay comparison \newline- No scalability testing \\
\hline 
\hfil \cite{apriani2021performance} & \centering General & \hfil 2021 & Comparing SPONGENT, PHOTON, and KECCAK in terms of processing time and memory requirements  & - Finding the best function in terms of hashing delay and memory demand & - Tested on Ethereum only \newline- No scalability testing \\
\hline 
\hfil \cite{nabeel2021security} & \centering General & \hfil 2021 & Hash function uses a Mersenne number to improve the diffusion property and achieve a variable-length hash digest & - High security \newline- Acceptable energy consumption & - No physical testing \newline- Low hashing speed (1.3 sec) \\
\hline 
\hfil \cite{fu2020study} & \centering General & \hfil 2020 & Pipeline hashing algorithm is used to enhance the calculation performance of the BC hashing function & - Dynamic hash algorithm \newline- Use of Carry-Save Adders & - Requires dedicated hardware \newline- Limited testing \\
\hline 
\end{tabular}
\end{adjustwidth}
\end{table}

\subsection{Lightweight Encryption/Decryption and Hashing}

Several versions of the Elliptic Curve Cryptography (ECC) and the SPONGENT hash function were utilized by the author of \cite{mershad2022proact} to perform the encryption and signing operations in the BC. 
The author defines two security levels: permanent-security, which is applied to transactions that are intended to be always secured, and temporary-security, which is applied to transactions that need to be secured for a limited duration. Transaction of the first type are encrypted using 256-bit ECC and signed using the SPONGENT-224 hash function, while those of the second types are encrypted and signed using a 64-bit ECC curve (\textit{shortECC}) and the SPONGENT-88 function. 

A DAG-based system powered with lightweight encryption and hashing for reducing the energy consumption of WSN nodes was proposed in \cite{revanesh2022dag}. In this system, the sensor nodes are clustered via the Emperor Penguin Colony (EPC) algorithm in order to minimize the hotspot problem and decrease their energy consumption. In addition, the BC data is encrypted using the lightweight block cipher XTEA with the Chaotic Map algorithm. With respect to hashing, the authors utilize the Blake-256 algorithm for message authentication. XTEA is a 64-bit cipher in which the block is split into two equal parts and a 128-bit encryption key is used. In order to avoid the high key generation processing time, the authors used the 3D chaotic map to calculate the encryption key. 

Khan \textit{et al.} \cite{khan2021aechain} propose a lightweight cryptography scheme that comprises lightweight encryption using the ASCON algorithm, which is a lightweight \textit{authenticated encryption} (AE) scheme that is used to encrypt the transactions; and the corresponding hashing algorithm of ASCON, which is ASCON-hash, to generate the transaction and block signatures. The proposed system relies on grouping resource-constrained nodes into clusters in which a CH participates in the BC. Each node gathers its readings into a transaction, generates a random nonce, and inputs the nonce and the data into the AE module. The latter uses a symmetric key to generate the encrypted data and a tag. The node enters these two into the hashing module to generate the transaction signature and then broadcasts the (transaction, tag, signature) to the network. The CH receives the transaction and validates it by entering the encrypted transaction data and the signature to the AE module along with the node’s symmetric key. The AE module generates the tag and the original transaction data. The CH enters these into the hashing module and compares the result with the signature. 

The authors of \cite{guruprakash2020ec} consider a network architecture in which resource-constrained nodes are clustered such that each cluster contains a CH that acts as a sink that gathers sensors’ data, aggregates them, and sends them to the BC miners in the cloud. The authors propose to maintain the integrity of BC data by making each sensor encrypt the transactions using EC-ElGamal, in which the elliptic curve law of addition is mixed with discrete logarithmic operations based on elliptic curves to establish a hybrid cryptosystem. 
With respect to BC hashing, the authors utilize the SHA-384 function in which the intermediate hash value is encrypted using the message block with a key generated from the Genetic algorithm. 

A hardware-based mining and verification solution is proposed in \cite{yan2020pcbchain} to relieve resource-constrained nodes from performing these heavy operations in software-based PoW. The authors assume a large network that is divided into subsystems, where each subsystem contains a full node that acts as the administrator and controller of the subsystem, and a large number of light nodes that are equipped with hardware components that enable them to perform lightweight mining and validation operations. The authors propose replacing the hashing puzzle in traditional PoW with register shifts in a \textit{nonlinear feedback shift register} (CNLFSR), and digital signatures by \textit{public physical unclonable function} (PPUF) authentication. When the full node wants to generate the new block of its subsystem, it divides the transactions between the light nodes that it controls, and configures the mining function of the CNLFSR at each node. The light node performs the mining by shifting the CNLFSR register values until the time reaches the predefault threshold. Next, the node uses the PPUF to sign the mining result and sends it to the full node. 

A summary of the solutions discussed in this section is shown in \autoref{table_Crypto3}.

\begin{table}[!t]
\begin{adjustwidth}{-1.2cm}{}
\centering
\caption{Summary of lightweight cryptography - encryption/decryption and hashing solutions.} 
\label{table_Crypto3} 
\centering
\begin{tabular}{|p {0.7 cm}|p {1.9 cm}|p {1.2 cm}|p {5.0 cm}|p {3.5 cm}|p {3.5 cm}|}
\hline
\textbf{Ref.} & \hfil \textbf{Application.} & \hfil \textbf{Year.} & \hfil{\textbf{Main idea}} & \hfil \textbf{Pros} & \hfil \textbf{Cons}  \\ 
\hline 
\hfil \cite{mershad2022proact} & \centering UAV networks & \hfil 2022 & 64-bit ECC (\textit{shortECC}) for encryption; SPONGENT-224 and SPONGENT-88 for hashing & - Low delay \newline- Low storage requirements & - Considerable computation overhead \newline- Limited testing \\
\hline 
\hfil \cite{revanesh2022dag} & \centering WSN networks & \hfil 2022 & Encryption using the XTEA block cipher with Chaotic Map algorithm; Blake-256 algorithm for hashing & - Low encryption and decryption delays \newline- Low energy overhead & - High computation overhead \newline- Scalability not tested \\
\hline 
\hfil \cite{khan2021aechain} & \centering General & \hfil 2021 & ASCON and ASCON-hash are used to encrypt the BC transactions and generate the digital signatures & - High security \newline- Low energy overhead & - Possibility of forking \newline- High communication overhead \\
\hline 
\hfil \cite{guruprakash2020ec} & \centering General & \hfil 2020 & EC-ElGamal with discrete logarithmic operations for encryption and SHA-384 with Genetic Algorithm for hashing & - Good security \newline- Acceptable delay & - Limited testing \newline- No role for resource-constrained nodes in BC \\
\hline 
\hfil \cite{yan2020pcbchain} & \centering General & \hfil 2020 & Register shifts in nonlinear feedback shift register for mining and public physical unclonable function for authentication & - Low energy overhead \newline- Low storage requirements & - Requires dedicated hardware \newline- Single point of failure (full node) \\
\hline 
\end{tabular}
\end{adjustwidth}
\end{table}

\begin{figure}[!t]
\centering
\includegraphics[width=6.5in]{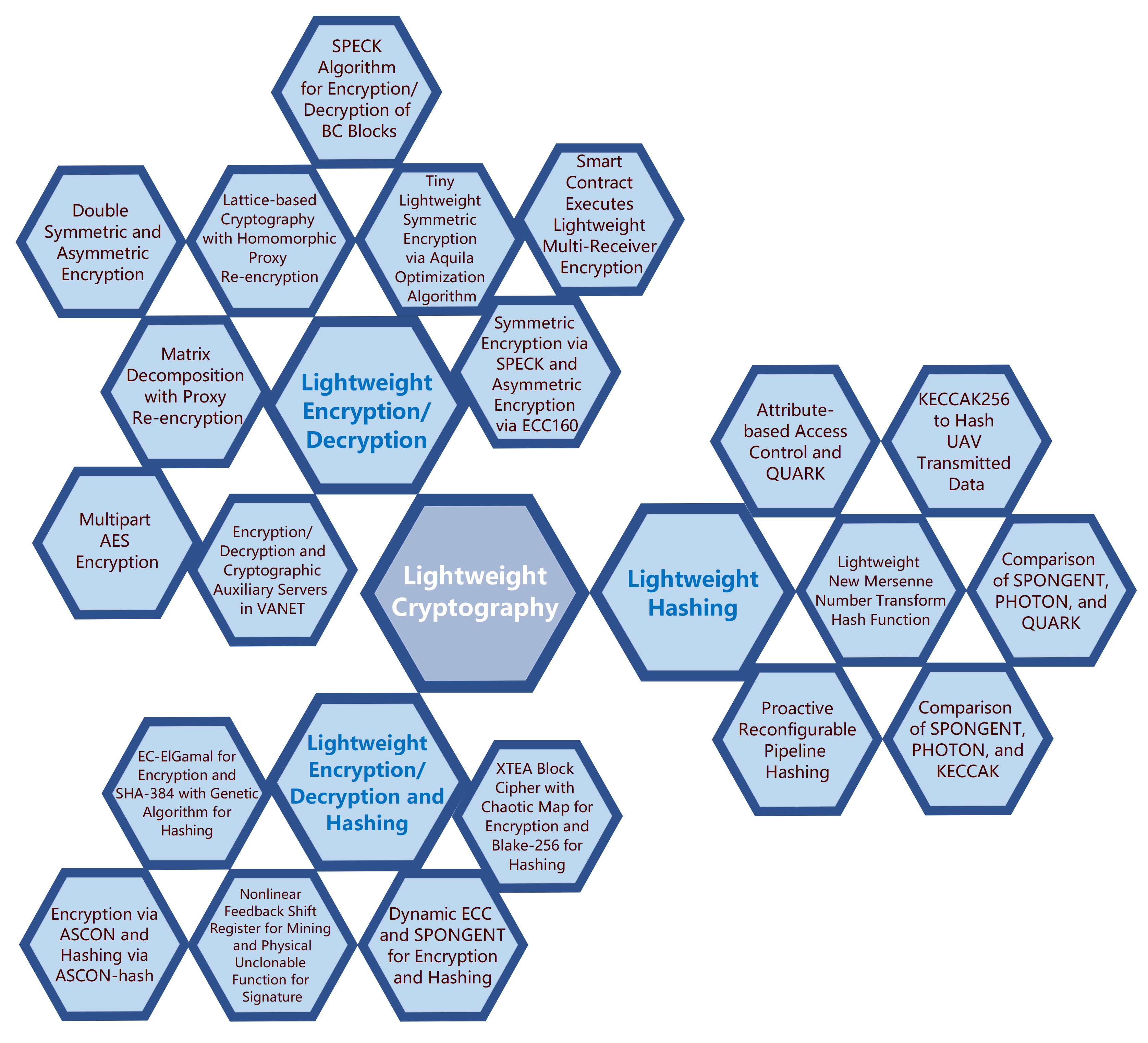}
\caption{Methods for implementing lightweight cryptography in the blockchain.}
\label{fig_Cry}
\end{figure}

An illustration of the main ideas of the lightweight cryptography systems that were discussed in this section is depicted in \autoref{fig_Cry}

\subsection{Analysis and Discussion}

Based on the objectives stated by the authors of the reviewed systems, the requirements of a powerful and efficient lightweight cryptographic system for the BC can be summarized as follows: 

\begin{enumerate}
\item 
\textbf{Security:} the cryptographic model should secure the BC data by ensuring its Integrity, Confidentiality, and Privacy. 
\item 
\textbf{Processing overhead:} resource-constrained nodes should be able to execute/perform the required BC cryptographic operations (encryption, decryption, hashing) with acceptable processing overhead.
\item 
\textbf{Energy overhead:} resource-constrained nodes should be able to execute/perform the required BC cryptographic operations with acceptable energy overhead.
\item 
\textbf{Space requirements:} if hardware-based, the cryptographic model should not require a large space that cannot be supported by limited-size resource-constrained nodes.
\item 
\textbf{Storage requirements:} the cryptographic model should not require an amount of memory space that cannot be supported by memory-limited resource-constrained nodes.
\item 
\textbf{Delay requirements:} the cryptographic model should perform the cryptographic operation within a predefined time limit and achieve a high throughput (processed bits per second) that is acceptable by the BC application. 
\end{enumerate}

\autoref{table_Cry} contains the results of the analysis that we performed on each of the systems that we discussed in this section to discover which of the lightweight cryptography requirements it satisfies. The symbols used in \autoref{table_Cry} are similar to those in the previous tables. 
Additionally, a couple of papers \cite{abed2021analysis, apriani2021performance} did not propose new systems. Rather, they compared the performance of hashing functions. Hence, they were not included in \autoref{table_Cry}.

\begin{table}[!t]
\begin{adjustwidth}{-0.5cm}{}
\centering
\caption{Mapping of lightweight cryptography solutions to the lightweight cryptography requirements.} 
\label{table_Cry} 
\begin{tabular}{|p {0.8 cm}|p {2.2 cm}|p {2.4 cm}|p {2.3 cm}|p {2.3 cm}|p {2.0 cm}|p {2.4 cm}|}
\hline
\textbf{Ref.} & \centering \textbf{Security} & \centering \textbf{Processing overhead} & \centering \textbf{Energy overhead} & \centering \textbf{Space} \textbf{req.} & \centering \textbf{Storage req.} & \centering\arraybackslash \textbf{Delay req.} \\ \hline 
\hfil \cite{avula2023efficient} & \hfil \yes  & \hfil $\downarrow$ & \hfil $\otimes$  & \hfil $\otimes$  & \hfil $\otimes$  & \hfil $\downarrow$ \\ \hline 
\hfil \cite{jnss2023novel} & \hfil $\mathsf{?}$ & \hfil $\downarrow$  & \hfil $\otimes$ & \hfil $\otimes$ & \hfil $\otimes$ & \hfil $\uparrow$ \\ \hline 
\hfil \cite{khashan2023efficient} & \hfil $\uparrow$  & \hfil $\downarrow$ & \hfil $\leftrightarrow$  & \hfil $\otimes$ & \hfil $\downarrow$ & \hfil $\leftrightarrow$ \\ \hline 
\hfil \cite{li2023privacy} & \hfil \yes  & \hfil $\leftrightarrow$ & \hfil $\otimes$ & \hfil $\otimes$ & \hfil $\otimes$ & \hfil $\uparrow$ \\ \hline 
\hfil \cite{hameedi2022improving} & \hfil $\leftrightarrow$ & \hfil $\otimes$ & \hfil $\leftrightarrow$ & \hfil $\otimes$ & \hfil $\leftrightarrow$ & \hfil $\uparrow$ \\ \hline 
\hfil \cite{xu2022vmt} & \hfil \yes & \hfil $\otimes$ & \hfil $\otimes$ & \hfil $\otimes$ & \hfil $\otimes$ & \hfil $\downarrow$ \\ \hline 
\hfil \cite{zhang2022fruit} & \hfil \yes & \hfil $\leftrightarrow$ & \hfil $\otimes$ & \hfil $\otimes$ & \hfil $\otimes$ & \hfil $\downarrow$ \\ \hline 
\hfil \cite{singh2020internet} & \hfil $\mathsf{?}$ & \hfil $\otimes$ & \hfil $\leftrightarrow$ & \hfil $\otimes$ & \hfil $\uparrow$ & \hfil $\uparrow$ \\
\hline
\hfil \cite{dwivedi2019decentralized} & \hfil \yes & \hfil $\downarrow$  & \hfil $\otimes$  & \hfil $\otimes$  & \hfil $\downarrow$  & \hfil $\otimes$ \\
\hline
\hfil \cite{ahamed2022distributed} & \hfil \yes & \hfil $\uparrow$ & \hfil $\otimes$ & \hfil $\otimes$ & \hfil $\uparrow$ & \hfil $\downarrow$ \\ \hline 
\hfil \cite{alshehri2022aac} & \hfil \yes & \hfil $\leftrightarrow$ & \hfil $\otimes$  & \hfil $\otimes$ & \hfil $\leftrightarrow$ & \hfil $\leftrightarrow$ \\ \hline 
\hfil \cite{nabeel2021security} & \hfil $\uparrow$  & \hfil $\leftrightarrow$ & \hfil $\leftrightarrow$ & \hfil $\otimes$ & \hfil $\uparrow$ & \hfil $\downarrow$ \\ \hline 
\hfil \cite{fu2020study} & \hfil \yes & \hfil $\uparrow$ & \hfil $\leftrightarrow$ & \hfil $\downarrow$ & \hfil $\downarrow$ & \hfil $\otimes$ \\ \hline 
\hfil \cite{mershad2022proact} & \hfil $\leftrightarrow$ & \hfil $\otimes$ & \hfil $\leftrightarrow$ & \hfil $\otimes$ & \hfil $\leftrightarrow$ & \hfil $\leftrightarrow$ \\
\hline
\hfil \cite{revanesh2022dag} & \hfil \yes & \hfil $\downarrow$ & \hfil $\uparrow$ & \hfil $\otimes$ & \hfil $\otimes$ & \hfil $\uparrow$  \\ \hline
\hfil \cite{khan2021aechain} & \hfil \yes & \hfil $\leftrightarrow$ & \hfil $\uparrow$ & \hfil $\leftrightarrow$ & \hfil $\leftrightarrow$ & \hfil $\uparrow$ \\
\hline
\hfil \cite{guruprakash2020ec} & \hfil \yes & \hfil $\downarrow$ & \hfil $\otimes$ & \hfil $\otimes$ & \hfil $\downarrow$ & \hfil $\leftrightarrow$ \\
\hline
\hfil \cite{yan2020pcbchain} & \hfil $\leftrightarrow$ & \hfil $\uparrow$ & \hfil $\uparrow$ & \hfil $\downarrow$ & \hfil $\uparrow$ & \hfil $\leftrightarrow$ \\
\hline
\multicolumn{7}{l}{\yes = Satisfied, \no = Not satisfied, $\otimes$=Not studied,  $\uparrow$=High, $\leftrightarrow$=Medium, $\downarrow$=Low, $\mathsf{?}$= Uncertainty } \\
\end{tabular}
\end{adjustwidth}
\end{table}

Among the five “lightweight” properties that we study, cryptography is the least explored within the lightweight BC context. Hence, this field still requires a lot of effort and experimentation to find the lightweight cryptography mechanisms that would be most suitable for integration into the BC when implemented within resource-constrained networks. From the studied papers, we deduce the following: 
\begin{itemize}
    \item 
\textbf{Hardware vs Software Implementation:} We notice that only some of the systems proposed for lightweight BC cryptography are hardware-based \cite{fu2020study, khan2021aechain, yan2020pcbchain}, while the others are software-based. In general, hardware-based implementations produce more efficient and lightweight performance than software ones, as they are optimized to the specified objectives. However, they require extra physical space within the resource-constrained node, which is not always affordable since many applications require the resource-constrained node to have a strict hardware design. Some of the systems proposed heavy cryptographic algorithms that were meant to be executed by 
edge and/or cloud nodes. However, these solutions are not suitable for resource-constrained nodes. 
\item 
\textbf{Modifying Traditional Systems:} Several systems attempted to modify the characteristics of cryptographic solutions to make them suitable for resource-constrained nodes. For example, 
utilizing different security-level schemes based on the security requirements of resource-constrained nodes \cite{jnss2023novel, fu2020study, mershad2022proact} 
However, these solutions require major modifications to the basic BC structure and additional overhead to maintain the security level and the details of each cryptography mechanism that is used. 
\item 
\textbf{Security vs Performance:} Some of the studied systems utilize cryptographic algorithms that are questionable from the security perspective. For example, the system proposed in \cite{singh2020internet} produces good results in terms of throughput and limited various overheads. However, its security is questionable since the SPECK cipher is known to have a too small security margin. Finally, other solutions succeed in strengthening the security of the resource-constrained devices, 
but add a lot of overhead that cannot be afforded by such devices \cite{avula2023efficient, khashan2023efficient, dwivedi2019decentralized}. 
\end{itemize}

In general, we notice that all the lightweight cryptography systems show trade-offs between the specified requirements. For example, some systems 
provide excellent security at the expense of high processing and storage. Other systems 
require low storage and energy consumption but need dedicated hardware space. Hence, we identify and discuss the characteristics of an efficient lightweight cryptography mechanism for resource-constrained networks as follows:

\begin{itemize}
\item 
From the performance perspective, a lightweight cryptography mechanism should require the resource-constrained node to perform limited processing, consume a small amount of energy, and produce its result as fast as possible with high throughput.
\item 
From the implementation perspective, the lightweight cryptography system should require a small amount of memory for its operations, and a small hardware space that can be embedded within a resource-constrained node.
\item 
In terms of security, the lightweight cryptography system should prove its ability to resist various attacks that target cryptosystems such as cryptanalysis, brute force, key-recovery, meet-in-the-middle, etc. 
\end{itemize}

\section{Lightweight Storage}
\label{Sec_LSt}

\begin{figure}[!t]
\centering
\includegraphics[width=6.7in]{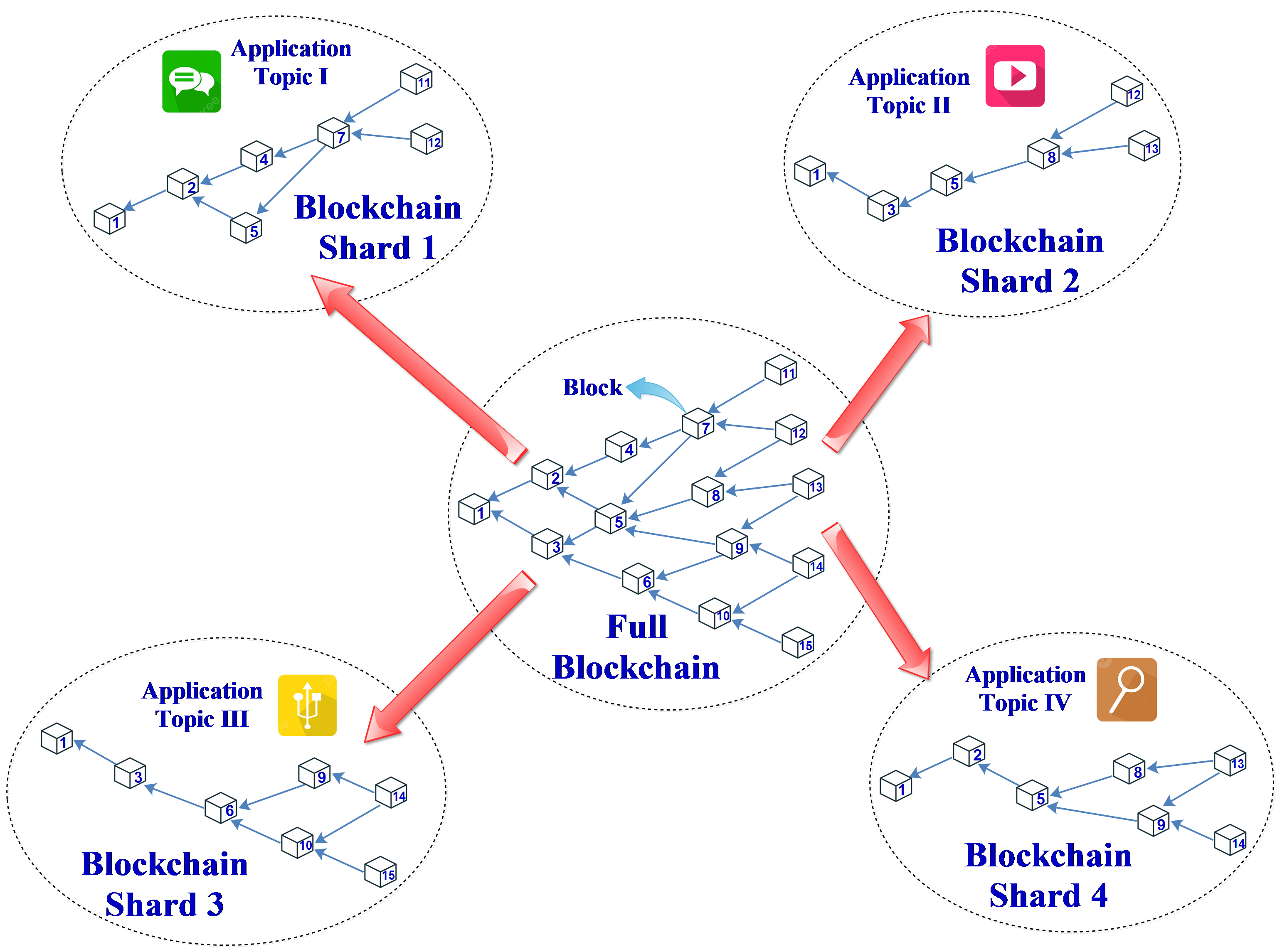}
\caption{Sample lightweight storage mechanism: Dividing the blockchain into shards such that each lightweight node can store specific shards as required by its applications.}
\label{fig_SLSt}
\end{figure}

The last “lightweight” factor that we study is storage, which is very important since several types of resource-constrained nodes can support a very limited amount of data within their storage systems. At the same time, the traditional BC system requires that each node should store the full BC in order to be able to retrieve and validate the BC transactions locally. Since the data produced by resource-constrained nodes can accumulate to become very large in size, storing the full BC becomes infeasible for a lightweight node. Hence, a lightweight storage system should be adopted to solve this issue. After studying the various lightweight storage solutions that have been proposed in the literature, we divide them into three main categories, which can be described as follows:
\begin{itemize}
\item 
\textbf{SPV-based solutions:} these systems adopt the SPV model used in Bitcoin in which resource-constrained nodes store the headers of the BC blocks and the body of specific blocks only \cite{zhao2019novel, fan2022scalable, zhao2022dht, yu2020virtual, liu2019mathsf, tian2019block}.
\item 
\textbf{Application-based solutions:} these systems propose that resource-constrained nodes should store selected BC transactions or blocks based on the application requirements \cite{akrasi2022adaptive, fan2022dlbn, ding2021lightweight, dlimi2021lightweight, singh2020odob, yang2020ldv, qu2018hypergraph}. For example, \autoref{fig_SLSt} illustrates how the blockchain is divided into shards based on the application data, such that each shard is stored by a group of lightweight nodes that implement the corresponding application.
\item 
\textbf{Compression-based solutions:} in these systems, resource-constrained nodes store a compressed version of the BC \cite{yu2023tinyledger, song2022block, wang2022airbc, du2021partitionchain, xu2021slimchain, kim2020selcom, xu2017epbc}.
\end{itemize}

Note that we focus in this section on studying “lightweight storage” systems that maintain the main BC characteristics. A large number of BC-based storage systems modify the BC in such a way that the main BC characteristics, such as immutability, do not exist anymore. We do not include these systems (such as the redactable BC) in our study. In addition, we do not consider systems that are not suitable for resource-constrained devices; for example, systems that apply sharding to divide the BC between the nodes. Although sharding reduces the overall storage requirements, it still requires the node to store part of the BC, which is not possible for lightweight nodes that have very limited storage capacity.

\subsection{SPV-Based Solutions}

Based on the concept of simplified payment verification (SPV) node, which stores the headers of the BC blocks only, the authors of \cite{zhao2019novel} propose the Enhanced SPV (ESPV) node, which stores the full latest blocks, and slices of the old blocks along with their headers. In order to determine the number of newest blocks that should be stored, the authors analyzed the Bitcoin BC and found that around 80\% of the Bitcoin requests target the last 3000 Bitcoin blocks. In addition, they found that the percentage of new transactions that can be verified by using the last 3000 blocks is around 95\%. Hence, the authors propose that each ESPV node can fully keep the last 3000 blocks. 
In addition, each node keeps the headers of old blocks for verification purposes. 
Although the idea proposed in \cite{zhao2019novel} is interesting, it is still impossible for some types of resource-constrained nodes to be ESPV since the storage requirements of the latter are in the terms of tens or few hundreds of GBs, which is usually greater than the storage space of many types of these nodes. The authors of \cite{fan2022scalable} enhance the ESPV model by proposing a block time window (instead of the fixed 3000 blocks) that changes dynamically based on the verification rate of the blocks in the window. Hence, the lightweight node can increase/decrease the window size when it wants to increase/decrease its local verification rate. 

A framework that combines distributed hash table (DHT) and clustering was proposed in \cite{zhao2022dht}. Here, the BC nodes are grouped into \textit{m} DHT clusters, where each cluster contains \textit{n} nodes. The BC transactions are divided between clusters such that the nodes of each cluster store 1/\textit{m} of the data. Within each cluster, each node keeps part of the data allocated to the cluster. Hence, the storage requirement of a node is reduced to \textit{s}/(\textit{m} x \textit{n}), where (\textit{s} - 1) represents the maximum number of faulty nodes. In order to attain efficient querying of transaction data among the clusters, the nodes in each cluster select a master node that is responsible for maintaining the index of data in the cluster. 

A virtual block group (VBG) is created in \cite{yu2020virtual} by combining \textit{n} blocks into a single entity that is stored by a specific set of nodes. When a new VBG is created, a hash value of the VBG is formed using the Merkle tree. The VBG hash value and the ID list of the nodes that store the VBG are added to a DHT in the form of a \textless key, value\textgreater \hspace{0.1em} pair. The DHT is then divided into pieces and distributed among the BC nodes via the Chord algorithm. Hence, each node will store the VBGs that are assigned to it and part of the DHT. Lightweight nodes can select some of the blocks in the VBG to store and depend on the other neighbor nodes (who store the same VBG) to retrieve the blocks that it does not store. 
In addition, all nodes in a VBG cluster store the full VBG metadata. 

Liu \textit{et al.} \cite{liu2019mathsf} propose an Unrelated Block Offloading Filter (UBOF) algorithm that is used to reduce the storage requirements in the LightChain system. The transactions in the BC are divided into two main types. The first type includes transactions that become obsolete when they are replaced with newer transactions. 
The second type includes transactions that become obsolete after a certain deadline. The authors propose two additional BC data structures: the first is \textit{UTXOTable} which is a dictionary that is used to add a new UTXO directly after the output that it references in the dictionary and mark this output in the dictionary as consumed. The second data structure is \textit{UTXONum}, which saves the number of remaining UTXOs in each block. 

A mixed block structure within a private BC is proposed by the authors of \cite{tian2019block}. In this system, each BC node stores some of the blocks and the headers of the remaining blocks. In addition, the authors propose that each block should be stored in a minimum set of nodes to ensure the block's validity and availability. In order to achieve that, the authors propose the Name-based PBFT (NPBFT) algorithm in which each node is assigned an arbitrary name. When a block is created, each node compares the hash of the block with the hashes of the names of all nodes. If the hash of its own name is the nearest to the block hash, it will take the responsibility of generating the next block. 

A summary of the solutions discussed in this section is shown in \autoref{table_Storage1}.

\begin{table}[!t]
\begin{adjustwidth}{-1.2cm}{}
\centering
\caption{Summary of lightweight storage - SPV-based solutions.} 
\label{table_Storage1} 
\centering
\begin{tabular}{|p {0.7 cm}|p {1.9 cm}|p {1.2 cm}|p {5.0 cm}|p {3.5 cm}|p {3.5 cm}|}
\hline
\textbf{Ref.} & \hfil \textbf{Application.} & \hfil \textbf{Year.} & \hfil{\textbf{Main idea}} & \hfil \textbf{Pros} & \hfil \textbf{Cons}  \\ 
\hline 
\hfil \cite{zhao2019novel} & \centering Bitcoin & \hfil 2019 & Enhanced SPV (ESPV) node stores the full latest blocks and slices of the bodies of old blocks  & - Low overhead latency \newline- Better verification than SPV & - Still requires considerable storage \newline- Low to average storage scalability \\
\hline 
\hfil \cite{fan2022scalable} & \centering Bitcoin & \hfil 2022 & Enhancing the ESPV model using a block time window that changes dynamically based on the verification rate of the blocks & - Flexible storage \newline- Storing shards of old blocks & - Additional storage overhead \newline- Poor scalability \\
\hline 
\hfil \cite{zhao2022dht} & \centering General & \hfil 2022 & Nodes in each cluster store part of the BC data; each node store part of the cluster data & - Combining DHT and clustering \newline- High query efficiency & - Requires large network \newline- Delay overhead not tested \\ \hline 
\hfil \cite{yu2020virtual} & \centering General & \hfil 2020 & Hash of virtual block group and IDs of nodes that store it are saved in DHT & - Use of DHT \newline- Acceptable storage overhead & - Low security \newline- Delay overhead not tested \\
\hline 
\hfil \cite{liu2019mathsf} & \centering General & \hfil 2019 & Deleting an obsolete block from local storage: when all transactions in the block expire or are replaced by newer transactions & - Better transaction usability \newline- Low computation overhead & - Still requires considerable storage \newline- Low scalability \\
\hline 
\hfil \cite{tian2019block} & \centering General & \hfil 2019 & Node stores blocks whose hash start with the same bits as the hash of its name & - Good security \newline- Low computation overhead & - Poor decentralization \newline- Requires large network \\
\hline 
\end{tabular}
\end{adjustwidth}
\end{table}

\subsection{Application-Based Solutions}

An adaptive approach for selecting the BC blocks that should be moved from the local storage of a lightweight node to the cloud was proposed in \cite{akrasi2022adaptive}. Here, a deep reinforcement learning (DRL) agent executes on each node and moves the blocks between the cloud and the node’s storage based on the node's needs and the network state in order to create a balance between the query latency and the remaining storage and guarantee optimal BC performance. 
The authors test two DRL algorithms: the Advantage Actor-Critic (A2C) and Proximal Policy Optimization (PPO), to solve the block selection problem. The simulations show that the two algorithms achieve 43\% and 48\% storage reduction. 
However, the computational overhead of the proposed system was not tested.

A double-layer blockchain network (DLBN) in which BC nodes are divided into consensus nodes (CN) and storage nodes (SN) was proposed in \cite{fan2022dlbn}. In the proposed system, network nodes are divided into storage units (SU), where each SU contains a single CN and a group of SNs that divide the BC among themselves and jointly maintain a copy of the complete BC. The nodes within an SU execute consensus and voting algorithms to manage task allocation and role rotation between CN and SN. The block header of the proposed BC is modified to include an index tree that is used by an SN to determine the SNs that store a query result. In addition, the Merkle tree is replaced with a Merkle search tree (MST) that merges the Merkle tree with a balanced binomial search tree. 

A lightweight BC for IoV in which RSUs are clustered into collections was proposed in \cite{ding2021lightweight}. Each collection contains a leader RSU (LRSU) that manages the block allocation within the collection. In the proposed system, vehicles store block headers while each RSU stores part of the BC based on its \textit{History Request Record}, which includes the requested transactions on the RSUs over a specified period of time. After a new block is generated, the LRSU executes the \textit{Block Allocation Algorithm}, which is a \textit{Heat-and-History-Request based Genetic Algorithm} (HHRGA) that takes the data reported by the RSUs in the collection as input and splits the block into several block lists that correspond to different RSU needs. 
The authors show that their system achieves a low latency (average of 60 ms) in a small network of 10 nodes. However, other important parameters, such as storage and communication overhead, energy consumption, and resilience to attacks were not tested.

A TOPIC subscription-based data storage mechanism labeled Assisted Selected Relevant Data in Local Ledger (ASRDLL) was proposed in \cite{dlimi2021lightweight} to decrease the data size of a lightweight node’s local BC. The proposed IoT-SmartChain system defines a roles hierarchy of the lightweight nodes based on their TOPIC subscriptions, which enables the node to save only the data it is interested in. Each transaction contains a TOPICS identifier, which defines the list of topics the transaction belongs to. Each lightweight node saves its interesting context in the configuration module, which includes the topics to which it wishes to subscribe. When the node receives a new block, it keeps the complete content of each transaction that belongs to one of its TOPICs and replaces the other transactions with their IDs, which allows the node to obtain the transaction from other nodes when needed. 

In \cite{singh2020odob}, a BC system for the Internet of Drones is created by decoupling the block header from the body and enabling each drone to save the block headers and the body of its own block only. 
Each time a drone or GCS creates a new transaction, a group of voters is selected to validate it. 
The voters verify the transactions using a consensus model similar to PBFT, then the drone’s block is updated with the new transaction. After updating its block, the drone calculates the new hash and broadcasts it to the network so that all the other drones update their headers’ BC. Meanwhile, each GCS updates its two blockchains. 

Based on the Directed Acyclic Graph (DAG), the authors of \cite{yang2020ldv} propose a lightweight storage method for Vehicular Social Networks (VSNs). In a DAG, transactions are stored as graph vertexes rather than in blocks. Since vehicles in VSNs generate huge amounts of data, the authors propose to divide this data into “topics of interest”, such that each vehicle will join the topics that it selects and store the transactions of these topics only. 
When the vehicle joins a topic, it downloads the latest transactions of that topic from the monitoring node of the topic, which is an edge node that is trusted to maintain all the topic transactions. 

The authors of \cite{qu2018hypergraph} propose using the Hypergraph theory to divide the BC into many subblockchains such that each node can store one or more subblockchains based on its storage capacity and application requirements. In this approach, each subblockchain is created as a hyperedge in the hypergraph. 
When the size of the data in a hyperedge becomes large, an algorithm is executed to split the hyperedge such that nodes can choose to save one of the resulting hyperedges only. Each node can keep track of data stored in other subblockchains by storing a linear independence matrix and a blockchain-list. 

A summary of the solutions discussed in this section is shown in \autoref{table_Storage2}.

\begin{table}[!t]
\begin{adjustwidth}{-1.2cm}{}
\centering
\caption{Summary of lightweight storage - application-based solutions.} 
\label{table_Storage2} 
\centering
\begin{tabular}{|p {0.7 cm}|p {1.9 cm}|p {1.2 cm}|p {5.0 cm}|p {3.5 cm}|p {3.5 cm}|}
\hline
\textbf{Ref.} & \hfil \textbf{Application.} & \hfil \textbf{Year.} & \hfil{\textbf{Main idea}} & \hfil \textbf{Pros} & \hfil \textbf{Cons}  \\ 
\hline 
\hfil \cite{akrasi2022adaptive} & \centering Industrial Internet of Things & \hfil 2022 & DRL agent on each node moves the BC blocks between the cloud and the node’s storage & - Efficient objective functions \newline- DRL algorithms & - Computational overhead not tested \newline- Query latency not tested \\
\hline 
\hfil \cite{fan2022dlbn} & \centering General & \hfil 2022 & BC nodes are divided into storage units that store part of the BC  & - Flexibility in dividing the blocks between SUs & - Storage overhead not tested \newline- High query latency \\
\hline 
\hfil \cite{ding2021lightweight} & \centering Vehicular networks & \hfil 2021 & Vehicles store block headers while each RSU stores part of the BC based on its \textit{History Request Record} & - Efficient HHRGA algorithm \newline- Low latency  & - Many important parameters were not tested \\
\hline 
\hfil \cite{dlimi2021lightweight} & \centering General & \hfil 2021 & IoT nodes are grouped based on their TOPICs subscriptions, and store the data of the TOPICs they are interested in & - Acceptable storage overhead \newline- Use of Topics Balancer Assistant & - Latency overhead not tested \newline- Scalability not tested \\
\hline 
\hfil \cite{singh2020odob} & \centering Internet of Drones & \hfil 2020 & A drone stores the block headers and a complete block that contains its own transactions & - Highly reduced storage \newline- Decoupling block header from body & - Weak decentralization \newline- High latency \\
\hline 
\hfil \cite{yang2020ldv} & \centering Vehicular Social Networks & \hfil 2020 & DAG data are divided into topics of interest; each vehicle stores the transactions of the topics it selects & - Flexible joining/leaving topic \newline- Transaction pruning mechanism & - Storage could still be high \newline- Possibility of malicious consensus \\
\hline 
\hfil \cite{qu2018hypergraph} & \centering General & \hfil 2018 & Each node stores one or more hyperedges based on its storage capacity and application requirements  & - Flexibility in splitting/combining hyperedges & - Storage could still be high \newline- High communication overhead \\
\hline 
\end{tabular}
\end{adjustwidth}
\end{table}

\subsection{Compression-Based Solutions}

Transaction data in the BC block are compressed into a succinct proof using the zero-knowledge proof algorithm in \cite{yu2023tinyledger}. Through the zero-knowledge proof algorithm, the block producer can compress the transaction data in the block into a succinct zero-knowledge proof that is used by the other nodes to ensure that the updated BC state is correct without storing old transaction data, which greatly reduces the storage requirements of lightweight nodes. 
The block of the proposed BC, TinyLedger, contains the height, hash, and timestamp of the block, the proof of consensus, the Merkle root of the global state, the Merkle root of the block transactions, the storage path of the transactions in the off-chain storage system, and the state transition proof. 

An approach for compressing the Bitcoin BC based on UTXO aggregation was proposed in \cite{song2022block}. Instead of block file pruning that is adopted in Bitcoin, the authors propose a method for periodically aggregating the UTXOs of the same user into a single UTXO that contains the full balance of the user. Each period of time, called an epoch, the miner that produces the new block performs the aggregation operation to produce an aggregation block that is broadcast to all nodes. Here, the miner aggregates the UTXOs of each user into a single UTXO by using a new Bitcoin script called pay-to-witness script-hash (P2WSH). This script enables the miner to collect the signatures of the user from the aggregated UTXOs and merge them using its signature into the output UTXO to prove the integrity of the aggregation process. 

A lightweight storage model that uses elimination and compression to decrease the storage overhead for UAV nodes was proposed in \cite{wang2022airbc}. In the proposed system, a UAV can be either a full or light node. Here, light nodes save only partial blocks based on two criteria: 1) expiration time; where the block creator specifies an expiration time for the block, and 2) reference count; where each transaction in the block can reference other transactions in old blocks. When all the transactions in a block have been referenced by newer transactions, the body of the block is deleted. 
The authors show that their elimination and compression technique reduces the total storage by 63\%, but at the expense of reduced security. 

Based on the BFT-Store concept, the authors in \cite{du2021partitionchain} propose the PartitionChain system. BFT-Store utilizes Reed-Solomon (RS) coding and the PBFT protocol to decrease the storage requirement per block from \textit{O}(\textit{n}) to \textit{O}(1) in a network of \textit{n} nodes. This is achieved by making each node stores chunks of encoded blocks. BFT-Store also implements an online re-encoding process that is executed when a node joins or leaves the network. In PartitionChain, the computational complexity is reduced by partitioning a block into smaller pieces before it is encoded, hence decreasing the block decoding complexity. 
When partitioning, a block is divided into \textit{r} primary-pieces, then each piece is further partitioned into smaller second-level-pieces. After partitioning, a large prime number \textit{p} is used to map each small piece into \textit{n} specific evaluations, and then distributed among the \textit{n} nodes. 

Xu \textit{et al.} \cite{xu2021slimchain} propose SlimChain, in which the short commitments of ledger states are maintained on-chain, while the stateful data is saved off-chain. In this system, transactions are replaced with their corresponding digests in the block body. In addition, the BC world states and Merkle tries are also stored off-chain, while the root hash of each Merkle trie is saved on-chain for validation purposes. SlimChain utilizes off-chain transaction execution, in which smart contracts are executed by off-chain storage nodes that provide proof of their execution integrity via the Intel SGX trusted execution environment (TEE). 
The experiments show that SlimChain can reduce the storage requirements at lightweight nodes by 97\%. However, the system latency is very high (can reach up to 10s), which is not suitable for many resource-constrained network applications.

The authors of \cite{kim2020selcom} propose compression and selection of blocks as a solution for lightweight nodes to store the BC. In the proposed system, two chains are present: main chain, which is stored by full nodes, and checkpoint chain, which is used by lightweight nodes to store part of the main chain and checkpoints, which represent the compression of the main chain at different instances. Each time a new block is created, it is added to the two chains. When the number of the most recent blocks at the lightweight node becomes equal to \textit{n\textsubscript{c}}, the node compresses these blocks using a block Merkle tree (BMT) scheme that hashes blocks instead of transactions and generates a checkpoint block. 
Next, the lightweight node keeps only the interesting blocks among the \textit{n\textsubscript{c}} blocks (which are the blocks that it needs to use) and deletes the remaining blocks. 

Xu \textit{et al.} \cite{xu2017epbc} propose a system in which a lightweight node stores a compressed summary of the BC. The summary of all the previous blocks is calculated and added to the block header each time a new block is created. A lightweight node needs to store only the latest summary of the BC and update it with each new block in order to verify any transaction/block that it obtains from the full nodes. 
The summary is created using a cryptography accumulator based on the RSA algorithm. 
When a lightweight node obtains a block and needs to verify it, it selects a random group of full nodes and requests the \textit{block proof} from them. When the lightweight node receives the \textit{block proof}, it validates it using the cryptography accumulator theory. 

A summary of the solutions discussed in this section is shown in \autoref{table_Storage3}.

\begin{table}[!t]
\begin{adjustwidth}{-1.2cm}{}
\centering
\caption{Summary of lightweight storage - compression-based solutions.} 
\label{table_Storage3} 
\centering
\begin{tabular}{|p {0.7 cm}|p {1.9 cm}|p {1.2 cm}|p {5.0 cm}|p {3.5 cm}|p {3.5 cm}|}
\hline
\textbf{Ref.} & \hfil \textbf{Application.} & \hfil \textbf{Year.} & \hfil{\textbf{Main idea}} & \hfil \textbf{Pros} & \hfil \textbf{Cons}  \\ 
\hline 
\hfil \cite{yu2023tinyledger} & \centering Multi-access edge computing & \hfil 2023 & Transactions in the block are compressed into a succinct proof using the zero-knowledge proof algorithm & - Efficient compression scheme \newline- High throughput & - High proof time \newline- Computational overhead not studied \\
\hline 
\hfil \cite{song2022block} & \centering Bitcoin & \hfil 2022 & The UTXOs of each user are periodically aggregated into a single UTXO via a dedicated Bitcoin script & - Aggregation block scheme \newline- High reduction in number of UTXOs & - Storage could still be high \newline- Query latency not studied \\
\hline 
\hfil \cite{wang2022airbc} & \centering UAV networks & \hfil 2022 & Lightnodes delete a block in which all transactions have expired or have been referenced by newer transactions & - Co-elimination strategy \newline- Low transaction latency & - Query delay overhead not studied \newline- Questionable security \\
\hline 
\hfil \cite{du2021partitionchain} & \centering General & \hfil 2021 & BC block is partitioned into smaller pieces before it is encoded using Reed-Solomon; each node stores chunks of encoded blocks & - Low storage overhead \newline- Low re-initialization time & - High computational overhead and query latency \\
\hline 
\hfil \cite{xu2021slimchain} & \centering General & \hfil 2021 & Transactions are saved off-chain and replaced with their digests in the block body & - Off-chain transaction execution \newline- Low storage overhead & - High transaction latency \newline- Requires dedicated hardware \\
\hline 
\hfil \cite{kim2020selcom} & \centering General & \hfil 2020 & Checkpoints of the main chain are periodically created by compressing the BC blocks via the block Merkle tree scheme & - Good compression scheme \newline- Selection of interesting blocks & - Storage could still be high \newline- Possibility of weak decentralization \\
\hline 
\hfil \cite{xu2017epbc} & \centering General & \hfil 2017 & A cryptography accumulator creates a summary of the BC blocks that is added to the new block header & - Very lightweight storage \newline- Efficient transaction verification & - Communication and delay overhead \newline- Questionable security \\
\hline 
\end{tabular}
\end{adjustwidth}
\end{table}

\begin{figure}[!t]
\centering
\includegraphics[width=6.5in]{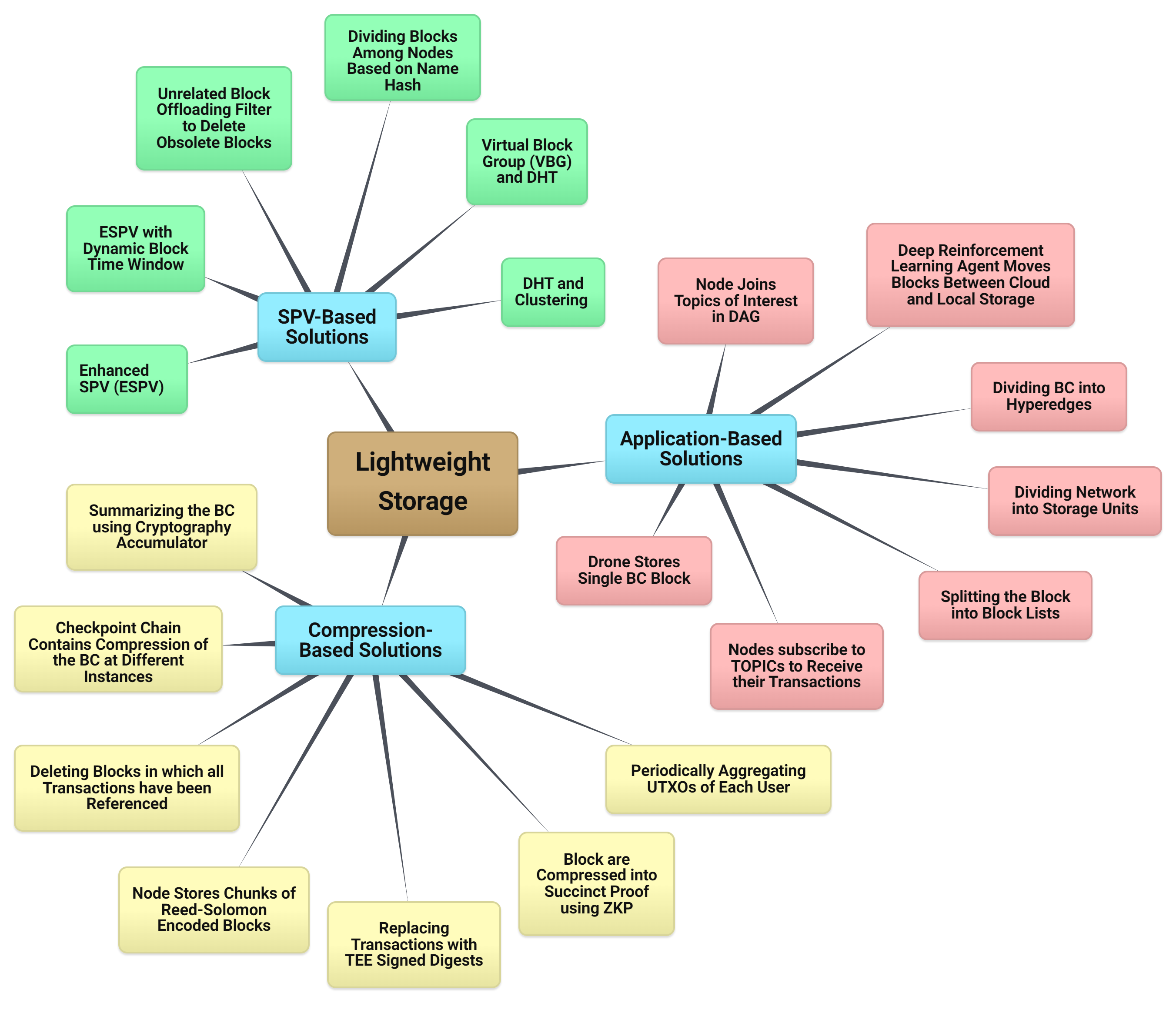}
\caption{Methods for lightweight storage in the blockchain.}
\label{fig_St}
\end{figure}

\autoref{fig_St} illustrates the main idea of each of the lightweight storage approaches that we studied in this section.

\subsection{Analysis and Discussion}

Different systems for reducing the size of BC data at the resource-constrained nodes have been proposed. 
The requirements that were deduced from the studied systems for an efficient and effective lightweight storage system are as follows:
\begin{enumerate}
\item 
\textbf{Applicability:} The lightweight node should be able to store the BC or BC-related data that enables it to perform its role in the utilized application, without negatively affecting its normal operations.
\item 
\textbf{Storage scalability:} The system should be storage scalable, i.e., the node should be able to store new BC blocks or BC-related data as the BC size increases, without negatively affecting its normal operations.
\item 
\textbf{Data validation:} The lightweight node should be able, by using the lightweight storage system, to obtain and verify the BC transactions that it needs in an efficient manner. 
\item 
\textbf{Overhead:} The lightweight storage system should not add significant communication, latency, or processing overhead to the lightweight node. 
\item 
\textbf{Decentralization:} The lightweight storage system should keep the network highly decentralized by maintaining sufficient redundancy of the BC data. 
\item 
\textbf{Availability:} The lightweight storage system should guarantee that all the BC data is available all the time. 
\item 
\textbf{Security:} The lightweight storage system should keep the BC data and communications secure against various BC attacks (such as data tampering, long range, collusion, and DoS-based attacks). In other words, attackers should not be able to exploit vulnerabilities in the storage mechanism to launch attacks on the BC. 
\end{enumerate}

Similar to the previous sections, we first study the lightweight storage requirements that each of these systems satisfies (\autoref{table_St}), then we discuss the missing aspects in the literature and the possible future advancements to this category. 

Note that in \autoref{table_St}, the symbol “$\bigCI$” is used with the \textit{Applicability} requirement to indicate that the corresponding system requires the existence of resource-constrained nodes that have a storage capacity of at least tens of GBs in order for the system to be applicable. Also, the symbol “\CheckedBox” indicates that the system is applicable for a certain application only. The other symbols in the table are similar to those used in the previous sections.

\begin{table}[!t]
\begin{adjustwidth}{-0.9cm}{}
\centering 
\caption{Mapping of lightweight storage solutions to the lightweight storage requirements.} 
\label{table_St} 
\begin{tabular}{|p {0.9 cm}|p {2.4 cm}|p {1.6 cm}|p {2.2 cm}|p {2.0 cm}|p {1.6 cm}|p {2.2 cm}|p {1.9 cm}|}
\hline
\textbf{Ref.} & \centering \textbf{Appl.} & \centering \textbf{Scal.} & \centering \textbf{Data Validation} & \centering \textbf{Overhead} &  \centering \textbf{Decent.} & \centering \textbf{Avail.} &\centering\arraybackslash \textbf{Security} \\ \hline 
\hfil \cite{zhao2019novel} & \centering $\bigCI$ & \centering $\leftrightarrow$ & \hfil \yes & \centering $\downarrow$ & \hfil \yes & \hfil \yes & \centering\arraybackslash \halfcheckmark \\
\hline
\hfil \cite{fan2022scalable} & \centering $\bigCI$  & \centering $\downarrow$ & \hfil \yes & \centering $\downarrow$  & \hfil \yes  & \hfil \yes & \centering\arraybackslash \halfcheckmark \\
\hline
\hfil \cite{zhao2022dht} & \centering \yes & \centering $\leftrightarrow$ & \hfil $\otimes$ & \centering $\downarrow$ & \hfil $\downarrow$ & \hfil $\leftrightarrow$  & \centering\arraybackslash \halfcheckmark \\
\hline
\hfil \cite{yu2020virtual} & \centering $\bigCI$  & \hfil $\downarrow$  & \hfil \yes  & \centering $\downarrow$  & \hfil \yes  & \hfil \yes  & \hfil \yes \\
\hline 
\hfil \cite{liu2019mathsf} & \centering $\bigCI$  & \hfil $\downarrow$  & \hfil \yes  & \centering $\leftrightarrow$  & \hfil \yes  & \hfil \yes  & \centering\arraybackslash \halfcheckmark \\
\hline 
\hfil \cite{tian2019block} & \centering $\bigCI$  & \hfil $\downarrow$  & \hfil \yes  & \centering $\leftrightarrow$  & \hfil $\downarrow$  & \hfil \yes  & \centering\arraybackslash \yes  \\
\hline 
\hfil \cite{akrasi2022adaptive} & \centering $\bigCI$  & \centering \yes & \hfil $\otimes$ & \centering $\downarrow$ & \hfil $\otimes$ & \hfil $\downarrow$ & \centering\arraybackslash $\otimes$ \\
\hline
\hfil \cite{fan2022dlbn} & \centering \yes & \centering $\downarrow$ & \hfil $\otimes$  & \centering $\otimes$  & \hfil $\downarrow$ & \hfil $\leftrightarrow$ & \centering\arraybackslash \halfcheckmark \\
\hline
\hfil \cite{ding2021lightweight} & \centering $\bigCI$ & \centering \yes & \hfil $\otimes$ & \centering $\otimes$ & \hfil $\leftrightarrow$ & \hfil $\leftrightarrow$ & \centering\arraybackslash $\otimes$ \\
\hline
\hfil \cite{dlimi2021lightweight} & \centering \yes & \centering $\downarrow$ & \hfil \no & \centering $\otimes$ & \hfil $\leftrightarrow$ & \hfil \yes & \centering\arraybackslash $\otimes$ \\
\hline
\hfil \cite{singh2020odob} & \hfil \yes  & \hfil \yes  & \centering $\otimes$  & \centering $\downarrow$  & \hfil \no  &\centering $\mathsf{?}$  & \centering\arraybackslash \halfcheckmark  \\
\hline 
\hfil \cite{yang2020ldv} & \centering \CheckedBox & \hfil \yes  & \hfil \yes  & \centering $\leftrightarrow$  & \centering $\downarrow$  & \centering $\leftrightarrow$  & \hfil \yes  \\
\hline 
\hfil \cite{qu2018hypergraph} & \centering $\bigCI$  & \hfil $\downarrow$  & \centering $\otimes$  & \hfil $\downarrow$  & \centering $\leftrightarrow$  & \hfil \yes  & \centering\arraybackslash \halfcheckmark  \\
\hline 
\hfil \cite{yu2023tinyledger} & \centering \yes & \centering \yes & \hfil \yes & \centering $\downarrow$ & \hfil $\downarrow$ & \hfil \yes & \centering\arraybackslash $\otimes$ \\
\hline
\hfil \cite{song2022block} & \centering $\bigCI$ & \centering $\leftrightarrow$ & \hfil \yes & \centering $\otimes$ & \hfil \yes & \hfil $\leftrightarrow$ & \centering\arraybackslash $\otimes$ \\
\hline
\hfil \cite{wang2022airbc} & \centering $\bigCI$ & \centering $\downarrow$ & \hfil $\otimes$ & \centering $\otimes$ & \hfil $\downarrow$ & \hfil $\leftrightarrow$ & \centering\arraybackslash \halfcheckmark \\
\hline
\hfil \cite{du2021partitionchain} & \centering \yes & \centering \yes & \hfil  \yes & \centering $\downarrow$ & \hfil $\otimes$ & \hfil $\downarrow$ & \centering\arraybackslash \halfcheckmark \\
\hline
\hfil \cite{xu2021slimchain} & \centering $\downarrow$  & \centering \yes & \hfil \yes & \centering $\downarrow$ & \hfil $\downarrow$ & \hfil \yes & \centering\arraybackslash $\otimes$ \\
\hline
\hfil \cite{kim2020selcom} & \hfil \yes  & \centering $\leftrightarrow$  & \hfil \yes  & \centering $\leftrightarrow$  & \centering $\downarrow$  & \centering $\downarrow$  & \centering\arraybackslash $\otimes$  \\
\hline 
\hfil \cite{xu2017epbc} & \hfil \yes  & \hfil \yes  & \hfil \yes  & \hfil $\downarrow$  & \hfil \no  & \hfil \no  & \centering\arraybackslash \halfcheckmark  \\
\hline 
\multicolumn{8}{l}{\small Appl. = Applicability, Scal. = Scalability, Decent.= Decentralization, Avail. = Availability} \\
\multicolumn{8}{l}{\yes = Satisfied, \no = Not satisfied, \halfcheckmark = Partially satisfied, \CheckedBox= Satisfied only for certain application} \\
\multicolumn{8}{l}{$\otimes$=Not studied,  $\uparrow$=High, $\leftrightarrow$=Medium, $\downarrow$=Low, $\mathsf{?}$= Uncertainty, $\bigCI$ = High storage capacity needed } \\

\end{tabular}
\end{adjustwidth}
\end{table}

After examining the various lightweight BC storage systems that have been proposed in the literature, we notice the following:
\begin{itemize}
    \item 
\textbf{Limited Storage:} Several systems reduce the storage requirements of traditional BC, but cannot be implemented in many resource-constrained devices. The reason is that most applications that utilize such devices (such as healthcare, grid, agriculture, intelligent transportation system (ITS), manufacturing, etc.) produce vast amounts of continuous data that will be saved in the BC. Assuming that resource-constrained nodes can store part of the BC data in such applications is not realistic, since the size of data produced could be in the TB scale. Hence, resource-constrained nodes will not be able to store part of the BC, as many of the studied systems 
assume. On the other hand, some of these systems propose that the resource-constrained node save a very limited part of the BC that contains its own data only (such as a single block). 
However, such systems remove one of the essential properties of the BC, which is 
decentralization, since the BC becomes stored on a limited number of nodes only (such as the cloud servers or edge/fog nodes). 
\item 
\textbf{Compression Limitations:} Other systems that apply various compression techniques to reduce the size of the BC data or transactions at the resource-constrained node assume that the latter can obtain the original data securely from a full BC node and verify them via the local compressed data. However, the security aspect was not studied very well in most of these systems. In addition, these approaches pose significant communication and processing overhead on the resource-constrained node. 
\item 
\textbf{Scalability Issues:} In some applications, old data in the BC becomes obsolete. In such cases,  
the resource-constrained node can delete this data and keep only the verification information that is required to validate the history of a transaction. However, such an approach suffers from the scalability problem: as the BC grows indefinitely, the storage capacity of the resource-constrained node will not be able to handle the verification information. 

\item 
\textbf{Blockchain Decentralization:} Most of the lightweight storage systems assume that the resource-constrained node depends on the full BC nodes to retrieve missing blocks. 
In some of these systems, the node keeps verification data (such as block headers) that enable it to verify the integrity of the obtained block locally. On the other hand, other systems assume that the resource-constrained node will obtain the verification data securely from other nodes. 
However, applying such approaches weakens the decentralization of BC data in the network, and makes the resource-constrained nodes dependable on the full BC nodes to acquire the BC data. If resource-constrained nodes need to access the BC very frequently, this could lead to significant communication overhead between the resource-constrained network and the full nodes, which could create security problems and lead to attacks (such as DoS). For example, an attacker could block the wireless connection between the resource-constrained node and the full node by continuously jamming the channel, which will disrupt the resource-constrained network operations. Even with no attacks, the BC access latency could be very high in situations where all resource-constrained nodes access the full nodes very frequently. 
\end{itemize}

Hence, a lightweight storage mechanism should attempt to resolve the trade-off between local storage amount from one side and overhead and security from the other side by considering the following factors:
\begin{itemize}
\item 
If the resource-constrained node has sufficient storage capacity and the deployed application does not produce large and frequent amounts of data, the resource-constrained node can store part of the BC data. 
However, if the application produces large amounts of data, then the node should store only a small part of the BC data according to its storage capacity and application needs, and depend on full nodes to obtain and verify the missing data. 
Here, scalability plays an important factor in the efficiency of the adopted storage mechanism, which should contain a method to keep the storage space at the node within a certain limit as the size of the BC increases. In addition, the system should guarantee that the resource-constrained node will obtain the missing data from the full nodes in a secure and efficient manner. 
\item 
A lightweight storage system should impose limited overhead on the resource-constrained node in terms of processing and communication. If the storage system will require the node to execute heavy computations 
or participate in a lot of communications, 
then the performance of the node could degrade in the long term. In addition, the storage system should ensure that the delay encountered by the node to obtain BC data from full nodes is acceptable by the deployed application. 
\item 
One of the most important properties of BC is decentralization. The BC data should be stored by a large number of nodes to ensure that an attacker will not be able to execute an attack and create a malicious branch of the BC. If few cloud/edge nodes store the full copy of the BC, while resource-constrained nodes store limited parts of it, the decentralization factor becomes weak, and an attacker could create serious problems by compromising more than half of the full nodes. Hence, a lightweight storage solution should keep the BC highly decentralized. In addition, the system should keep copies of the full BC distributed within the resource-constrained network in such a way that ensures that all the data is available to the resource-constrained nodes at all times, in case the connection to the full nodes fails for any reason. 
\item 
Finally, a lightweight storage system should prove to be immune to various BC attacks, such as data tampering, 51\%, collusion, etc. 
In addition, the lightweight storage system should not contain vulnerabilities that allow attackers to launch new attacks. For example, many compression-based systems 
do not discuss how the BC summary at resource-constrained node will be secured against attacks, since a successful attack on the BC summary will compromise the BC application at the node.  
\end{itemize}

\section{Summary and Future Directions}

The “lightweight blockchain” has emerged as one of the most important research topics in the field of resource-constrained network security. In order to be lightweight, the BC system should be less resource-demanding from several perspectives. In this paper, we identified five main categories that should be adjusted in a traditional BC framework in order to make it suitable for resource-constrained systems. 
A large number of papers have proposed lightweight BC systems by considering one of these categories only. Very few systems considered two or more of these categories in their lightweight BC models. We believe that a lightweight BC should be built by considering each of these five categories, designing and implementing a solution that satisfies the lightweight requirements of the category, and integrating the methods of the five solutions together. Hence, we identified at the beginning of each section the requirements to achieve a lightweight BC solution in that category, and analyzed the corresponding BC models that have been proposed so far. Based on our analysis of what has been achieved so far in each category, we highlighted the missing/weak aspects that should be focused on in future research in order to create an efficient and secure lightweight BC system for resource-constrained networks. By combining our inferences from the five categories, we believe that the following points are the most important for future research directions in the lightweight BC field.

First, we believe that the “Lightweight Architecture” and “Lightweight Storage” features are closely linked and researchers should focus on designing solutions that integrate these two lightweight aspects together. 
Such a solution can adopt the successful features from the “lightweight architecture” systems that were discussed in \autoref{Sec_LAr}, and integrate into them a storage mechanism that allows the resource-constrained node to store part of the BC data and securely verify the remaining data that it obtains from full BC nodes. For example, a mixed vertical and horizontal splitting approach can be applied to divide the main BC into multiple sub-blockchains based on the deployed application, such that resource-constrained nodes store only the sub-blockchains that contain the data they require. In addition, a clustering approach can be applied to group each set of resource-constrained nodes together, such that each cluster stores part of each sub-blockchain using one of the mechanisms that were studied in Section \ref{Sec_LSt}. 

In order to secure the BC network from malicious nodes, a lightweight authentication system should be implemented to register the light node in the BC, associate its pseudonym with its real identity while keeping this information private to authorities, and provide it with cryptographic credentials. Each time the light node joins or rejoins the network (for example, after a sleep period), it executes a BC smart contract to authenticate itself and obtain a new session key that it uses to securely communicate with the CH or with another light node. In addition, the session key is renewed and a new session is opened if the session extends for a long period. Such an authentication mechanism should use a lightweight cryptographic scheme to reduce the light node’s resource consumption. Exploiting the lightweight cryptography algorithm that the node will use for encryption and hashing will avoid the need to install/deploy multiple cryptographic systems within the node. Finally, the lightweight authentication system should be designed such that it does not require a lot of communication and results in fast authentication, i.e., the authentication delay is minimized. 

With respect to consensus, we suggest that a lightweight consensus protocol for resource-constrained networks should exploit the existence of the light nodes to assign to them simple consensus tasks that do not require a lot of resources such that the correctness and decentralization of the consensus process are enhanced. However, light nodes should not be assigned heavy mining tasks that require a lot of resources such as solving a hash puzzle. For example, 
a lightweight consensus algorithm can be built by having each CH in the clustered resource-constrained network 
periodically generate a block that contains the transactions of the light nodes within the cluster. Instead of generating the blocks sequentially, as proposed in the literature, 
blocks can be generated in parallel by the CHs and added to the local BC according to a specific order that is predefined by a block orderer. 
Each time a new block is generated by a CH, it is broadcast to the 
full BC nodes in the edge/cloud. In order to speed up the consensus process and reduce the communication overhead, a “Hierarchical Consensus” approach, similar to \cite{de2022hierarchical}, can be applied. For example, consensus can be executed in parallel between the CHs as a set and the edge/cloud nodes as another set, with the two sets of nodes exchanging the consensus results when they are ready and agreeing on a final consensus decision. Hence, each group of nodes (such as CHs, edge servers, or cloud servers) can form a consensus set and execute the consensus algorithm within the set, while set leaders exchange the consensus results of the sets. Such a model should increase the consensus correctness since a larger number of nodes participate in the consensus process, reduce the delay and increase the throughput since the blocks are produced and agreed on in parallel, and increase the consensus scalability by splitting the consensus nodes in a set into multiple sets if their number highly increases. As explained before, such a mechanism could lead to new types of consensus attacks. Hence, it should be carefully analyzed to discover the vulnerabilities that it contains and implement the corresponding solutions.

Finally, we discuss the future directions that should guide the research on lightweight cryptography for the BC. As highlighted in \autoref{Sec_LCr}, it remains a challenge to fabricate a tiny size hardware circuit that implements a lightweight cryptography algorithm to encrypt/decrypt/hash data with limited overhead (i.e., small delay, memory usage, and power consumption) and high performance. A hardware-based lightweight cryptography solution for resource-constrained nodes should take into consideration that many types of these nodes have strict size constraints. 
Future lightweight cryptography research should focus on designing hardware implementations of lightweight cryptography algorithms that can be fabricated within the boards of such nodes, while ensuring high performance and excellent security, as discussed in \autoref{Sec_LCr}. In addition, we believe that several other aspects of lightweight cryptography for the BC should be tackled in the near future to advance this field of research. The most important among these issues are: 

\begin{itemize}
\item 
\textbf{Unified Cryptosystem:} Among the “lightweight cryptography” mechanisms that have been proposed so far, there is no system that can be utilized as a unified lightweight cryptography solution for all of the BC cryptographic operations: authentication, digital signing, and encryption/decryption of transactions. Future BC systems for resource-constrained networks should consider utilizing a single lightweight cryptography solution for all three operations. 
\item 
\textbf{Post-quantum lightweight cryptography:} Securing the BC operations in a quantum world should be considered by researching the implementation aspects of lightweight post-quantum cryptographic primitives to make them ready for deployment on resource-constrained devices. 
A survey of the post-quantum cryptography techniques for resource-constrained networks was presented in \cite{kumari2022post}. 
\item 
\textbf{Need for cryptography standards for the lightweight BC:} Systems proposed so far use various cryptography models that are efficient in certain applications but not suitable for others. There is a need for a standard cryptography model for the lightweight BC that would be suitable for most resource-constrained applications.
\item 
\textbf{Lightweight algorithms:} A large number of lightweight cryptographic models have been proposed in recent years, such as PRESENT \cite{bogdanov2007present}, LBlock \cite{wu2011lblock}, LED \cite{guo2011led}, HIGHT \cite{hong2006hight}, SPARKLE \cite{beierle2020lightweight}, Romulus \cite{iwata2020duel}, etc. Very few of these algorithms have been tested in the lightweight BC context (such as \cite{khan2021aechain}). Future works should focus on comparing the performance of these algorithms when applied within the lightweight BC to assess their suitability for various resource-constrained applications.
\end{itemize}

\section{Conclusion}

Resource-constrained devices are those that by design have restricted storage and processing capabilities, and execute their tasks with minimal power input while staying cost-effective. Due to their regular utilization as smart equipment deployed in several fields, in harsh environments or/and difficult-to-reach locations, they typically operate on batteries that are designed to keep the balance between the device lifetime and the cost of device replacement. Many challenges can obstruct the successful deployment of an application within a set of resource-constrained devices, including interoperability, power/processing capabilities, scalability, availability, and most importantly, security. 

Safeguarding resource-constrained devices and the networks to which they connect can be a difficult task due to the diversity of devices and providers, the complexity of resource provisioning, and the necessity of building reliable communications. A large number of security mechanisms have been proposed and implemented for resource-constrained networks, each of which has its own characteristics and merits. Among these solutions, the blockchain stands as a framework that can endow several benefits to these networks, such as data immutability, high decentralization, and shared network management. However, the high resource demands of the BC become a hurdle to its deployment in resource-constrained environments. Hence, lightweight BC solutions must be considered in order to make the BC applicable for resource-constrained devices.

In this paper, we surveyed the various lightweight blockchains systems that have been proposed in the last years. We identified five categories within the lightweight BC, which are architecture, authentication, consensus, cryptography, and storage. For each category, we outlined the lightweight requirements that were outlined in the literature work related to that category. The lightweight requirements of each category define the characteristics and conditions that the BC system should satisfy in order to be considered lightweight in that category. Also for each category, we analyzed in depth each of the lightweight BC systems that have been proposed so far and highlighted the requirement that it satisfies. 
Finally, we discussed the features and qualities that are needed for building a future BC system that is lightweight from the category perspective. In the last section, we summarized our findings and laid the foundation for a comprehensive future BC system that is lightweight in terms of the requirements of all five categories. Our findings and suggestions will open the way for future research that enhances the current BC systems for resource-constrained networks and reach more mature BC solutions that can be deployed for a large variety of resource-constrained applications.

\bibliographystyle{elsarticle-num}
\bibliography{main}

\end{document}